\pdfoutput=1
\documentclass{lmcs}
\pdfoutput=1
\usepackage[utf8]{inputenc}

\usepackage{lastpage}
\lmcsdoi{19}{4}{17}
\lmcsheading{}{\pageref{LastPage}}{}{}%
{Oct.~18,~2022}{Nov.~29,~2023}{}


\usepackage{hyperref}

\usepackage{tikz}
\usetikzlibrary{positioning, automata}

\usepackage{amssymb}
\usepackage{booktabs}
\usepackage[labelfont=normal,labelformat=simple]{subcaption}
\usepackage{csquotes}
\usepackage{forest}

\RequirePackage{ifthen}
\RequirePackage{xparse}
\RequirePackage{xcolor}
\RequirePackage{xspace}
\RequirePackage{etoolbox}
\RequirePackage{tabularx}
\RequirePackage{ltablex}
\RequirePackage{amsmath}
\RequirePackage{pgffor}

\provideboolean{PackageBabelAvailable}
\ifcsdef{bbl@loaded}{\setboolean{PackageBabelAvailable}{true}}{}

\ifbool{PackageBabelAvailable}{%
	\RequirePackage{translations}%
}{}

\makeatletter             %
\newcommand{\providecounter}[1]{%
	\ifcsdef{c@#1}%
		{\setcounter{#1}{0}}%
		{\newcounter{#1}}%
}

\provideboolean{@NNUseRTL}
\provideboolean{@NNDetailed}
\provideboolean{@NNInListMode}
\provideboolean{@NNPrintGroup}
\providecounter{@NNCurrentGroup}
\providecounter{@NNNumberOfAllGroups}
\providecounter{@NNNumberOfEntriesInGroup0}%
\providecounter{@NNNumberOfAllEntries}%

\relax

\ifbool{PackageBabelAvailable}{%
	\DeclareTranslationFallback{Table of notation}{Table of notation}%
	\DeclareTranslation{German}{Table of notation}{Notationstabelle}%
	\DeclareTranslationFallback{symbol}{symbol}%
	\DeclareTranslation{German}{symbol}{Symbol}%
	\DeclareTranslationFallback{macro name}{macro name}%
	\DeclareTranslation{German}{macro name}{Makroname}%
	\DeclareTranslationFallback{uses}{uses}%
	\DeclareTranslation{German}{uses}{Verw.}%
	\DeclareTranslationFallback{description}{description}%
	\DeclareTranslation{German}{description}{Beschreibung}%
	\DeclareTranslationFallback{no entries}{no entries}%
	\DeclareTranslation{German}{no entries}{keine Einträge}%
}{}

\ifbool{PackageBabelAvailable}%
	{\newcommand{\gettranslation}[1]{\GetTranslation{#1}}}%
	{\newcommand{\gettranslation}[1]{#1}}

\newcommand{\@NNLAYOUTArgument}[1]{\textcolor{blue}{#1}}
\newcommand{\@NNLAYOUTZeroUses}[1]{\textcolor{red}{#1}}
\newcommand{\@NNLAYOUTColumnHeader}[1]{\textbf{#1}}
\newcommand{\@NNLAYOUTTable}[1]{%
	\notbool{@NNUseRTL}%
		{{\Large\textbf{#1}}\hrulefill}%
		{\hrulefill{\Large\textbf{#1}}}%
}
\newcommand{\@NNLAYOUTGroup}[1]{%
	\notbool{@NNUseRTL}%
		{{\large\textbf{#1}}\dotfill}%
		{{\dotfill{\large\textbf{#1}}}}%
}

\newcommand{\@NNDefaultArgumentA}{1}
\newcommand{\@NNDefaultArgumentB}{2}
\newcommand{\@NNDefaultArgumentC}{3}
\newcommand{\@NNDefaultArgumentD}{4}
\newcommand{\@NNDefaultArgumentE}{5}
\newcommand{\@NNDefaultArgumentF}{6}
\newcommand{\@NNDefaultArgumentG}{7}
\newcommand{\@NNDefaultArgumentH}{8}
\newcommand{\@NNDefaultArgumentI}{9}
\let\@NNArgumentA\@NNDefaultArgumentA
\let\@NNArgumentB\@NNDefaultArgumentB
\let\@NNArgumentC\@NNDefaultArgumentC
\let\@NNArgumentD\@NNDefaultArgumentD
\let\@NNArgumentE\@NNDefaultArgumentE
\let\@NNArgumentF\@NNDefaultArgumentF
\let\@NNArgumentG\@NNDefaultArgumentG
\let\@NNArgumentH\@NNDefaultArgumentH
\let\@NNArgumentI\@NNDefaultArgumentI

\newcommand{\@NNShowIf}[3]{\ifthenelse{#2 > #3}{}{{\@NNLAYOUTArgument{#1}}}}

\newcommand{\@NNPrintTableHeader}{%
	\notbool{@NNDetailed}%
	{%
		\notbool{@NNUseRTL}%
		{%
			\@NNLAYOUTColumnHeader{\gettranslation{symbol}} &%
			\@NNLAYOUTColumnHeader{\gettranslation{description}} \\%
		}%
		{%
			\@NNLAYOUTColumnHeader{\gettranslation{description}} & %
			\@NNLAYOUTColumnHeader{\gettranslation{symbol}} \\ %
		}%
	}%
	{%
		\notbool{@NNUseRTL}%
		{%
				\@NNLAYOUTColumnHeader{\gettranslation{symbol}} &%
				\@NNLAYOUTColumnHeader{\gettranslation{macro name}} &%
				\@NNLAYOUTColumnHeader{\gettranslation{uses}} &%
				\@NNLAYOUTColumnHeader{\gettranslation{description}}\\%
		}%
		{%
				\@NNLAYOUTColumnHeader{\gettranslation{description}}&%
				\@NNLAYOUTColumnHeader{\gettranslation{uses}} &%
				\@NNLAYOUTColumnHeader{\gettranslation{macro name}} &%
				\@NNLAYOUTColumnHeader{\gettranslation{symbol}} \\%
		}%
	}%
}

\newcommand{\@NNPrintCommand}[1]{%
	\letcs{\NNnum}{@NNArgumentsOfEntry#1}%
	\ensuremath{%
		\csname @NNCommandOfEntry#1\endcsname%
		{\@NNShowIf{\@NNArgumentA}{1}{\NNnum}}%
		{\@NNShowIf{\@NNArgumentB}{2}{\NNnum}}%
		{\@NNShowIf{\@NNArgumentC}{3}{\NNnum}}%
		{\@NNShowIf{\@NNArgumentD}{4}{\NNnum}}%
		{\@NNShowIf{\@NNArgumentE}{5}{\NNnum}}%
		{\@NNShowIf{\@NNArgumentF}{6}{\NNnum}}%
		{\@NNShowIf{\@NNArgumentG}{7}{\NNnum}}%
		{\@NNShowIf{\@NNArgumentH}{8}{\NNnum}}%
		{\@NNShowIf{\@NNArgumentI}{9}{\NNnum}}%
	}%
}

\newcommand{\@NNPrintCommandName}[1]{%
	\texttt{%
		\csname @NNMacroNameOfEntry#1\endcsname%
		\expandafter\ifthenelse{\csname @NNArgumentsOfEntry#1\endcsname < 1}{}{%
			[\csname @NNArgumentsOfEntry#1\endcsname]}%
		}%
}

\newcommand{\@NNPrintEntry}[1]{%
	\@nameuse{@NNArgDescsOfEntry#1}%
	\notbool{@NNDetailed}{%
		\notbool{@NNUseRTL}%
			{\@NNPrintCommand{#1} & \csname @NNDescriptionOfEntry#1\endcsname \\}%
			{\csname @NNDescriptionOfEntry#1\endcsname & \@NNPrintCommand{#1} \\}%
	}%
	{%
		\notbool{@NNUseRTL}%
		{%
			\@NNPrintCommand{#1} & %
			\@NNPrintCommandName{#1} & %
			\csname @NNNumberOfUsesOfEntry#1\endcsname & %
			\csname @NNDescriptionOfEntry#1\endcsname \\ %
		}%
		{%
			\csname @NNDescriptionOfEntry#1\endcsname & %
			\csname @NNNumberOfUsesOfEntry#1\endcsname & %
			\@NNPrintCommandName{#1} & %
			\@NNPrintCommand{#1} \\ %
		}%
	}%
}

\newcommand{\@NNPrintGroup}[1]{%
	\setboolean{@NNPrintGroup}{true}
	\ifbool{@NNDefaultGroup#1}%
		{%
			\ifthenelse{\expandafter\value{@NNNumberOfEntriesInGroup#1} < 1}%
				{\setboolean{@NNPrintGroup}{false}}%
				{}%
		}%
		{%
			\noindent\expandafter\@NNLAYOUTGroup{\csname @NNNameOfGroup#1\endcsname}%
			\ifthenelse{\expandafter\value{@NNNumberOfEntriesInGroup#1} < 1}%
				{\setboolean{@NNPrintGroup}{false}\gettranslation{no entries}}%
				{}%
		}%
	\ifbool{@NNPrintGroup}%
	{%
		\keepXColumns%
		\ifbool{@NNDetailed}{%
			\notbool{@NNUseRTL}%
				{\begin{tabularx}{\textwidth}{lllX}}%
				{\begin{tabularx}{\textwidth}{Xlll}}%
		}%
		{%
			\notbool{@NNUseRTL}%
				{\begin{tabularx}{\textwidth}{lX}}%
				{\begin{tabularx}{\textwidth}{Xl}}%
		}%
		\@NNPrintTableHeader %
		\forlistcsloop{\@NNPrintEntry}{@NNEntriesInGroup#1} %
		\end{tabularx}%
	}{}%
}

\newcommand{\@NNGroups}{}

\newcommand{\@NNNewGroup}[3]{%
	\stepcounter{@NNNumberOfAllGroups}%
	\setcounter{@NNCurrentGroup}{\value{@NNNumberOfAllGroups}}%
	\listeadd{\@NNGroups}{\the@NNCurrentGroup}%
	\csedef{@NNNameOfGroup\the@NNCurrentGroup}{#1}%
	\csedef{@NNOutputLevelOfGroup\the@NNCurrentGroup}{#3}%
	\expandafter\provideboolean{@NNDefaultGroup\the@NNCurrentGroup}%
	\expandafter\setboolean{@NNDefaultGroup\the@NNCurrentGroup}{#2}%
	\expandafter\providecounter{@NNNumberOfEntriesInGroup\the@NNCurrentGroup}%
}

\newcommand{\@NNPrepareEntry}[1]{%
	\stepcounter{@NNNumberOfAllEntries}%
	\stepcounter{@NNNumberOfEntriesInGroup\the@NNCurrentGroup}%
	\providecounter{@NNCounter#1}%
	\csdef{@NNNumberOfUsesOfEntry\the@NNNumberOfAllEntries}%
	{%
		\ifthenelse{\value{@NNCounter#1} > 0 }%
			{\arabic{@NNCounter#1}}%
			{\@NNLAYOUTZeroUses{\arabic{@NNCounter#1}}}%
	}%
}

\newcommand{\@NNAddEntry}[5]{%
	\listcseadd{@NNEntriesInGroup\the@NNCurrentGroup}{\the@NNNumberOfAllEntries}%
	\csedef{@NNCommandOfEntry\the@NNNumberOfAllEntries}{#1}%
	\csedef{@NNMacroNameOfEntry\the@NNNumberOfAllEntries}{#2}%
	\csedef{@NNArgumentsOfEntry\the@NNNumberOfAllEntries}{#3}%
	\csedef{@NNDescriptionOfEntry\the@NNNumberOfAllEntries}{#4}%
	\csdef{@NNArgDescsOfEntry\the@NNNumberOfAllEntries}{#5}%
}

\newcommand{\notationSetLayoutTable}[1]{\renewcommand{\@NNLAYOUTTable}[1]{#1}}
\newcommand{\notationSetLayoutGroup}[1]{\renewcommand{\@NNLAYOUTGroup}[1]{#1}}
\newcommand{\notationSetLayoutColumnHeader}[1]{\renewcommand{\@NNLAYOUTColumnHeader}[1]{#1}}
\newcommand{\notationSetLayoutArgument}[1]{\renewcommand{\@NNLAYOUTArgument}[1]{#1}}
\newcommand{\notationSetLayoutZeroUses}[1]{\renewcommand{\@NNLAYOUTZeroUses}[1]{#1}}

\newcommand{\notationarg}[2]{%
	\ifthenelse{#1 = 1}{\renewcommand{\@NNArgumentA}{#2}}{%
	\ifthenelse{#1 = 2}{\renewcommand{\@NNArgumentB}{#2}}{%
	\ifthenelse{#1 = 3}{\renewcommand{\@NNArgumentC}{#2}}{%
	\ifthenelse{#1 = 4}{\renewcommand{\@NNArgumentD}{#2}}{%
	\ifthenelse{#1 = 5}{\renewcommand{\@NNArgumentE}{#2}}{%
	\ifthenelse{#1 = 6}{\renewcommand{\@NNArgumentF}{#2}}{%
	\ifthenelse{#1 = 7}{\renewcommand{\@NNArgumentG}{#2}}{%
	\ifthenelse{#1 = 8}{\renewcommand{\@NNArgumentH}{#2}}{%
	\ifthenelse{#1 = 9}{\renewcommand{\@NNArgumentI}{#2}}%
		{\PackageWarning{newnotation}{notationarg: Invalid argument number}{Must be an integer between 1 and 9.}}%
	}}}}}}}}%
}

\DeclareDocumentCommand{\tableofnotation}{O{0}}{%
	\setboolean{@NNInListMode}{true}%
	{%
		
		\bigskip\noindent%
		\@NNLAYOUTTable{\gettranslation{Table of notation}}

	}%
	\renewcommand*{\do}[1]{%
		\ifthenelse{\@nameuse{@NNOutputLevelOfGroup##1} < #1}{}{%
			\@NNPrintGroup{##1}%
		}%
	}%
	\dolistloop{\@NNGroups}%
	\setboolean{@NNInListMode}{false}%
}

\DeclareDocumentCommand{\detailedtableofnotation}{O{0}}{%
	\setboolean{@NNDetailed}{true}%
	\tableofnotation[#1]%
	\setboolean{@NNDetailed}{false}%
}

\DeclareDocumentCommand{\notationnewgroup}{O{0} m}{%
	\@NNNewGroup{#2}{false}{#1}%
}

\newcommand{\notationnewtable}{%
	\renewcommand{\@NNGroups}{}%
	\@NNNewGroup{}{true}{0}%
}

\notationnewtable %

\newcommand{\notationsavetable}[1]{%
	\csedef{@NNTableCurrentGroup#1}{\the@NNCurrentGroup}%
	\forlistloop{\listcsadd{@NNTable#1}}{\@NNGroups}%
}

\newcommand{\notationloadtable}[1]{%
	\ifcsundef{@NNTable#1}{\PackageError{newnotation}{notationloadtable: Unknown table '#1'.}{Misspelled name?}}{}%
	\notationnewtable%
	\expandafter\setcounter{@NNCurrentGroup}{\csname @NNTableCurrentGroup#1\endcsname}%
	\forlistcsloop{\listcsadd{@NNGroups}}{@NNTable#1}%
}

\newcommand{\newnotationclass}[2]{%
	\csdef{@NNClass#1}{#2}%
}

\DeclareDocumentCommand{\newnotation}{s o m O{0} o m O{} O{}}{%
	\ifdef{#3}%
		{\PackageError{newnotation}{newnotation: Command '\string#3' already defined.}{Misspelled name?}}{}%
	\@NNPrepareEntry{\string#3}%
	\IfNoValueTF{#5}{%
		\newrobustcmd{#3}[#4]{%
			\IfNoValueTF{#2}%
				{\IfBooleanTF{#1}%
					{\ensuremath{#6}}%
					{\ensuremath{{#6}}}}%
				{\IfBooleanTF{#1}%
					{\ensuremath{\@nameuse{@NNClass#2}{#6}}}%
					{\ensuremath{{\@nameuse{@NNClass#2}{#6}}}}}%
			\ifthenelse{\boolean{@NNInListMode}}{}{\protect\stepcounter{@NNCounter\string#3}}%
			\xspace%
		}%
	}{%
	\newrobustcmd{#3}[#4][#5]{%
			\IfNoValueTF{#2}%
				{\IfBooleanTF{#1}%
					{\ensuremath{#6}}%
					{\ensuremath{{#6}}}}%
				{\IfBooleanTF{#1}%
					{\ensuremath{\@nameuse{@NNClass#2}{#6}}}%
					{\ensuremath{{\@nameuse{@NNClass#2}{#6}}}}}%
			\ifthenelse{\boolean{@NNInListMode}}{}{\protect\stepcounter{@NNCounter\string#3}}%
			\xspace%
		}%
	}%
	\@NNAddEntry{#3}{\string#3}{#4}{#7}{#8}%
}
\DeclareDocumentCommand{\newrelation}{s o m O{0} o m O{} O{}}{%
	\ifdef{#3}%
		{\PackageError{newnotation}{newrelation: Command '\string#3' already defined.}{Misspelled name?}}{}%
	\@NNPrepareEntry{\string#3}%
	\IfNoValueTF{#5}{%
		\newrobustcmd{#3}[#4]{%
			\IfNoValueTF{#2}%
				{\IfBooleanTF{#1}%
					{\ensuremath{\mathrel{#6}}}%
					{\ensuremath{\mathrel{{#6}}}}}%
				{\IfBooleanTF{#1}%
					{\ensuremath{\mathrel{\@nameuse{@NNClass#2}{#6}}}}%
					{\ensuremath{\mathrel{\@nameuse{@NNClass#2}{#6}}}}}%
			\ifthenelse{\boolean{@NNInListMode}}{}{\protect\stepcounter{@NNCounter\string#3}}%
			\xspace%
		}%
	}{%
	\newrobustcmd{#3}[#4][#5]{%
			\IfNoValueTF{#2}%
				{\IfBooleanTF{#1}%
					{\ensuremath{\mathrel{#6}}}%
					{\ensuremath{\mathrel{{#6}}}}}%
				{\IfBooleanTF{#1}%
					{\ensuremath{\mathrel{\@nameuse{@NNClass#2}{#6}}}}%
					{\ensuremath{\mathrel{\@nameuse{@NNClass#2}{#6}}}}}%
			\ifthenelse{\boolean{@NNInListMode}}{}{\protect\stepcounter{@NNCounter\string#3}}%
			\xspace%
		}%
	}%
	\@NNAddEntry{#3}{\string#3}{#4}{#7}{#8}%
}

\DeclareDocumentCommand{\renewnotation}{s o m O{0} o m O{} O{}}{%
	\ifundef{#3}%
		{\PackageError{newnotation}{renewnotation: Command '\string#2' not defined.}%
		{Misspelled name?}}{}%
	\@NNPrepareEntry{\string#3}%
	\IfNoValueTF{#5}{%
		\renewrobustcmd{#3}[#4]{%
			\IfNoValueTF{#2}%
				{\IfBooleanTF{#1}%
					{\ensuremath{#6}}%
					{\ensuremath{{#6}}}}%
				{\IfBooleanTF{#1}%
					{\ensuremath{\@nameuse{@NNClass#2}{#6}}}%
					{\ensuremath{{\@nameuse{@NNClass#2}{#6}}}}}%
			\ifthenelse{\boolean{@NNInListMode}}{}{\protect\stepcounter{@NNCounter\string#3}}%
			\xspace%
		}%
	}{%
	\renewrobustcmd{#3}[#4][#5]{%
			\IfNoValueTF{#2}%
				{\IfBooleanTF{#1}%
					{\ensuremath{#6}}%
					{\ensuremath{{#6}}}}%
				{\IfBooleanTF{#1}%
					{\ensuremath{\@nameuse{@NNClass#2}{#6}}}%
					{\ensuremath{{\@nameuse{@NNClass#2}{#6}}}}}%
			\ifthenelse{\boolean{@NNInListMode}}{}{\protect\stepcounter{@NNCounter\string#3}}%
			\xspace%
		}%
	}%
	\@NNAddEntry{#3}{\string#3}{#4}{#7}{#8}%
}

\makeatother            %
\newcommand{\atm}{\ensuremath{A}}
\newnotation{\valuation}{\vartheta}
\newrelation{\partto}{\rightharpoonup}

\newrelation{\join}{\bowtie}

\newcommand{\mathsc}[1]{{\normalfont\textsc{#1}}}

\newcommand{\plus}{\scalebox{0.6}{$+$}}

\newnotationclass{queryclass}{\mathsf}
\newnotationclass{cal}{\mathcal}
\newnotationclass{bb}{\mathbb}
\newnotationclass{bf}{\mathbf}
\newnotationclass{cop}{\textnormal}
\newnotationclass{sf}{\mathsf}
\newnotationclass{decision-problem}{\mathsc}
\newnotationclass{complexity-class}{\mathsc}

\notationnewgroup{Basics}
\newnotation[cal]{\calA}{A}
\newnotation[cal]{\calB}{B}
\newnotation[cal]{\calC}{C}
\newnotation[cal]{\calV}{V}
\newnotation[cal]{\calW}{W}

\newnotation[bf]{\bfx}{x}

\newnotation[bb]{\N}{N}
\newnotation{\NN}{\N_0}

\notationnewgroup{Query Classes}
\newnotation[queryclass]{\CQ}{CQ}
\newnotation[queryclass]{\ACQ}{ACQ}
\newnotation[queryclass]{\HCQ}{HCQ}
\newnotation[queryclass]{\QHCQ}{QHCQ}
\newnotation[queryclass]{\CCQ}{CCQ}
\newnotation[bb]{\ViewClass}{V}
\newnotation[bb]{\QueryClass}{Q}
\newnotation[bb]{\RewritingClass}{R}

\notationnewgroup{Decision Problems}
\newnotation{\rewrprob}[3]{\mathsc{Rewr}(#1,#2,#3)}
\newnotation{\rewrprobk}[4][k]{\mathsc{Rewr}^{#1}(#2,#3,#4)}
\newnotation[decision-problem]{\coverdescprob}[2]{CovDesc(#1, #2)}
\newnotation[decision-problem]{\selprojequiv}[1]{SelProjEquiv(#1)}

\notationnewgroup{Query Operators}
\newnotation[cop]{\atoms}{atoms}
\newnotation[cop]{\vars}{vars}
\newnotation[cop]{\head}{head}
\newnotation[cop]{\body}{body}
\newnotation[cop]{\ar}{ar}
\newnotation[cop]{\bvars}{bvars}

\newnotation[sf]{\dom}{dom}[set of database entries]
\newnotation[cal]{\schema}{S}
\newnotation[sf]{\var}{var}

\newnotation{\varsof}[1]{\vars(#1)}
\newnotation{\headof}[1]{\head(#1)}
\newnotation{\bodyof}[1]{\body(#1)}
\newnotation{\bvarsof}[1]{\bvars(#1)}
\newnotation{\bvarsofrel}[2][Q]{\bvars_{#1}(#2)}

\newnotation[cop]{\id}{id} %

\newnotation{\graphof}[1]{G(#1)}

\newnotation{\homi}[3]{h_{#1}\colon {#2} \rightarrow {#3}}

\newnotation[sf]{\canonicalrewr}{canon}
\newnotation{\candof}[1]{\canonicalrewr(#1)}

\notationnewgroup{Complexity Classes}
\newnotation[complexity-class]{\classP}{P}
\newnotation[complexity-class]{\classNP}{NP}
\newnotation[complexity-class]{\classcoNP}{coNP}

\notationnewgroup{From Gaetano}
\newnotationclass{op}{\mathsf}
\newnotationclass{cal}{\mathcal}
\newnotationclass{bar}{\bar}
\newnotationclass{rel}{\texttt}
\newnotationclass{problem}{\mathsc}

\newrelation{\contained}{\sqsubseteq}

\newnotation{\Q}{Q}
\newnotation{\R}{R}
\newnotation{\V}{V}
\newnotation{\app}{\alpha}

\newnotation[cal]{\ClauseAtoms}{C}
\newnotation[cal]{\VarAtoms}{X}
\newnotation[cal]{\NegAtoms}{N}
\newnotation{\ClauseAtomSet}[2][]{\ClauseAtoms^{#1}_{\text{#2}}}
\newnotation{\VarAtomSet}[2][]{\VarAtoms^{#1}_{\text{#2}}}
\newnotation{\NegAtomSet}[2][]{\NegAtoms^{#1}_{\text{#2}}}

\newnotation{\formula}{f}
\newnotation[bar]{\nx}{x}
\newnotation{\lit}[1]{\ell_{#1}}
\newnotation{\wfalse}{w_0}
\newnotation{\wtrue}{w_1}
\newnotation{\wproj}{u}
\newnotation{\Clauses}[1]{\ClauseAtomSet{#1}}
\newnotation{\ClausesFalse}{\Clauses{0}}
\newnotation{\ClausesTrue}{\Clauses{1}}
\newnotation{\Neg}[1]{\NegAtomSet{$#1$}}
\newnotation{\VarNeg}[1]{\VarAtomSet{$#1$}}
\newnotation{\LitPos}[1]{\VarAtomSet[+]{$#1$}}
\newnotation{\LitNeg}[1]{\VarAtomSet[-]{$#1$}}
\newnotation{\tass}{\beta}

\newnotation[rel]{\RelForm}{Form}
\newnotation[rel]{\RelClause}{C}
\newnotation[rel]{\RelNeg}{Neg}

\newnotation[problem]{\ThreeSat}{3SAT}
\forestset{
	variable tree/.style={
		child anchor=north,
	},
	query/.style={
		rectangle,
		draw,
		child anchor=north,
	},
	subquery/.style={
		parent anchor=south,
		child anchor=north,
		inner sep=1pt,
		regular polygon,
		regular polygon sides=3,
	},
}

\newlength{\targetwidth}
\newcommand{\makesamewidth}[3][c]{%
  \settowidth{\targetwidth}{#2}%
  \makebox[\targetwidth][#1]{#3}%
}
\newlength{\targetheight}
\newcommand{\makesamewidthandheight}[4][c]{%
  \settoheight{\targetheight}{#3}%
	\makesamewidth[#1]{#2}{\vbox to \targetheight{\vfil}#4}%
}

\newcommand{\consistent}{consistent\xspace}
\newcommand{\consistency}{consistency\xspace}
\newcommand{\covergraph}{cover graph\xspace}

\newlist{enumdesc}{enumerate}{1}
\setlist[enumdesc]{
	left=0pt .. \parindent,
	itemindent=*,
	align=left,
	labelwidth=*,
	labelsep=2mm,
	font=\normalfont\bfseries,
	label={Case~\arabic*:},
	ref={{Case~\arabic*}},
}

\begin{document}

\title{Rewriting with Acyclic Queries: Mind Your Head}
\titlecomment{An abridged conference version of this
  paper without full proofs and with less detailed explanations has been published in the proceedings of the 25th International Conference on Database Theory \cite{geck_et_al:LIPIcs.ICDT.2022.8}.
Besides \autoref{result:equivalence-expansion}, the results in this version and the conference version are identical.}

\author[G.~Geck]{Gaetano Geck\lmcsorcid{0000-0002-8946-9440}}

\author[J.~Keppeler]{Jens Keppeler}

\author[T.~Schwentick]{Thomas Schwentick\lmcsorcid{0000-0002-1062-922X}}

\author[C.~Spinrath]{Christopher Spinrath}

\address{TU Dortmund University, Germany}
\email{\{gaetano.geck, jens.keppeler, thomas.schwentick, christopher.spinrath\}@tu-dortmund.de}

\begin{abstract}
The paper studies the rewriting problem, that is, the decision
problem whether, for a given 
conjunctive query~$Q$ and a set~$\mathcal{V}$ of views, there is a
conjunctive query~$Q'$ over~$\mathcal{V}$ that is equivalent to~$Q$,
for cases where the query, the views, and/or the desired
rewriting are acyclic or even more restricted.

It shows that, if~$Q$ itself is acyclic, an acyclic rewriting
exists if there is any rewriting. An analogous statement also holds for
free-connex acyclic, hierarchical, and q-hierarchical queries.

Regarding  the complexity of the rewriting problem, the paper identifies a
border between tractable and (presumably) intractable variants of the
rewriting problem: for schemas of bounded arity, the acyclic rewriting problem is
$\mathsc{NP}$-hard, even if both~$Q$ and the views in~$\mathcal{V}$ are
acyclic or hierarchical. However, it becomes tractable if the views are free-connex acyclic
(i.e., in a nutshell, their body is (i) acyclic and (ii) remains acyclic if their head is
added as an additional atom).
 \end{abstract}

\maketitle

\section{Introduction}
The   \emph{query rewriting problem} asks, for a
query~\Q and a set~\calV of views, whether there is a
query~$\Q'$ over~\calV that is equivalent to~\Q, and to find such a \emph{rewriting}  $\Q'$.

We emphasise that in the literature, various notions of rewriting are studied and that the one above is sometimes called \emph{exact} \cite{DBLP:conf/icdt/CalvaneseGLV05} or \emph{equivalent} \cite{DBLP:series/synthesis/2019Afrati}. Since we are in this paper exclusively interested in exact rewritings, we use the term rewriting in the sense stated above.

Over the last decades, query rewriting has been studied in a wide range of settings with varying database models (relational, XML, graph, \dots), query languages, semantics (set, bag, \dots) and more. A particular well-known setting is formed by conjunctive queries ($\CQ$) under set semantics, which captures some core aspects of SQL queries. This is also the setting studied in this paper.
We refer to \cite{DBLP:series/synthesis/2019Afrati,DBLP:journals/ftdb/ChirkovaY12,DBLP:journals/vldb/Halevy01} for an overview of the extensive literature.

In this paper we are interested in a modified version of the query rewriting problem which does not merely ask for \emph{any} rewriting but rather for rewritings that are structurally simple and allow for efficient evaluation. This is relevant in a setting where access to the database is only possible through the views, in which case the use of an efficient rewriting is obvious.
It can also be interesting in a scenario where multiple queries have to be evaluated: by evaluating some of the queries and considering their results as views, some other queries might be rewritable into acyclic queries over the database extended by these views, and therefore they could possibly be evaluated more efficiently.

The forms of structural simplicity that we consider in this paper are acyclic ($\ACQ$) and hierarchical queries ($\HCQ$), and their slightly stronger versions free-connex acyclic ($\CCQ$) and q-hierarchical queries ($\QHCQ$).
	For a brief discussion of the origin and applications of these classes, we refer to \autoref{section:relationship-characterization}.

We are interested in two kinds of questions.
\begin{enumerate}[(1)]
\item Under which circumstances is it guaranteed that a structurally simple rewriting exists, if  there exists a rewriting at all?\label{question:intro-a}
\item What is the complexity to decide whether such a rewriting exists and to compute one?\label{question:intro-b}
\end{enumerate}
We study these questions depending on the structure of the given views and the given query and we consider the same simplicity notions.

In particular, we study the decision problem $\rewrprob{\ViewClass}{\QueryClass}{\RewritingClass}$ that asks whether for a given set of views from $\ViewClass$ and  a query from $\QueryClass$, there is a rewriting from $\RewritingClass$, for various classes $\ViewClass$, $\QueryClass$ and $\RewritingClass$, with an emphasis on the case where $\RewritingClass$ is $\ACQ$ or a subclass. In the following we refer to this case as the \emph{acyclic rewriting problem}.
	To the best of our knowledge, there has been no previous work dedicated to the study of questions~\ref{question:intro-a} and~\ref{question:intro-b}.

The answer to Question~\ref{question:intro-a} turns out to be very simple and also quite encouraging: in the case that $\QueryClass=\ACQ$, whenever a rewriting exists, there is also an acyclic rewriting. And the same is true for the three subclasses of \ACQ.
	That is, for every query in $\CCQ$, $\HCQ$, or $\QHCQ$, whenever a rewriting exists, there is also a rewriting in $\CCQ$, $\HCQ$, or $\QHCQ$, respectively.
	Thus, in these cases a rewriting with efficient evaluation exists, if there is a rewriting at all.
	This answer to Question~\ref{question:intro-a} simplifies the study of Question~\ref{question:intro-b}, since it implies that for $\QueryClass\in \{\ACQ,\CCQ,\HCQ,\QHCQ\}$ and $\ViewClass\subseteq\CQ$ the decision problems $\rewrprob{\ViewClass}{\QueryClass}{\CQ}$ and $\rewrprob{\ViewClass}{\QueryClass}{\QueryClass}$---and, thus, their complexities---coincide.

	The study of Question~\ref{question:intro-b} reveals that the complexity of the acyclic rewriting problem
may depend on two parameters: the arity of the underlying schema and the arity of the views. We denote the restriction of  $\rewrprob{\ViewClass}{\QueryClass}{\RewritingClass}$ to database schemas of arity at most $k$ by $\rewrprobk{\ViewClass}{\QueryClass}{\RewritingClass}$, and we indicate by $\ViewClass^k$ if the arity of views is at most $k$.

\begin{table}
	\caption{%
		Complexity results for the (acyclic) rewriting problem.
		Note that $\QHCQ\subseteq \CCQ \subseteq \ACQ$ and $\QHCQ\subseteq \HCQ\subseteq \ACQ$ hold.
		The head arity of a view $V$ is the arity of its head atom $\headof{V}$.
		The weak head arity of a view is defined in \autoref{def:weak-head-arity}.
		In a nutshell, a view with weak head arity $\ell$ can be \enquote{split} into (multiple) views with head arity at most $\ell$.%
	}%
	\resizebox{\linewidth}{!}{%
\begingroup
\renewcommand{\arraystretch}{1.5}
\begin{tabular}{@{}cccccc@{}}%
	\toprule
	\parbox{0.8cm}{\centering $\ViewClass$\\[-.3\baselineskip]{\scriptsize Views}}
		& \parbox{0.8cm}{\centering $\QueryClass$\\[-.3\baselineskip]{\scriptsize Query}}
		& \parbox{1.25cm}{\centering
           $\RewritingClass$\\[-.3\baselineskip]{\scriptsize
	Rewriting}}
		& \parbox{1.6cm}{\centering\footnotesize Restriction of views}
		& \parbox{4.85cm}{\centering $\rewrprobk{\ViewClass}{\QueryClass}{\RewritingClass}$ {\scriptsize for every $k\in\N$}\\[-.3\baselineskip] {\scriptsize (bounded arity db-schema)}}
		& \parbox{4cm}{\centering $\rewrprob{\ViewClass}{\QueryClass}{\RewritingClass}$\\[-.3\baselineskip]{\scriptsize (unbounded arity db-schema)}}	\\
	\midrule
	$\CQ$ & $\ACQ$	& $\CQ$ &
		\parbox{2cm}{\centering\scriptsize{}Boolean views} &
		\multicolumn{2}{c}{\classNP-complete for $k\ge 3$ (\autoref{proposition:np-hard-simple})}
	\\
	\addlinespace
	$\ACQ$ & $\CQ$	& $\CQ$ &
		\parbox{2cm}{\centering\scriptsize{}Boolean views} &
		\multicolumn{2}{c}{\classNP-complete for $k\ge 3$ (\autoref{proposition:np-hard-simple})}
	\\
	\addlinespace
	$\ACQ$ & $\ACQ$	& $\ACQ$ &
		\parbox{2cm}{\centering\scriptsize{}no restriction} &
		\multicolumn{2}{c}{\classNP-complete for $k\ge 3$ (\autoref{thm:arewriting-np-hard})}
	\\
	\addlinespace
    $\ACQ$ & $\ACQ$ & $\ACQ$ &
    	\parbox{2cm}{\centering\scriptsize{}head arity $\le \ell$\\$\ell\in\N$}
    	& \multicolumn{2}{c}{in polynomial time (\autoref{corollary:arewrprobk-tractable})}
    \\
    \addlinespace
	$\ACQ$ & $\ACQ$ & $\ACQ$ &
        \parbox{2.75cm}{\centering\scriptsize{}weak head arity $\le \ell$\\$\ell\in\N$}& \multicolumn{2}{c}{in polynomial time (\autoref{theorem:tractable:weak-head-arity})}
    \\
	\addlinespace
	$\CCQ$ & $\ACQ$ & $\ACQ$ & \parbox{2cm}{\centering\scriptsize{}no restriction} &
        in polynomial time (\autoref{theorem:tractable:free-connex})	& open
  	\\
	\addlinespace
	$\HCQ$ & $\HCQ$ & $\HCQ$ & \parbox{2cm}{\centering\scriptsize{}no restriction} &
		\multicolumn{2}{c}{\classNP-complete for $k\ge 3$ (\autoref{corollary:hcq-np-complete})}
	\\
	\addlinespace
	$\QHCQ$	& $\HCQ$ & $\HCQ$ & \parbox{2cm}{\centering\scriptsize{}no restriction} &
		in polynomial time (\autoref{corollary:qhcq-views-tractable}) & open
	\\
	\addlinespace
	$\QHCQ$	& $\QHCQ$ & $\QHCQ$	& \parbox{2cm}{\centering\scriptsize{}no restriction} &
		in polynomial time (\autoref{corollary:qhcq-views-tractable})	& open
	\\
	\bottomrule
\end{tabular}
\endgroup
	}%
	\label{table:overview-acyclic-rewriting-results}%
\end{table}
Our main findings regarding the complexity of the acyclic rewriting problem are as follows (see \autoref{table:overview-acyclic-rewriting-results} for an overview).
\begin{itemize}
\item If the query and the views are acyclic and the arity of the views is bounded by some fixed $k$, then the acyclic rewriting problem is tractable, that is, for every $k$ the decision problem  $\rewrprob{\ACQ^k}{\ACQ}{\ACQ}$ is in polynomial time (\autoref{corollary:arewrprobk-tractable}). Furthermore, an acyclic
  rewriting can be computed in polynomial time (if it exists). This follows easily with the help of the canonical rewriting approach (see \cite[Proposition~5.1]{DBLP:journals/tods/NashSV10}) and our answer to Question~\ref{question:intro-a}.
\item If the query and the views are acyclic and the arity of the views can be unbounded, then the acyclic rewriting problem is intractable, even if the database schema has bounded arity, more precisely: $\rewrprobk[3]\ACQ\ACQ\ACQ$ is \classNP-complete (\autoref{thm:arewriting-np-hard}).
\item If the query is acyclic, the views are free-connex acyclic, and the arity of the database schema is bounded by some fixed $k$, then the acyclic rewriting problem is tractable, that is, for every $k$: $\rewrprobk{\CCQ}{\ACQ}{\ACQ}$ is in polynomial time (\autoref{theorem:tractable:free-connex}).
\end{itemize}

Many  results use a characterisation of rewritability, based on the notion of \emph{cover partitions}, that might be interesting in its own right. Similar notions have been used in earlier work, but ours is particularly suited for the study of exact (equivalent) rewritings.

Our paper is organised as follows: we introduce general basic notions in \autoref{sec:preliminaries} and notions and observations related to rewritings in \autoref{section:rewriting-problem}.
The characterisation of rewritability in terms of cover partitions is given in \autoref{section:characterisation}. The answer to Question~\ref{question:intro-a} and the complexity of the acyclic rewriting problem for acyclic queries and views are studied in \autoref{section:towards-acyclic-rewritings}.
The feasibility for free-connex acyclic views (and an additional class) is presented in \autoref{section:tractable:free-connex}. \autoref{section:tractable} shows that the answer to Question~\ref{question:intro-a} is the same for hierarchical and q-hierarchical queries and views. \autoref{section:relationship-characterization} discusses related work and, in particular, the relationship of our characterisation of rewritability with results in the literature. \autoref{section:conclusion} concludes.

\section{Preliminaries}
\label{sec:preliminaries}
In this section, we fix some notation and recall the basic concepts from relational databases that are relevant for this paper.
Let~\NN denote the set of non-negative integers.
Let~\dom and~\var be countably infinite sets of \emph{data values} (also called \emph{constants}) and \emph{variables}, respectively.
We use the natural extensions of mappings of variables
onto tuples and sets without
notational distinction.
That is, for a mapping~$f\colon X \to Y$, we write~$f(X')$ for $\{f(x) \mid x \in X'\}$ and $f(\bfx)$ for $(f(x_1),\dots,f(x_k))$ for sets $X' \subseteq X$ and tuples $\bfx = (x_1,\dots,x_k) \in X^k$, respectively.
The \emph{composition} $f \circ g$ of two functions $g\colon X \to Y$ and $f\colon Y \to Z$ is the function $(f\circ g)\colon X \to Z$ defined by $(f\circ g)(x) = f(g(x))$ for all $x\in X$.
By \id we denote the identity mapping (on any set of variables).

\paragraph*{Databases.}%
Databases and queries are formulated over \emph{database schemas}. %
A \emph{database schema~\schema} is a finite set of relation schemas,
each represented by a symbol~$R$ and associated with a fixed arity
$\ar(R) \in \NN$.\footnote{We refer to database schemas usually by \emph{schema}.}
A \emph{fact} or \emph{$R$-fact} $R(a_1,\dots,a_r)$ comprises a relation symbol~$R$ with some arity~$r$ and data values~$a_1,\dots,a_r$.
A \emph{database}~$D$ over schema~\schema is a finite set of $R$-facts for $R \in \schema$.

\paragraph*{Queries.}
  In general, queries map  databases to relations.
In this paper, we consider conjunctive queries and restrictions of
them. Conjunctive queries are queries that can be expressed
syntactically as  conjunctions of relation atoms as follows.

An \emph{atom} is of the form $R(x_1,\dots,x_r)$ with a relation symbol~$R$ with arity~$r$, and variable set $\{x_1,\dots,x_r\}$.
We denote the variable set of an atom $\atm$ by $\varsof{\atm}$.
Analogously to
  facts, an atom with relation symbol~$R$ is called an \emph{$R$-atom}
  if we want to stress the associated relation (symbol). More
  generally, \schema-atoms are $R$-atoms for some symbol~$R$ in
  schema~\schema.%

We represent a \emph{conjunctive query} (\emph{CQ} for short) over schema~\schema as a rule of the form
\begin{math}
	A \gets A_1,\dots,A_m
\end{math}
whose \emph{body} $\{A_1,\dots,A_m\}$ consists of a positive number of atoms and whose \emph{head} is a single atom~$A$ such that the following two conditions are satisfied. First, atoms $A_1,\dots,A_m$ refer to relation symbols from~\schema and atom~$A$, on the contrary, does not. Second, the query is \emph{safe}, that is, every variable in the head occurs in at least one atom of the body.
Let~\headof{Q} and~\bodyof{Q} denote the head and body of a query~$Q$,
respectively. Variables that occur in the head of a query are called
\emph{head variables}; all other variables are called \emph{quantified
  variables}.
	A query without quantified variables is called a \emph{full} query.
	Queries with two or more body atoms that refer to the
  same relation are said to have \emph{self-joins}.
The \emph{arity} of a query $Q$ is the arity of its head atom.
Furthermore, a query $Q$ is a Boolean query if the arity of $\headof{Q}$ is $0$.

Like for sets and tuples, we also use the natural extension of mappings of variables on atoms without difference in notation: for a mapping $f: \var \to Y$, let~$f(A)$ denote $R(f(x_1),\dots,f(x_r))$ for an atom $A = R(x_1,\dots,x_r)$.

A \emph{valuation} is a mapping $\valuation: \var \partto \dom$. A database $D$ \emph{satisfies} a set $\calA$ of atoms under a valuation $\valuation$,  if $\valuation(\calA)\subseteq D$, that is for every atom $R(x_1,\ldots,x_r)$ in $\calA$ the fact $R(\valuation(x_1),\ldots,\valuation(x_r))$ is contained in $D$.
The \emph{result} of query~$Q$ on database~$D$ is defined as
\begin{displaymath}
	Q(D) = \{
		\valuation(\headof{Q})
		\mid
		 \text{$\valuation$ is a valuation and $D$ satisfies $\bodyof{\Q}$ under \valuation}
	\}.
      \end{displaymath}

\paragraph*{Relationships between queries.}
Queries over the same schema can be compared with respect to the results they define. Let~$Q_1$ and~$Q_2$ be queries. We say that~$Q_1$ is \emph{contained} in~$Q_2$ (notation: $Q_1 \contained Q_2$) if $Q_1(D) \subseteq Q_2(D)$ for every database~$D$. We say that~$Q_1$ and~$Q_2$ are \emph{equivalent} (notation: $Q_1 \equiv Q_2$) if~$Q_1 \contained Q_2$ and~$Q_2 \contained Q_1$.

It is well-known that a conjunctive query~$Q_1$ is contained in a conjunctive query~$Q_2$ if and only if there is a homomorphism from the latter query into the first~\cite{DBLP:conf/stoc/ChandraM77}. Such a \emph{homomorphism} is a mapping~$h\colon \varsof{Q_2} \to \varsof{Q_1}$ such that
\begin{enumerate}[(1)]
	\item $h(\bodyof{Q_2}) \subseteq \bodyof{Q_1}$ and \label{def:hom-condition-a}
	\item $h(\headof{Q_2}) = \headof{Q_1}$ hold.
\end{enumerate}
We call~$h$ a \emph{body homomorphism} if it fulfils Condition~\ref{def:hom-condition-a}.

	A conjunctive query $Q_1$ is \emph{minimal} if there is no conjunctive query $Q_2$ such that $Q_2 \equiv Q_1$ and $|\bodyof{Q_2}| < |\bodyof{Q_1}|$ holds.

\paragraph*{Structurally simple queries.}
  Despite their simplicity and restricted expressibility, several interesting problems are intractable for conjunctive queries in general. Therefore, different fragments have been studied in the literature. In this paper, we are particularly concerned with \enquote{acyclic} queries, which allow, for instance, evaluation in polynomial time. We also consider
three subclasses of acyclic queries, namely, free-connex acyclic, hierarchical, and q-hierarchical queries.

\paragraph*{Acyclic and free-connex acyclic queries.}
A \emph{join tree} for a query~$Q$ is a tree whose vertices are the atoms in the query's body that satisfies the following path property: for two atoms $A,A' \in \bodyof{Q}$ with a common variable~$x$, all atoms on the path from~$A$ to~$A'$ contain~$x$.
A query~$Q$ is \emph{acyclic} if it has a join tree.
It is \emph{free-connex acyclic} if $Q$ is acyclic and the Boolean query whose body is $\bodyof{Q} \cup \{\headof{Q}\}$ is acyclic as well \cite{DBLP:conf/csl/BaganDG07,HAL:braultbaron:tel-01081392}.

\paragraph*{Hierarchical queries.}
For a fixed query~$Q$ and some variable~$x$ in~$Q$, let $\atoms(x)$ denote the set of atoms in~\bodyof{Q} in which~$x$ appears.
\begin{defiC}[\cite{DBLP:conf/pods/DalviS07a}, {\cite[Definition~3.1]{DBLP:conf/pods/BerkholzKS17}}]\label{definition:qhierarchical}
	A conjunctive query $Q$ is \emph{hierarchical} if, for all variables $x,y$ in~$Q$, one of the following conditions is satisfied.
	\begin{enumerate}[label={(\arabic*)}]
		\item\label{hier:x-contained-by-y}
			$\atoms(x)\subseteq\atoms(y)$
		\item\label{hier:y-contained-by-x}
			$\atoms(x)\supseteq\atoms(y)$
		\item\label{hier:empty-intersection}
			$\atoms(x)\cap\atoms(y) = \emptyset$
	\end{enumerate}
 	Thus, the \enquote{hierarchy} is established by the query's variables and the sets of atoms that contain them.

	A conjunctive query $Q$ is \emph{q-hierarchical} if it is hierarchical and for all variables $x,y\in\varsof{Q}$ the following is satisfied:
	if $\atoms(x)\subsetneq\atoms(y)$ holds and $x$ is in the head of $Q$, then $y$ is also in the head of $Q$.
\end{defiC}
For brevity, we denote by~\CQ, \ACQ, \CCQ, \HCQ, and \QHCQ the
classes of conjunctive queries in general and those conjunctive
queries that are acyclic, free-connex acyclic, hierarchical or
q-hierarchical, respectively.
We note that each q-hierarchical query is free-connex acyclic \cite[Proposition~4.25]{DBLP:conf/sigmod/IdrisUV17} and
each hierarchical query is acyclic.%
\footnote{The latter inclusion is mentioned in, e.g.\ \cite{DBLP:conf/pods/Hu019,DBLP:conf/pods/0002NOZ20}; it also follows readily from the former inclusion. For acyclicity the head of a query is of no concern, and the Boolean variant of a hierarchical query is always q-hierarchical by definition, and, thus, free-connex acyclic which implies the existence of a join tree for the body of the original query.}

\paragraph*{Views and rewritings.} A view $V$ over a schema $\schema$ is just a query over schema~\schema.\footnote{They are called views due to their special role which  distinguishes them from \enquote{normal} queries.}
A set $\calV=\{V_1,\ldots,V_k\}$ of views induces, for each database~$D$ over schema~\schema, the \emph{\emph{\calV-defined} database}
\begin{math}
	\calV(D) = V_1(D) \cup \dots \cup V_k(D).
\end{math}

In this paper, we consider only views that are conjunctive queries. We
furthermore assume that every view defines its own relation in the
derived database. We denote the set of relation schemas induced by the heads of
the views in $\calV$ by $\schema_\calV$.

In a nutshell, a rewriting of a query $Q$ over a set $\calV$ of views is a query over $\schema_\calV$ that is
meant to yield, for each database $D$, the same result over $\calV(D)$
as $Q$ does over $D$.

\begin{defi}
	Let $Q$ be a query, $\calV$ a set of views, and $Q'$ a query
	over   $\schema_\calV$.
	We call $Q'$ a \emph{$\calV$-rewriting of $Q$} if, for every
	database $D$, it holds that $Q'(\calV(D))=Q(D)$.
\end{defi}

Formally, we only consider rewritings over    $\schema_\calV$. This is
not a restriction, since database relations can be replicated as view relations.
In the literature, rewritings often only aim at
approximations of the query \cite{DBLP:journals/vldb/PottingerH01}. However, in this paper, we are only interested
in \emph{equivalent} and \emph{complete} rewritings, where $Q'\circ\calV\equiv Q$ holds.
Here $Q'\circ \calV$ denotes the query with query result $\Q'(\calV(D))$ for every database $D$.

We illustrate the above definitions by an example.
\begin{exa}\label{ex:beginningex}
	Let~\calV be a set of two views
	\begin{align*}
		V_1(x_1,w_1) &\ \gets\ P(v_1,v_1',x_1),\;R(x_1,w_1),\;S(w_1), \\
		V_2(y_2,z_2) &\ \gets\ S(y_2),\;T(y_2,z_2)
	\end{align*}
	over schema $\schema = \{P,R,S,T\}$ and let~$Q$ be the query
	\begin{displaymath}
		H(x,y,y') \gets P(u,u',x), R(x,w), S(w), T(w,y), T(w,y').
	\end{displaymath}
For each database~$D$, query~$Q$ yields the same result on~$D$ as the following query~$Q'$
	\begin{displaymath}
		H(x,y,y') \gets V_1(x,w), V_2(w,y), V_2(w,y')
	\end{displaymath}
	on $\calV(D)$. Therefore, $Q'$ is a
        $\calV$-rewriting of $Q$.
      \end{exa}
\section{The Rewriting Problem}\label{section:rewriting-problem}

In this section, we define the rewriting problem, recall some known
results about its complexity and  discuss some relevant concepts that
were used to tackle it.

\begin{defi}
	Let~\ViewClass, \QueryClass , and~\RewritingClass be classes of conjunctive queries.
	The \emph{rewriting problem} for~\ViewClass, \QueryClass , and~\RewritingClass,  denoted $\rewrprob{\ViewClass}{\QueryClass}{\RewritingClass}$ asks, upon input of a
	query $Q\in\QueryClass$ and a  (finite) set
        $\calV\subseteq\ViewClass$ of views,  whether there is a
        $\calV$-rewriting of~$Q$ in the class~\RewritingClass. We write
        $\rewrprobk{\ViewClass}{\QueryClass}{\RewritingClass}$  for the restriction,
        where the arity of the database schema is bounded by $k$.
      \end{defi}

In general, the  rewriting problem for conjunctive queries is
$\classNP$-complete.

\begin{thmC}[{\cite[Theorem~3.10]{DBLP:conf/pods/LevyMSS95}}]\label{theo:np-complete}
	$\rewrprob{\CQ}{\CQ}{\CQ}$ is \classNP-complete.
\end{thmC}

There is a straightforward, albeit in general inefficient, way of finding
a rewriting for a conjunctive query and views. In fact, it can be
shown that, for given $Q$ and $\calV$, a rewriting $Q'$ over $\calV$
exists,
	if and only if a certain canonical query $\candof{Q,\calV}$ exists and is
	a rewriting.

This \emph{canonical candidate} $\candof{Q,\calV}$ can be obtained by
evaluating $\calV$ on the canonical database $\candof{Q}$, which in
turn is defined as the set of body atoms of $Q$ viewed as facts, in
which the variables are considered as fresh constants.
The canonical candidate has the same head as $Q$ and its body is just
$\calV(\candof{Q})$.\footnote{Here the variables are
  considered as variables, not as constants.}
If the canonical candidate turns out to be a rewriting, then we often
call it \emph{canonical rewriting}.

An important detail should be mentioned here: the canonical
  candidate does not always exist, e.g.,\ if
  $\calV(\candof{Q})$ does not contain all head variables of $Q$ or it is
  outright empty. In that case, there is no rewriting.

\begin{propC}[{\cite[Proposition~5.1]{DBLP:journals/tods/NashSV10}}]\label{prop:canonical-rew}
	Let $Q$ be a conjunctive query and $\calV$ a set of views.
	If there is a $\calV$-rewriting of $Q$, then the canonical candidate $\candof{Q,\calV}$ is such a rewriting.
      \end{propC}

For conjunctive queries without self-joins this had been shown
      in~\cite[Lemma~7]{DBLP:journals/tcs/ChekuriR00}.

\begin{exa}\label{example:canonicalrewriting}
		Let us consider the query $Q$ defined by
	\[
	H(x,y,z) \gets C(x,y,z),\;R(x,y),\;S(y,z),\;T(z,x)
	\]
	and the views
	\begin{align*}
	V_1(u_1,v_1,w_1) \gets C(u_1,v_1,w_1)\ \ \text{ and }\ \
	V_2(x_2,y_2,z_2,u_2) \gets R(x_2,y_2),\;S(y_2,z_2),\;T(z_2,u_2).
	\end{align*}
		Evaluating the views $V_1$ and $V_2$ on the canonical database
        $\candof{Q}$ of $Q$ yields the result $\{V_1(x,y,z), V_2(x,y,z,x)\}$.
	Thus, the query $Q'$ defined by
        \[
          H(x,y,z) \gets V_1(x,y,z), V_2(x,y,z,x)
        \]
        is
        the canonical candidate, which happens to be an actual $\calV$-rewriting.
\end{exa}

	In this example it is easy to see that the canonical candidate is a $\calV$-rewriting,
without considering any (let alone all) databases.
In fact, in this particular case, this can be established with the
help of an expansion of the rewriting.
Before we define expansions, we need one more definition.
\begin{defi}[View Application]\label{def:body-homomorphism}
	An \emph{application} of a view~$V$ is a substitution~$\app\colon\!\varsof{V}\, {\to}\allowbreak \var$ that does not unify any quantified variable of~$V$ with another variable of $V$.
\end{defi}
This definition reflects the fact that a rewriting can access a view
only through its head and therefore cannot unify any quantified variables.

We say that a collection $\alpha_1,\ldots,\alpha_m$ of applications for views
$V_1,\ldots,V_m$  fulfils  \emph{quantified variable
  disjointness}, if
for
    all $i,j\in\{1,\ldots,m\}$, each quantified variable $x$ of $V_i$, and
    each variable $y$ of
    $V_j$ with $i\neq j$, it holds
    $\alpha_i(x) \not= \alpha_j(y)$.
      That is, beyond what is already required by the definition of an
      application, a view application never maps a quantified variable
      of a view to a variable in the range of any other view application.
	This ensures that, for all $i,j\in\{1,\ldots,m\}$ with $i\neq j$, the bodies $\alpha_i(\bodyof{V_i})$ and $\alpha_j(\bodyof{V_j})$ only share variables from $\varsof{\alpha_i(\headof{V_i})}\cap \varsof{\alpha_j(\headof{V_j})}$.

	An expansion of a \(\calV\)-rewriting~\(Q'\) is, intuitively, obtained by inlining the bodies of the views from \(\calV\) in \(Q'\).

\begin{defi}[Expansion]\label{definition:expansion}
	Let~\calV be a set of views over a schema $\schema$ and let~$Q'$ be a query with $\bodyof{Q'} = \{A'_1,\dots,A'_m\}$ over the schema $\schema_\calV$.
	Furthermore, let, for each $i \in \{1,\dots,m\}$,  \(V_i\) be the view in \(\calV\) and~$\app_i$ be
        an application for \(V_i\) such that $A'_i =
        \app_i(\headof{V_i})$, such that $\alpha_1,\ldots,\alpha_m$
        fulfil quantified variable
  disjointness.

	The \emph{expansion  of~$Q'$ w.r.t.\ $\calV$ and $\app_1,\ldots,\app_m$} is the
        query that has the same head as~$Q'$ and body
		\begin{math}
          \bigcup_{i=1}^m 	\app_i(\bodyof{V_i}).
        \end{math}
\end{defi}

Since the applications $\app_1,\ldots,\app_m$ in \autoref{definition:expansion} are uniquely
determined up to renaming of quantified variables, we
usually do not mention them explicitly  and just speak of an expansion of a query~$Q'$ w.r.t.\ $\calV$.

Furthermore, because an expansion \(Q''\) of a \(\calV\)-rewriting
\(Q'\) is, intuitively, obtained by inlining the bodies of the views, the expansion \(Q''\) is \enquote{equivalent} to \(Q'\) in the following sense.
\begin{lem}\label{result:equivalence-expansion}
	Let \(\calV\) be a set of views, \(Q'\) be a query over the schema \(\schema_\calV\), and \(Q''\) be an expansion of \(Q'\) w.r.t.\ \(\calV\).
	Then \(Q''(D) = Q'(\calV(D))\) holds for every database $D$.
\end{lem}
\autoref{result:equivalence-expansion} has been proved, e.g., as part of the proof of \cite[Theorem~3.5]{DBLP:series/synthesis/2019Afrati}.
For the sake of convenience, we provide a short proof using our notation in the following.
\begin{proof}[Proof of \autoref{result:equivalence-expansion}]
	Following \autoref{definition:expansion}, let \(\{A'_1,\ldots,A'_m\}\) be the body of \(Q'\), and let \(\app_1,\ldots,\app_m\) be applications fulfilling quantified variable disjointness, and \(V_1,\ldots,V_m\in \calV\) views such that \(A'_i = \app_i(\headof{V_i})\) holds for all \(i\in\{1,\ldots,m\}\) and \(\bodyof{Q''} = \bigcup_{i=1}^m \app_i(\bodyof{V_i})\) holds.

	We first show the inclusion \(Q''(D) \subseteq Q'(\calV(D))\).
	Let \(D\) be a database and \valuation be a valuation such
        that $D$ satisfies \(\valuation(\bodyof{Q''}) =
        \valuation(\bigcup_{i=1}^m \app_i(\bodyof{V_i}))\) under \valuation.
	By the definition of an expansion, \(\valuation(\headof{Q''}) = \valuation(\headof{Q'})\) holds.
	Thus, it suffices to show that \(\valuation(\bodyof{Q'})\subseteq \calV(D)\) holds witnessing that \(\valuation(\headof{Q'})\in Q'(\calV(D))\).

	Since \(\valuation(\bodyof{Q''}\subseteq
        D\),   it holds \(\valuation(\app_i(\bodyof{V_i})) \subseteq
        D\), for all \(i\in\{1,\ldots,m\}\).
	Interpreting \(\valuation\circ\app_i\) as a valuation, it follows that \(\valuation(\app_i(\headof{V_i})) \in V_i(D)\).
	But \(\app_i(\headof{V_i}) = A'_i\) for all \(i\in\{1,\ldots,m\}\) and, therefore, we can conclude that \(\valuation(\bodyof{Q'}) = \valuation(\{A'_1,\ldots,A'_m\}) \subseteq \calV(D)\).

	For the inclusion \(Q''(D) \supseteq Q'(\calV(D))\) let
        \valuation be a valuation
such
        that $\calV(D)$ satisfies \(\bodyof{Q'}\) under \valuation.
Thus \(\valuation(A'_i) \in \calV(D)\) and therefore
\(\valuation(\app_i(\headof{V_i}))\in V_i(D)\) holds, for all \(i\in
\{1,\ldots,m\}\). 
The latter implies that there are valuations \(\mu_i\) such that
\(\valuation(\app_i(\headof{V_i})) = \mu_i(\headof{V_i})\)
such that \(\mu_i(\bodyof{V_i}) \subseteq D\). That is,  the \(\mu_i\) map \(\bodyof{V_i}\) into \(D\) and agree with \(\valuation\circ \app_i\) on all head variables of \(V_i\).
	Moreover, since the application \(\app_i\) does not unify any quantified variables with other variables, the valuation \(\valuation\) can be extended to a valuation \(\valuation_i^{\plus}\) such that \(\valuation_i^{\plus}(\app_i(\bodyof{V_i}))\subseteq D\).

	Lastly, thanks to the quantified variable disjointness of the \(\app_i\), the extended mappings \(\valuation_i^{\plus}\) can be combined into a valuation \(\valuation^{\plus}\) which maps all \(\app_i(\bodyof{V_i})\) into \(D\).
\end{proof}

We further note that the expansion of a rewriting $Q'$ can
be directly compared with a query $Q$ since it is over the same schema.
	In fact, we will frequently use the following result which follows readily from \autoref{result:equivalence-expansion}.
	\begin{propC}[{\cite[Theorem~3.5]{DBLP:series/synthesis/2019Afrati}}]
		Let \(Q\) be a query over a schema \(\schema\), \(\calV\) be a set of views over \(\schema\), \(Q'\) be a query over the schema \(\schema_\calV\), and \(Q''\) be an expansion of \(Q'\) w.r.t.\ \(\calV\).

		The expansion \(Q''\) of \(Q'\) is equivalent to \(Q\) if and only if \(Q'\) is a \(\calV\)-rewriting of \(Q\).
	\end{propC}

In \autoref{example:canonicalrewriting},
the query \(Q'\) is a \(\calV\)-rewriting of \(Q\), because the views are full queries, and, thus, there is only one (unique) expansion of $\Q'$ w.r.t.\ $\calV$ which even coincides with $Q$.

	If there is no a
priori bound on the arity of views, then the canonical
candidate can be of exponential size in $|Q|+|\calV|$.
However, if there is a rewriting, then there always is one of
polynomial size.
In fact, at most~$n$ view applications are needed if the body of query~$Q$ contains~$n$ atoms
\cite[Lemma~3.5]{DBLP:conf/pods/LevyMSS95}.

\begin{exa}
	Consider the query \(H() \gets S(x,y), R(x), R(y)\) and the views
	\[
		V_1(x_1,y_1) \gets S(x_1,y_1) \text{ and } V_2(x_{2,1},\ldots,x_{2,n}) \gets R(x_{2,1}),\ldots,R(x_{2,n}).
	\]
	Evaluating the views on the canonical database \(\candof{Q}\) yields the result
	\[
		\{V_1(x,y)\}\cup \{V_2(a_1,\ldots,a_n) \mid a_1,\ldots,a_n\in \{x,y\}\}.
	\]
	This result, and thus, the body of the canonical candidate, has exponential size.
	There is, however, a simple \(\{V_1,V_2\}\)-rewriting which has as many view atoms as \(Q\) has atoms, namely,
	\[ H() \gets V_1(x,y), V_2(x,\ldots,x), V_2(y,\ldots,y).\]
\end{exa}

  Therefore, an \classNP-algorithm for $\rewrprob{\CQ}{\CQ}{\CQ}$ can
  \enquote{guess} a rewriting of polynomial size and test whether it yields
  an expansion that is equivalent to the given query. Since equivalence of \CQ can be tested in \classNP,
 this indeed shows that $\rewrprob{\CQ}{\CQ}{\CQ}$ is in \classNP.

  Due to their potentially
  exponential size, canonical candidates are, in general, not useful
  as a starting point for polynomial time algorithms.
  However, in some cases their size is only polynomial, in particular,
  if the views have bounded arity. In such cases, the above characterisation
can sometimes be used to obtain efficient algorithms. In particular,
this holds
for acyclic views of bounded arity and acyclic queries.
For a class $\ViewClass$ of views, we write $\ViewClass^k$ for the restriction to views of arity at most $k$.

\begin{prop}\label{proposition::bounded-head-acyclic}
  For every $k\ge 0$, $\rewrprob{\ACQ^k}{\ACQ}{\CQ}$ is in polynomial time.
\end{prop}
\begin{proof}
    We first observe that, if it exists, the canonical candidate always obeys $Q\contained \candof{Q,\calV}\circ \calV$.
		Indeed, a homomorphism from an expansion of $\candof{Q,\calV}$ to $Q$ can be obtained by combining the valuations that yield $\calV(\candof{Q})$ from $\candof{Q}$ as follows.
		To improve readability and avoid technical clutter we identify the variables in $Q$ with their corresponding constants in $\candof{Q}$ in the following.
		Notably, this means that we identify $\candof{Q}$ with $\bodyof{Q}$, i.e.\ we write $\candof{Q} = \bodyof{Q}$.

		Let $A'_1,\ldots,A'_m$ be the atoms of $\candof{Q,\calV}$ and, for each $i\in\{1,...,m\}$, let $\app_i$ be an application and $V_i$ be a view such that $A'_i = \app_i(\headof{V_i})$, such that $\app_1,\ldots,\app_m$ fulfil quantified variables disjointness.
		Furthermore, since the $A'_i$ are also contained in $\calV(\candof{Q})$ by definition of $\candof{Q,\calV}$, there are valuations $\valuation_i$ such that $A_i' = \valuation_i(\headof{V_i})$ and $\valuation_i(\bodyof{V_i}) \subseteq \candof{Q}$ holds.
		In particular, we have that $\valuation_i(\headof{V_i}) = \app_i(\headof{V_i})$ for all $i \in \{1,\ldots,m\}$.

		Let $y$ be a variable occuring in $\app_i(\bodyof{V_i})$ and $x$ such that $y = \app_i(x)$ for some $i\in \{1,\ldots,m\}$.
		Then the desired homomorphism $h$ maps $y$ to $\valuation_i(x)$.
		Note that $h$ is the identity on variables occuring in $\app_i(\headof{V_i})$ because $\app_i(\headof{V_i}) = \valuation_i(\headof{V_i})$ and, thus, $y = \app_i(x) = \valuation_i(x) = h(y)$.
		Hence, $h$ is well-defined, since, if a variable $y$ occurs in $\app_i(\bodyof{V_i})$ and $\app_j(\bodyof{V_j})$ with $i\neq j$ then it also occurs in $\app_i(\headof{V_i})$ and $\app_j(\headof{V_j})$ thanks to quantified variable disjointness.

		It is easy to see that $h$ is a homomorphism from the expansion of $\candof{Q,\calV}$ to $Q$, because \[h(\app_i(\bodyof{V_i})) = \valuation_i(\bodyof{V_i}) \subseteq \candof{Q}\] holds for all $i\in\{1,\ldots,m\}$ and $h$ is the identity on all variables occuring in the head of $\candof{Q,\calV}$.

	Therefore, to test for rewritability, it suffices to check whether $\candof{Q,\calV}\circ \calV\contained Q$ holds.

	Since the views are acyclic and their arity is bounded by a constant, the polynomial-sized canonical candidate $\candof{Q,\calV}$ can be computed in polynomial time.\footnote{Evaluating and testing containment for acyclic conjunctive queries is in polynomial time \cite{DBLP:conf/vldb/Yannakakis81,DBLP:books/aw/AbiteboulHV95}.}
	It then suffices to test whether a homomorphism from $Q$ to an expansion of $\candof{Q,\calV}$ exists.
	This test can be done in polynomial time because the query \(Q\) is acyclic.

	Thus, overall, it can be tested efficiently whether $\candof{Q,\calV}\circ \calV\contained Q$ holds if the query and views are acyclic and the arity of the views is bounded by a constant.

	Hence, the statement follows due to \autoref{prop:canonical-rew}.
 \end{proof}
Chekuri and Rajaraman have shown that the rewriting problem is in $\classP$ for acyclic queries without self-joins and arbitrary views using a similar idea~\cite[Theorem~5]{DBLP:journals/tcs/ChekuriR00}.

However, even for Boolean views and databases over a small fixed schema, it does not
suffice to restrict only the query or only the views to acyclic queries.

\begin{prop}\label{proposition:np-hard-simple}
	$\rewrprobk{\CQ^0}{\ACQ}{\CQ}$ and $\rewrprobk{\ACQ^0}{\CQ}{\CQ}$ are
        \classNP-complete, for every $k\ge 3$. This even holds for Boolean queries, view sets
        with only one Boolean view and a schema with  two
        relations of maximum arity 3.
\end{prop}
\begin{proof}
	The upper bound holds thanks to \autoref{theo:np-complete}.
        For the lower bound, we first show,  by a reduction from the three-colouring
        problem, that it is \classNP-hard to test, for an
        \emph{acyclic} Boolean conjunctive query $Q_1$ and a Boolean
        conjunctive query $Q_2$, whether
        $Q_1\contained Q_2$ holds.\footnote{We think that this is
          folklore knowledge but could not find a reference for it.}

	To this end, let $G$ be an undirected graph without isolated vertices and self-loops (i.e.\ edges of the form $(v,v)$).
	With each vertex $v$ of $G$ we associate a variable $x_v$.
	We define $Q_2$ to be the Boolean conjunctive query with head $H()$ that consists of all atoms $E(x_v,x_w)$ such that $(v, w)$ is an edge in $G$.
	The Boolean query $Q_1$ is defined by the rule
	\[ H() \leftarrow E(b,r), E(r, y), E(y, b), E(r, b), E(y, r), E(b, y), C(b,r,y) \]
	where $b, r, y$ are variables representing the colours blue, red, and yellow.
	Observe that $Q_1$ is acyclic, because all variables are contained in the \enquote{cover-atom} $C(b,r,y)$.

          Towards the correctness of the reduction, if $Q_1 \contained Q_2$
	holds, there is a homomorphism $h$ from $Q_2$ to $Q_1$. This homomorphism represents a valid colouring, since two variables $x_v$ and $x_w$ cannot be mapped to the same variable in $Q_1$ if there is an edge $(v,w)$ in $G$.
	Conversely, every valid colouring of $G$ gives rise to a
        homomorphism from $Q_2$ to $Q_1$ and therefore witnesses $Q_1 \contained Q_2$.

        Since it is well-known that  $Q_1\contained Q_2$ if and only
        if $Q_1 \equiv Q_2 \land Q_1$, we conclude in the second step
        that the equivalence problem for Boolean conjunctive queries
        $Q_1,Q_2$ is \classNP-hard, even if $Q_1$  is acyclic.

	Finally, we reduce the equivalence problem with acyclic $Q_1$ to the rewriting problem.
	Let thus $Q_1$ and $Q_2$ be Boolean conjunctive queries.
	The input instance $(Q_1,Q_2)$ for the equivalence problem is mapped to an input instance for the rewriting problem by assigning $Q_1$ the role of the query and $Q_2$ the role of (the sole) view.

	The canonical rewriting, if it exists, consists only of the single atom $Q_2()$, since $Q_2$ is a Boolean query.
	Thus, there is a $Q_2$-rewriting of $Q_1$ if and only if $Q_1$ is equivalent to $Q_2$.
This reduction establishes that $\rewrprobk[3]{\CQ^0}{\ACQ}{\CQ}$ is
\classNP-hard. For the \classNP-hardness of $\rewrprobk[3]{\ACQ^0}{\CQ}{\CQ}$,
the roles of $Q_1$ and $Q_2$ are simply swapped in the last reduction.

By composing the three reductions, the
proposition is established.
 \end{proof}

\section{A Characterisation}\label{section:characterisation}
In this section we give a characterisation of rewritability of a query
$Q$ with respect to a set $\calV$ of views. It is in terms  of a
partition of the atoms of $\bodyof{Q}$, where each set of the partition can be
matched with a view in a specific way. We refer to such partitions as
\emph{cover partitions} and the matches as \emph{cover descriptions}.
The characterisation is very
similar to other such characterisations in the literature~\cite{DBLP:conf/sigmod/GouKC06, DBLP:series/synthesis/2019Afrati}, in
particular to \enquote{MiniCon Descriptions}~\cite{DBLP:journals/vldb/PottingerH01}. However, in its specific form
and notation it is tailored for
our needs in the subsequent sections. We will discuss the relationship
of our characterisation with others further below in \autoref{section:relationship-characterization}.

Next we define the notions of
\emph{cover descriptions} and \emph{cover partitions}.
Let $Q$ be a conjunctive query.
For a set $\calA \subseteq \bodyof{Q}$ we define
$\bvarsofrel{\calA}$ as the set of \emph{bridge variables} of $\calA$,
that occur in $\calA$ as well as in the head of $Q$ or in some atom
of $Q$ not in $\calA$.
More formally,
 $\bvarsofrel{\calA}=\varsof{\calA}\cap(\varsof{\headof{Q}}\cup \varsof{\bodyof{Q}\setminus \calA})$.

\begin{exa}
	Consider the query $Q$ defined by
	$ H(x,y,z) \gets R(x,u), S(u,y,w), T(y,w,z)$.
	For the set $\calA = \{R(x,u), S(u,y,w)\}$ of atoms from the body of~$Q$, we have $\bvarsofrel{\calA} = \{x,y,w\}$ because~$x$ and~$y$ are head variables of~$Q$ and because~$w$, and also~$y$, occurs in the atom $T(y,w,z)$ that does not belong to~$\calA$. The variable~$u$ in~$\calA$ is not a bridge variable since it is quantified and does not occur in any atom outside of~$\calA$.
\end{exa}

\begin{defi}[Cover Description]\label{definition:cover-desc}
  A \emph{cover description} $C$ for a query $Q$ is a tuple
  \begin{math}
    (\calA,V, \app,\psi)
  \end{math}
  where
  \begin{itemize}
  \item $\calA$ is a subset of $\bodyof{Q}$,
  \item $V$ is a view,
  \item  $\app$ is an application of $V$,  and
    \item $\psi$ is a mapping from $\varsof{\app(V)}$ to $\varsof{Q}$,
    \end{itemize}
    such that
	\begin{enumerate}[label={(\arabic*)}]
		\item \label{achar:cond2}\label{def:cover-desc:cond:phi-bhomo}
		$\calA\subseteq
			\alpha(\bodyof{V})$,
		\item \label{achar:cond4}\label{def:cover-desc:cond:bvars}
			$\bvarsofrel{\calA}\subseteq
			\alpha(\varsof{\headof{V}})$,
                      \item \label{achar:cond1}\label{def:cover-desc:cond:psi-bhomo}
			$\psi$ is a  body homomorphism from $\app(V)$ to
			$Q$, and
		\item \label{achar:cond3}\label{def:cover-desc:cond:identity}
		 $\psi$ is the identity on
			$\varsof{\calA}$.
  \end{enumerate}
\end{defi}

Intuitively, the conditions of \autoref{definition:cover-desc} testify that the atoms of $\Q$ in $\calA$ can be \enquote{represented} or \enquote{covered} by a $V$-atom in a rewriting of $\Q$---giving rise to the term cover description:
Condition~\ref{def:cover-desc:cond:phi-bhomo} means that every atom in $\calA$ has a matching atom in $V$.
Condition~\ref{def:cover-desc:cond:psi-bhomo} ensures that every database which satisfies the query also satisfies the view (under application $\app$).
Conditions~\ref{def:cover-desc:cond:bvars} and~\ref{def:cover-desc:cond:identity} establish an \enquote{interface} to combine multiple cover descriptions in a straightforward and compatible manner such that, overall, a rewriting is described.%

\begin{exa}\label{example:cover-desc}
	Consider the query $Q$ given by the rule
	\[ H(x,y,z) \gets R(x,y,z), T(x,v), F(v), E(w), S(w,z) \]
	and the following views.
	\begin{align*}
		V_1(x_1,y_1,w_1) \gets~& R(x_1,y_1,u_1), T(x_1,v_1), F(v_1), E(w_1), S(w_1,u_1) \\
		V_2(x_2,y_2,z_2) \gets~& R(x_2,y_2,z_2), F(v_2)	\\
		V_3(w_3, z_3) \gets~& S(w_3, z_3), E(w_3)
	\end{align*}
	The tuple $C_1 = (\calA_1,V_1,\app_1,\psi_1)$ with
	$\calA_1 = \{T(x,v), F(v)\}$,
	\begin{align*}
		\app_1 =&~\{ x_1 \mapsto x, y_1 \mapsto y', u_1\mapsto u', v_1 \mapsto v, w_1 \mapsto w' \}\textnormal{, and}\\
		\psi_1 =&~\{ x \mapsto x, y' \mapsto y, u' \mapsto z, v \mapsto v, w' \mapsto w \}
	\end{align*}
	is a cover description for $Q$ with $\bvarsofrel{\calA_1}=\{x\}$.
Although $\psi_1$ could be chosen as \id (by adapting
        $\app_1$  accordingly),  we will see in
        \autoref{example:cover-partition}, that this is not always desirable.
\end{exa}

Now we can simply
characterise rewritability of
a query $Q$  by the existence of a partition of
$\bodyof{Q}$ whose subsets have cover descriptions with views from
$\calV$.

\begin{defi}\label{definition:cover-partition}
  Let $Q$ be a query and $\calV$ be a set of views.
  A collection $\calC = C_1,\ldots,C_m$ of cover descriptions $C_i =
  (\calA_i,V_i,\app_i,\psi_i)$ for $Q$ with $V_i\in \calV$
  is a \emph{cover partition} for $Q$ over $\calV$ if the sets $\calA_1,\ldots,\calA_m$ constitute a partition
          of $\bodyof{Q}$.

\end{defi}

  We call a cover partition \emph{\consistent} if variables of any
  $\alpha_j(V_j)$ are in the range of any other $\alpha_i$ only if
  they also appear in $\bvarsofrel{\calA_j}$.
We note that, since each $\alpha_i$ is a view
application, in a \consistent cover partition, the applications obey
	quantified
variable disjointness.\footnote{
		For the sake of contradiction, assume a quantified variable $x$ and a variable $y$ are mapped to the same variable $z$ by two different view applications $\app_i$ and $\app_j$, respectively.
		Then, due to consistency, $z$ is in $\bvarsof{\calA_i}$ and, thus, in $\app_i(\headof{V_i})$ due to \autoref{definition:cover-desc}\ref{def:cover-desc:cond:bvars}.
		It follows that $\app_i$ unifies $x$ with a head variable.
		But this is a contradiction to $\app_i$ being a view application.
}

\begin{exa}[Continuation of \autoref{example:cover-desc}]\label{example:cover-partition}
	Let $C_1 = (\calA_1,V_1,\app_1,\psi_1)$ be the cover description defined in \autoref{example:cover-desc}.
	In addition, we consider the cover descriptions $C_2$ and $C_3$ with $C_i = (\calA_i,V_i,\app_i,\psi_i)$ for $i\in\{2,3\}$ where
	\begin{align*}
		\calA_2 =&~\{R(x,y,z)\}, && \calA_3 =\{E(w), S(w,z)\},\\
		\app_2 =&~\{x_2 \mapsto x, y_2 \mapsto y, z_2\mapsto z, v_2\mapsto v\}, && \app_3 =\{w_3 \mapsto w, z_3 \mapsto z\},
	\end{align*}
	and $\psi_2=\psi_3=\id$.
	The cover descriptions $C_1,C_2,C_3$ constitute a cover partition for $Q$ over $\{V_1,V_2,V_3\}$.
	It is, however, not \consistent, since $v$ is in the range of $\app_1$ and $\app_2$, but not in $\bvarsof{\calA_1}$.
	Replacing $\app_2$ and $\psi_2$ by mappings $\hat{\app}_2$ and $\hat{\psi}_2$ with $\hat{\app}_2(v_2) = \hat{v}$ and  $\hat{\psi}_2(\hat{v}) = v$ that agree with $\app_2$ and $\psi_2$ on all other variables, respectively, yields a \consistent cover partition.
          Note that there is no \consistent cover partition with a cover description of the
          form $ (\calA_1,V_1,\app'_1,\id)$
          because necessarily $\alpha'_1(w_1)=w$ would hold and thus $w$
          would be in the range of $\app'_1$ and
          $\app_3$.
\end{exa}

A consistent cover partition $\calC$ induces a query $Q_{\calC}$ and
an expansion $Q'_{\calC}$ as follows.
We note first that each variable in
$\headof{Q}$ occurs in at least one of the sets $\calA_i$ and thus in
some set $\bvarsofrel{\calA_i}$. Therefore,
Condition~\ref{achar:cond4} of \autoref{definition:cover-desc} guarantees that each
head variable of $Q$ occurs in some set $\app_i(\headof{V_i})$ and thus
a cover partition $\calC = C_1,\ldots,C_m$ with $C_i =
(\calA_i,V_i,\app_i,\psi_i)$ induces a query $Q_{\calC}$
with
      \begin{align*}
		  \headof{Q_{\calC}} = \headof{Q} \textnormal{ and }
	\bodyof{Q_{\calC}} = \{ \app_i(\headof{V_i}) \mid 1\le i\le m \}.
\end{align*}
Likewise, and thanks to quantified variable
  disjointness, it induces an expansion $Q'_{\calC}$ with
\begin{align*}
	\headof{Q'_{\calC}} = \headof{Q} \textnormal{ and }
	\bodyof{Q'_{\calC}} = \{ \app_i(\bodyof{V_i}) \mid 1\le i\le m \}.
\end{align*}

Next, we show that the existence of a cover partition indeed
characterises rewritability.

\begin{thm}\label{proposition:char-ext}\label{theorem:char-ext}
  Let $Q$ be a minimal conjunctive query and $\calV$ be a set of views. The
  following three statements are equivalent.
	\begin{enumerate}[label={(\alph*)}]
	  \item $Q$ is $\calV$-rewritable.\label{theorem:char-ext:rewriting}
	  \item There is a cover partition~\calC for $Q$ over $\calV$.\label{theorem:char-ext:partition}
	  \item There is a consistent cover partition~\calC for $Q$ over $\calV$.\label{theorem:char-ext:consistent}
  \end{enumerate}
	If~\calC is a consistent cover partition, then $Q_{\calC}$ is a $\calV$-rewriting of $Q$.
		If~\(\calC\) is a cover partition, then there is a consistent cover partition with the same partition of \(\bodyof{Q}\) as \(\calC\).
\end{thm}

\begin{proof}
  We first show that (c) implies (a). To this end, let $\calC = C_1,\ldots,C_m$ be a
      \consistent  cover partition for
	$Q$ over $\calV$, where, for each $i$,
        $C_i=(\calA_i,V_i,\app_i,\psi_i)$. %

        We prove that the query $Q_{\calC}$ induced by $\calC$ is a
        $\calV$-rewriting of $Q$: first, since $\calA_1, \dots, \calA_m$ partition
        $\bodyof{Q}$ and thanks to
		Condition~\ref{achar:cond2}, \id 
		is a homomorphism from $Q$ into the expansion $Q'_{\calC}$.

        Therefore, it suffices to show that the union of the mappings $\psi_1,\dots,\psi_m$ is a homomorphism~$h'$ from $Q'_\calC$ into~$Q$.

        We first show that $h'$ is well-defined: let us assume that a 
        variable $z$ 
        occurs in $\alpha_i(V_i)$ and $\alpha_j(V_j)$ with $i\not=j$. Thanks 
  to \consistency, 
        $z$ is in $\bvarsof{\calA_i}$ and $\bvarsof{\calA_j}$ and
		therefore $\psi_i(z)=\psi_j(z)$ thanks to Condition~\ref{achar:cond3}.

        That $h'$ is a homomorphism follows
        easily, because each $\psi_i$ is a body homomorphism
		by Condition~\ref{achar:cond1} and  $h'$ is the identity on
	$\headof{Q}$ by Condition~\ref{achar:cond3}.

\bigskip
Next, we show that (a) implies (b). Let us assume that $Q$ has a
\calV-rewriting~$Q_R$ with an expansion~$Q_E$ and let 
 the equivalence of~$Q$ and~$Q_E$ be witnessed by homomorphisms~$h$
 from~$Q$ to~$Q_E$ and  $h'$ from~$Q_E$ to~$Q$.
 
Since~$Q$ is minimal, we
can assume, thanks to
\autoref{lemma:characterization-completeness-automorphism}, that $h$ is injective  and $h'$ is the inverse of~$h$ on the atoms of
$h(\bodyof{Q})$. Since $h$ is injective,  we can further assume,
without loss of generality, that~$h$ is the identity mapping on the
variables in~$Q$ and that $h'(x)=x$ for every such variable.\footnote{This can be easily achieved by renaming the variables of
  $Q_R$ and $Q_E$ appropriately.} 

That is, we have $\headof{Q} = \headof{Q_E} = \headof{Q_R}$ as well as
$\bodyof{Q} \subseteq \bodyof{Q_E}$.  

Let $\app_1,\dots,\app_m$ be applications of views $V_1,\dots,V_m$ that result in expansion~$Q_E$, that is,
\begin{displaymath}
	\bodyof{Q_E} =
	\bigcup_{i=1}^m\bodyof{\app_i(V_i)},
\end{displaymath}
where the applications fulfil the quantified variable disjointness
property. These applications induce a partition
$\calA_1,\dots,\calA_m$  of~\bodyof{Q_E} in the following way:
For every $i\in\{1,\ldots,m\}$ we define
\begin{align*}
	\calA_i = \{ A\in\bodyof{Q} \mid &~\textnormal{$i$ is minimal
                                           such that }A\in \app_i(\bodyof{V_i})\}.
\end{align*}

Now, let $C_i = (\calA_i,V_i,\app_i,h'_i)$, where $h'_i$ is the
restriction of~$h'$ to the variables in~$\app_i(V_i)$, for every $i
\in \{1,\dots,m\}$.  To show that 
$C_1,\dots,C_m$ yield a cover partition, it only remains to
show that each~$C_i$ is a cover description. 

To this end, we show that all four conditions in
\autoref{definition:cover-desc} are satisfied for $C_i$, for every $i \in
\{1,\dots,m\}$. Condition~\ref{def:cover-desc:cond:phi-bhomo} holds by the definition of~$\calA_i$. 
Condition~\ref{def:cover-desc:cond:psi-bhomo} is true because each
$h'_i$ is a restriction of the body homomorphism $h'$. %
Condition~\ref{def:cover-desc:cond:identity} follows since
$h'$ is the identity on all variables in~\bodyof{Q} and thus
also on \emph{all} variables of $\calA_i \subseteq \bodyof{Q}$.

Hence, it only remains to show that
Condition~\ref{def:cover-desc:cond:bvars} is satisfied. To this end,
let~$x$ be an arbitrary bridge variable in some~$\calA_i$.
If~$x$ occurs in a subset~$\calA_j$ where $j \neq i$, then it is a head variable in both~$\app_i(V_i)$ and~$\app_j(V_j)$ because of the quantified variable disjointness of $\app_1,\dots,\app_m$.
Otherwise, $\calA_i$ is the only subset containing variable~$x$, which thus has to be a head variable of~$Q$ in order to be a bridge
variable in~$\calA_i$. Because of $\headof{Q} = \headof{Q_R}$, it is
then also a head variable of~$Q_R$. This, in turn, implies that~$x$ is
a head variable of~$\app_i(V_i)$ because the quantified variables of
$\app_i(V_i)$ do \emph{not} occur in~$Q_R$.

\bigskip
Finally, we show that (b) implies (c).  Let thus $\calC = C_1,\ldots,C_m$ be a
      cover partition for
	$Q$ over $\calV$, where, for each $i$,
        $C_i=(\calA_i,V_i,\app_i,\psi_i)$. Let $z$ be a variable that
        occurs in some $\app_i(V_i)$, but not in $\bvarsof{\calA_i}$,
        and also in some $\app_j(V_j)$ with $j\not=i$. Since
        $z\not\in\bvarsof{\calA_i}$ we have
        $z\not\in\varsof{\calA_j}$. We define $\app'_j$ like $\app_j$
        but, for some fresh variable $z'$, we set $\app'_j(x)=z'$
        whenever  $\app_j(x)=z$. Accordingly, we define $\psi'_j$ like
        $\psi_j$ but with $\psi'_j(z')=\psi_j(z)$. It is easy to verify
        that $C'_j=(\calA_j,V_j,\app'_j,\psi'_j)$ is a cover
        description, as well. By repeating this process, a \consistent
        cover partition  for
	$Q$ over $\calV$ can be obtained.
 \end{proof}

The previous proof used the following lemma. It is similarly stated in the proof of Lemma~2 in
\cite{DBLP:conf/ijcai/ChenGLP20}, see also the proof of Lemma~9 in the full version~\cite{DBLP:journals/corr/abs-2007-14169} of \cite{DBLP:conf/ijcai/ChenGLP20} for the proof idea.
We give its short proof to
keep the paper self-contained.
\begin{lem}\label{lemma:characterization-completeness-automorphism}
	Let $Q_1$ be a minimal conjunctive query, $Q_2$ a conjunctive
        query equivalent to $Q_1$ and  $\homi{1}{Q_1}{Q_2}$ a
        homomorphism. Then, there is a homomorphism
        $\homi{2}{Q_2}{Q_1}$ that is the inverse of $h_1$ on $h_1(\bodyof{Q_1})$.
\end{lem}
\begin{proof}
	Since $Q_1$ and $Q_2$ are equivalent, there is a homomorphism $h_2'\colon Q_2\rightarrow Q_1$.

	The mapping $h_2'\circ h_1$ is an
        automorphism\footnote{We note that compositions of homomorphisms are
          applied from right to left.} on $Q_1$, because $Q_1$ is minimal.
          Since the automorphisms of $Q_1$ constitute a group,
there is some $k > 0$ such that $(h_2'\circ h_1)^k$  is the identity on $Q_1$.
	We choose $h_2 = (h_2'\circ h_1)^{k-1}\circ h'_2$.
	Clearly, $h_2$ is a homomorphism from $Q_2$ to $Q_1$ and $h_2$
        is the inverse of $h_1$ on $h_1(Q)$. 
\end{proof}
 
\begin{rem}\label{remark:minimality}
	We note that the requirement in \autoref{theorem:char-ext} for $Q$ to be minimal is only seemingly a restriction in our setting, since we apply it only to acyclic queries.
	If $Q$ is acyclic but not minimal, an equivalent minimal query $Q'$ can be computed in polynomial time by iteratively removing atoms from its body \cite{DBLP:conf/stoc/ChandraM77,DBLP:journals/tcs/ChekuriR00}.
	Moreover, it is guaranteed that the minimal query $Q'$ is also acyclic (we believe this to be folklore, it follows readily from more general results, cf.\ for instance \cite{DBLP:journals/sigmod/BarceloPR17}).

          The same is true for free-connex acyclic queries: every
          homomorphism from $Q$ to $Q'$ is also a homomorphism from
          $\bodyof{Q} \cup \{\headof{Q}\}$ to
          $\bodyof{Q'}\cup\{\headof{Q'}\}$ and vice versa, since the
          relation symbol of $\headof{Q} = \headof{Q'}$ does not occur
          in $\bodyof{Q'}$.  Thus, $Q'$ is minimal if and only if the
          Boolean query whose body is $\bodyof{Q'}\cup\{\headof{Q'}\}$
          is minimal.  Therefore, if $\bodyof{Q} \cup \{\headof{Q}\}$
          is acyclic, so is $\bodyof{Q'} \cup \{\headof{Q'}\}$.  In
          other words, if $Q$ is free-connex acyclic, then $Q'$ is
          free-connex acyclic as well.

          For hierarchical and q-hierarchical queries the same holds:
          it is easy to see that removing atoms does not change the
          conditions in their respective definitions.
\end{rem}

\section{Towards Acyclic Rewritings}\label{section:towards-acyclic-rewritings}
In this section, we turn our focus to the main topic of this paper:
acyclic rewritings and the decision problem that asks whether such a
rewriting exists.
We study the complexity of $\rewrprob{\ViewClass}{\QueryClass}{\ACQ}$ for the case that \ViewClass
and~\QueryClass are the class of conjunctive queries and for various
subclasses. It will be helpful to analyse the case that
$\QueryClass=\ACQ$ and $\ViewClass=\CQ$ first.

\subsection{A Characterisation of Acyclic Rewritability for Acyclic Queries}

The following example illustrates that,
even if
an acyclic rewriting exists,   the
        canonical rewriting need not be acyclic. Furthermore, it may
		be that each \enquote{sub-rewriting} of the canonical
        candidate is cyclic or not a rewriting, and thus none of them
        is an  acyclic rewriting.

\begin{exa}\label{example:cyclic-rewriting}
	Consider the query $Q$ given by the rule
	\[ H() \gets R_1(x), R_2(y), S(x,z), T_1(z), T_2(y) \]
	and the following views $\calV = \{V_1,V_2,V_3\}$.
	\begin{align*}
		V_1(u_1, v_1) \gets R_1(u_1), R_2(v_1)\quad\quad
		V_2(u_2, v_2) \gets S(u_2, v_2)\quad\quad
		V_3(u_3, v_3) \gets T_1(u_3), T_2(v_3)
	\end{align*}
	The canonical candidate $Q_R = H() \gets V_1(x,y), V_2(x,z),
        V_3(z,y)$ is a \calV-rewriting, but it is \emph{cyclic}. Each
        query whose body is a proper subset of the body of $Q_R$ is
        not a \calV-rewriting for~\Q. However, an \emph{acyclic}
        \calV-rewriting for~\Q exists. One such rewriting is the
        query~$\Q'_R$.
	\[ H() \gets V_1(x,y), V_2(x, z), V_3(z, y'), V_3(z', y) \]
\end{exa}

      Even though \autoref{example:cyclic-rewriting} suggests that general and
      acyclic rewritings can be seemingly unrelated, it turns out that
      there is a close connection between them. In fact, we will show next
      that an acyclic \calV-rewriting
      exists if and only if an arbitrary \calV-rewriting
      exists. Furthermore, from an arbitrary rewriting an
			acyclic rewriting can always be constructed.

      Towards a proof, we
        study the relationship between a query~\Q and the
        view applications that can occur in any rewriting of~\Q more
        closely, to determine the circumstances under which a
        decomposition of view atoms is possible.

\begin{exa}[Continuation of \autoref{example:cyclic-rewriting}]\label{cyclic-rewriting-escaleted-ii}
		We first have a closer look at the $V_3$-atoms in the
        rewritings of the previous example.
         	The two \(V_3\)-atoms in \(Q'_R\) can be understood as a (sub-)query with body \(\{V_3(z,y'), V_3(z',y)\}\) and head variables \(z,y\), whereas the \(V_3\)-atom in \(Q_R\) can be understood as a (sub-)query with body \(\{V(z,y)\}\) and head variables \(z,y\).
			The expansions of these two (sub-)queries are equivalent.
          That is, the acyclic rewriting can be obtained
                 from the canonical rewriting by replacing an atom by a set of atoms
                 that is equivalent with respect to view
                 expansions.\footnote{We note that one could also replace the
                   $V_1$-atom in $Q_R$.}
\end{exa}

In \autoref{example:cyclic-rewriting} the acyclic rewriting was
obtained from the canonical rewriting by replacing a view atom by a
set that contained one view atom for each connected component of (the
hypergraph induced by) $V_3$.
The following example shows that the required
modifications can be more involved.
\begin{exa}\label{cyclic-rewriting-escaleted}
	Consider the query $Q$ given by the rule
	\[ H(x,y,z) \gets R(x,y,z), E_1(x), E_2(y), E_3(w), S(w,z) \]
	and the following views.
	\begin{align*}
		V_1(x_1,y_1,w_1) \gets~& R(x_1,y_1,v_1), E_1(x_1), E_3(w_1), S(w_1,v_1) \\
		V_2(x_2,y_2,z_2) \gets~& R(x_2,y_2,z_2), E_2(y_2)	\\
		V_3(w_3, z_3) \gets~& S(w_3, z_3), E_3(w_3)
	\end{align*}
	The canonical candidate
	\begin{math}
        H(x,y,z) \gets V_1(x,y,w), V_2(x,y,z), V_3(w, z)
	\end{math}
	is a cyclic $\{V_1,V_2,V_3\}$-rewriting of $Q$.
	In contrast to \autoref{example:cyclic-rewriting}, all view
        bodies are connected. Therefore, replacing a view atom by a
set of representatives of  connected components of some views does not
yield an acyclic rewriting.
	However, there is an acyclic $\{V_1,V_2,V_3\}$-rewriting of
        $Q$, namely
	\begin{math}
          H(x,y,z) \gets V_1(x, y', w'), V_2(x,y,z), V_3(w,z).
	\end{math}
\end{exa}

Now we turn to the main result of this section.

\begin{thm}\label{proposition:exacyclicrewr}\label{theorem:exacyclicrewr}
	Let \(Q\) be a conjunctive query and \(\calV\) be a set of views.
	\begin{enumerate}[(a)]
		\item If \(Q\) is acyclic and \(\calV\)-rewritable,
                  then it has is an acyclic \(\calV\)-rewriting.
		\item If \(Q\) is free-connex acyclic and
                  \(\calV\)-rewritable, then it has  a free-connex acyclic \(\calV\)-rewriting.
	\end{enumerate}
\end{thm}

\begin{proof}
Since $Q$ is acyclic, we can assume that it is minimal (cf.\ \autoref{remark:minimality}).
Moreover, it has a join tree $J_Q$, and, thanks to
    \autoref{theorem:char-ext}, since $Q$ is $\calV$-rewritable, there is a
        \consistent cover partition
         $\calC = C_1,\ldots,C_m$ for $Q$ over $\calV$ and the query $Q_\calC$
        is a $\calV$-rewriting of $Q$. For each $i$, let 
        $C_i=(\calA_i,V_i,\app_i,\psi_i)$.

		We show first that $Q_{\calC}$ is acyclic if each set
                $\calA_i$ is connected in $J_Q$.
		Afterwards we show that a \consistent cover partition with that
                property can always be constructed from $\calC$.
      
	To this end, we show how to construct a  join tree $J$ for
        $\Q_\calC$: we first cluster, for each $j$, the nodes for
        $\calA_j$  in $J_Q$ together into one node that is labelled by  $\alpha_j(\headof{V_j})$.
	Since the $\calA_j$ are connected in $J_Q$, the resulting graph $J$ is a tree.

To verify that $J$ is a join tree, let us consider two nodes $u,v$ of $J$
labelled by $\alpha_j(\headof{V_j})$ and $\alpha_k(\headof{V_k})$, respectively, and $x$ be a variable that appears in $\alpha_j(\headof{V_j})$ \emph{and} $\alpha_k(\headof{V_k})$.
	Thanks to \consistency of $\calC$, variable~$x$ appears in $\bvarsofrel{\calA_j}$ and in $\bvarsofrel{\calA_k}$.
	It follows that $x$ appears in two atoms $A\in \calA_j$ and $A'\in \calA_k$.
	Moreover, since $J_Q$ is a join tree, $x$ appears in every node on the (shortest) path from $A$ to $A'$ in $J_Q$.
	Again thanks to \consistency, $x$ is in $\bvarsofrel{\calA_\ell}$ for every
        $\alpha_\ell(\headof{V_\ell})$ along the corresponding contracted path
        in $J$ from $u$ to $v$.
        In particular, $x$ appears in all sets $\alpha_\ell(\headof{V_\ell})$ on the path from $u$ to $v$.
	Thus, $u$ and $v$ are $x$-connected.
	We conclude that $J$ is a join tree for $\Q_\calC$.
	Hence, $\Q_\calC$ is acyclic.

	If $Q$ is also free-connex acyclic, there is a join tree $J_Q^+$ for $\bodyof{Q}\cup \{\headof{Q}\}$.
	As we show later, we can assume that the sets $\calA_i$ of the cover partition $\calC$ are also connected in $J_Q^+$.
	Clustering the nodes\footnote{The node labelled \headof{\Q} is not clustered with any other node, since it does not occur in any $\calA_i$.} of $J_Q^+$ analogously as for $J_Q$ yields a join tree for $\bodyof{Q_\calC}\cup \{\headof{Q_\calC}\}$ since $\headof{Q_\calC} = \headof{Q}$ and all head variables in a set $\calA_i$ also occur in the new label $\alpha_i(\headof{V_i})$ thanks to Condition~\ref{def:cover-desc:cond:bvars} of \autoref{definition:cover-desc}.

        We now show how, from a \consistent cover partition $\calC$, we can construct  a \consistent cover partition  $\calC' = C'_1,\ldots,C'_m$
        such that each set $\calA'_i$ is
        connected.  To this end, 	let us assume that some set
        $\calA_j$   is not connected in $J_Q$. 	Let $\calB_j\subseteq
        \calA_j$ be a maximal connected subset and let $\calA_j' = \calA_j\setminus \calB_j$.
	We observe that each variable $x$ that appears in $\calB_j$ and
        $\calA_j'$, is also in $\bvarsofrel{\calA_j}$, since $x$ has
        to occur in at least one other atom in $J_Q$ (otherwise,
        $\calB_j$ and $\calA_j'$ would be connected in $J_Q$).
	Thus, $\bvarsofrel{\calB_j}\subseteq \bvarsofrel{\calA_j}$.
	We conclude that $C_{\calB_j} = (\calB_j,V_j,\app_j,\psi_j)$ is a
        cover description.
		Likewise, $C_j' = (\calA_j',V_j,\app_j,\psi_j)$ is a
                cover description. By repeated application of  this
                modification step, we eventually obtain a cover partition
                $\calC' = C'_1,\ldots,C'_m$
        in which  each set $\calA'_i$ is
        connected.
		If $\Q$ is free-connex, then $\calC'$ can be further refined as described above but w.r.t.\ $J_Q^+$ instead of $J_Q$.
		Refining the cover partition iteratively w.r.t.\ $J_Q$ and $J_Q^+$ yields a cover partition $\calC'' = C''_1,\ldots,C''_p$ in which each set $\calA''_i$ is connected in $J_Q$ and $J_Q^+$.
		The number of iterations is bounded by the number of atoms of $Q$.

		Thanks to \autoref{theorem:char-ext}, $\calC''$ can be
        turned into a consistent cover partition (and the required
				renamings of variables do not affect the connectedness).
 \end{proof}

\autoref{theorem:exacyclicrewr} delivers good news as well as bad news.
The good news is that, since the proofs of \autoref{theorem:exacyclicrewr}
and \autoref{theorem:char-ext} are constructive, we altogether have a procedure to construct an acyclic rewriting
from an arbitrary rewriting $Q_R$ if the query $Q$ is acyclic.
Furthermore, if a homomorphism from an expansion of $Q_R$ to $Q$ can be computed in polynomial time,
the overall procedure can be performed in polynomial time.
In particular, we get the following collary.
\begin{cor}\label{corollary:arewrprobk-tractable}
For every $k$,  $\rewrprob{\ACQ^k}{\ACQ}{\ACQ}$ is in polynomial time, and an acyclic
  rewriting can be computed in polynomial time (if it exists).
\end{cor}
That the decision problem is in polynomial time follows immediately
from \autoref{proposition::bounded-head-acyclic} and
\autoref{theorem:exacyclicrewr}.
	Moreover, the canonical candidate for an acyclic query $Q$ and a set $\calV$ of acyclic views with bounded arity can be computed in polynomial time, since the size of $\calV(\candof{Q})$ is bounded by a polynomial and query evaluation for acyclic queries is in polynomial time~\cite{DBLP:conf/vldb/Yannakakis81,DBLP:books/aw/AbiteboulHV95}.
	In case the canonical candidate is a rewriting, valuations witnessing $\calV(\candof{Q})$ can be computed in polynomial time~\cite{DBLP:conf/ngits/KimelfeldS06} and combined into a homomorphism from an expansion of the rewriting to $Q$.
	Thus, an acyclic rewriting can then be efficiently constructed by the above procedure.

However, this leaves open the complexity of
$\rewrprob{\ACQ}{\ACQ}{\ACQ}$ (as well as that of
$\rewrprob{\ACQ}{\ACQ}{\CQ}$), and of their restrictions to schemas
of bounded arity.
The bad news is that, since $\rewrprob{\ViewClass}{\ACQ}{\CQ}$ and
$\rewrprob{\ViewClass}{\ACQ}{\ACQ}$  are basically the same problem, lower bounds on $\rewrprob{\ViewClass}{\ACQ}{\CQ}$
transfer to $\rewrprob{\ViewClass}{\ACQ}{\ACQ}$.
In fact, we get the
following, due to \classNP-hardness of  $\rewrprobk{\CQ}{\ACQ}{\CQ}$ in the general case
(\autoref{proposition:np-hard-simple}).\footnote{We note that \autoref{corollary:arewr-np-hard-simple} also follows from the proof of \autoref{proposition:np-hard-simple}, since the canonical candidate constructed in that proof is trivially acyclic. Indeed, \classNP-hardness of $\rewrprob{\ACQ}{\CQ}{\ACQ}$ is also implied.}

\begin{cor}\label{corollary:arewr-np-hard-simple} For every
  $k\ge 3$,
  the problem $\rewrprobk{\CQ}{\ACQ}{\ACQ}$ and, therefore, also
  $\rewrprobk{\CQ}{\CQ}{\ACQ}$ is \classNP-hard.
\end{cor}

In the next part of this section, we resolve the complexity of
$\rewrprob{\ACQ}{\ACQ}{\CQ}$ and $\rewrprob{\ACQ}{\ACQ}{\ACQ}$ and  their restrictions to schemas
of bounded arity.

\subsection{The Complexity of Acyclic Rewritability for Acyclic
  Queries}
It may be tempting to assume that, since acyclic queries are so
well-behaved in general, it should be tractable to decide whether for
an acyclic query and a set of acyclic views there exists an acyclic
rewriting.
However, as we show next, this is (probably) not the case, and this
surprising finding even holds for the even better behaved hierarchical
queries as well.

\begin{thm}%
	\label{thm:arewriting-np-hard}%
	The problems $\rewrprobk\ACQ\ACQ\CQ$ and $\rewrprobk\ACQ\ACQ\ACQ$, as well as
        $\rewrprobk\HCQ\HCQ\CQ$ are \classNP-complete, for $k\ge
        3$.\footnote{
            These results hold for schemas of unbounded arity, as
            well.
} The lower bounds even hold for instances with a single view.
      \end{thm}

      Of course, \autoref{thm:arewriting-np-hard} immediately implies \classNP-hardness of
 \(\rewrprobk\ViewClass\QueryClass\ACQ\), for all pairs \(\ViewClass, \QueryClass\) of classes with \(\ACQ\subseteq\ViewClass\subseteq\CQ\) and \(\ACQ\subseteq\QueryClass\subseteq\CQ\).

      We will see in \autoref{section:hierarchical} that \autoref{theorem:exacyclicrewr} has an
      analogue for hierarchical queries from which it can be concluded
      that deciding the existence of a hierarchical rewriting,
      given a hierarchical query and hierarchical views, is still
      \classNP-complete (cf.\ \autoref{corollary:hcq-np-complete}).

\autoref{thm:arewriting-np-hard} will easily follow from
\classNP-hardness of a seemingly simpler problem.
From the characterisation in \autoref{theorem:char-ext}, we
already know that deciding the existence of a rewriting is the same as
deciding the existence of a cover partition. We show next that it is
even \classNP-hard to decide the existence of a \emph{cover
  description}, given a query, a set of atoms, and a single view.

\begin{defi}
	Let $\ViewClass\subseteq\CQ$ and $\QueryClass\subseteq\CQ$ be classes of conjunctive queries.
	The \emph{cover description problem} for~\ViewClass and~\QueryClass, denoted $\coverdescprob{\ViewClass}{\QueryClass}$ asks, upon input of a query $Q\in\QueryClass$, a subset $\calA \subseteq \bodyof{Q}$ and a view $V \in \ViewClass$, whether mappings~\app and~$\psi$ exist such that $(\calA,V,\app,\psi)$ is a cover description.
\end{defi}
\begin{thm}%
	\label{theorem:coverdesc-np-hard}%
	\label{theorem:acyclic-np-hard}%
	$\coverdescprob\ACQ\ACQ$ is \classNP-hard. Indeed, even $\coverdescprob\HCQ\HCQ$ is \classNP-hard and even if the input is restricted to $\calA = \bodyof{\Q}$.
\end{thm}
\begin{proof}
	\newnotation{\relForm}[1][\wproj]{\RelForm(\wfalse,\wtrue,#1)}%
	\newnotation{\relClause}[2]{\RelClause_{#1}(#2,\wproj)}%
	\newnotation{\relClauseF}[1]{\relClause{#1}{\wfalse,\wtrue}}%
	\newnotation{\relClauseT}[1]{\relClause{#1}{\wtrue,\wfalse}}%
	\newnotation{\relNeg}[1]{\RelNeg_{#1}(\wfalse,\wtrue,\wproj)}%
	\newnotation{\relNegA}[1]{\RelNeg_{#1}(\wtrue,\wfalse,\wproj)}%
	We reduce problem~\ThreeSat to $\coverdescprob\HCQ\HCQ$.
	This lower bound directly translates to $\coverdescprob\ACQ\ACQ$ since every hierarchical query is acyclic.
	For a formula~\formula in 3CNF, we describe how a query~\Q and a view~\V can be derived in polynomial time such that both \Q~and~\V are hierarchical and such that~\formula is satisfiable if and only if
	there are mappings $\app$ and~$\psi$ such that $(\bodyof{\Q},V,\app,\psi)$ is a cover description for~\Q.
		In the following we will use the term \emph{proposition} for propositional variables of a formula in order to easily distinguish them from variables occurring in queries and views.

	\paragraph*{Construction.}
	Let $\formula = \formula_1 \land \dots \land \formula_k$ be a propositional formula in~3CNF over propositions $x_1,\dots,x_n$ in clauses $\formula_1,\dots,\formula_k$, where~$\formula_j = (\lit{j,1} \lor \lit{j,2} \lor \lit{j,3})$ for each $j \in \{1,\dots,k\}$.

	We start with the construction of query~\Q. This query is Boolean, with head $H()$, and uses only three variables $\wfalse$, $\wtrue$ and $\wproj$, where the first two are intended to represent the truth values false and true.
	The structure of~\Q is depicted in \autoref{fig:hardness:query-structure}. The body of~\Q is defined as the union
	\begin{displaymath}
		\{\relForm\}
		\uplus \ClausesFalse
		\uplus \ClausesTrue
		\uplus \Neg{1}
		\uplus \dots
		\uplus \Neg{n}
	\end{displaymath}
	of the following sets of atoms.
	\begin{itemize}
		\item Set~\ClausesFalse contains an atom $\relClauseF{j}$ for each clause~$\formula_j$ in~\formula.

			\smallskip\noindent
			Intuitively, these atoms represent \emph{unsatisfied} clauses as opposed to the next atoms that represent \emph{satisfied} clauses. Note that these atoms differ in the order of variables~\wfalse and~\wtrue only.

		\item Set~\ClausesTrue contains an atom $\relClauseT{j}$ for each clause~$\formula_j$ in~\formula.

		\item There is a set~$\Neg{i}$ for each proposition~$x_i$ in formula~\formula.
			The sets differ only in the name of the relation they address and contain two atoms~\relNeg{i} and~\relNegA{i} each.
	\end{itemize}
	Query~\Q is hierarchical since every atom in each of the sets above refers to all three variables~\wfalse, \wtrue and~\wproj. Thus, query~\Q is acyclic in particular.

	\begin{figure}[t]
		\newcommand{\RESIZE}[1]{%
			\makesamewidthandheight%
				{\LitNeg{n}}%
				{\LitPos{n}}%
				{#1}%
		}%
		\begin{subfigure}[t]{.49\textwidth}
			\centering
			\resizebox{\textwidth}{!}{%
			\begin{forest}
				for tree={query}
				[\relForm
					[\RESIZE{\ClausesFalse},subquery]
					[\RESIZE{\ClausesTrue},subquery]
					[\RESIZE{\Neg{1}},subquery]
					[\RESIZE{\;\dots\;},subquery,draw=none,no edge]
					[\RESIZE{\Neg{n}},subquery]
				]
			\end{forest}
			}%
		\caption{Structure of query~$\boldsymbol\Q$, induced by formula~$\boldsymbol\formula$.}
		\label{fig:hardness:query-structure}
		\end{subfigure}
		\begin{subfigure}[t]{.49\textwidth}
			\centering
			\resizebox{\textwidth}{!}{%
			\begin{forest}
				for tree={query}
				[\relForm
					[\RESIZE{\ClausesFalse},subquery]
					[\RESIZE{\ClausesTrue},subquery,phantom]
					[\RESIZE{\VarNeg{1}},subquery
						[\RESIZE{\LitPos{1}},subquery]
						[\RESIZE{\LitNeg{1}},subquery]
					]
					[{\;\dots\;},subquery,draw=none,no edge]
					[\RESIZE{\VarNeg{n}},subquery
						[\RESIZE{\LitPos{n}},subquery]
						[\RESIZE{\LitNeg{n}},subquery]
					]
				]
			\end{forest}
			}%
		\caption{Structure of view~$\boldsymbol\V$, induced by formula~$\boldsymbol\formula$.}
		\label{fig:hardness:view-structure}
		\end{subfigure}
		\caption{Structure of the query~$\boldsymbol\Q$ and the view~$\boldsymbol\V$, induced by formula~$\boldsymbol\formula$.}
	\end{figure}

	We now proceed with the construction of view~\V, which refers to the same relations as query~\Q. Like query~\Q, the view uses variables~\wfalse, \wtrue and~\wproj but also additional variables for the propositions $x_1,\dots,x_n$ in formula~\formula. For each proposition~$x_i$, there are two variables~$x_i$ and~$\nx_i$, intended to represent the positive and negated literal over the proposition, respectively. With the exception of variable~\wproj, all other variables are in the head
	\begin{math}
		H(\wfalse,\wtrue,x_1,\nx_1,\dots,x_n,\nx_n)
	\end{math}
	of view~\V. The body of~\V, whose structure is depicted in \autoref{fig:hardness:view-structure}, is defined as the union
	\begin{displaymath}
		\{\relForm\}
		\uplus \ClausesFalse
		\uplus
		\big(
			\VarNeg{1}
			\uplus
			\LitPos{1}
			\uplus
			\LitNeg{1}
		\big)
		\uplus
		\dots
		\uplus
		\big(
			\VarNeg{n}
			\uplus
			\LitPos{n}
			\uplus
			\LitNeg{n}
		\big)
	\end{displaymath}
	of sets of atoms, where the first two sets~$\{\relForm\}$ and~\ClausesFalse are the same as above.
	The bodies of query~\Q and view~\V thus differ in the sets~\ClausesTrue and $\Neg{1}, \dots, \Neg{n}$ of atoms on the one hand as well as $\VarNeg{1},\dots,\VarNeg{n}$ and $\LitPos{1},\dots,\LitPos{n}$ and $\LitNeg{1},\dots,\LitNeg{n}$ on the other hand.
	Sets $\Neg{1}, \dots, \Neg{n}$ and $\VarNeg{1}, \dots, \VarNeg{n}$ are intended to guide the assignment of propositions, while sets $\LitPos{1}, \dots, \LitPos{n}$ and $\LitNeg{1}, \dots, \LitNeg{n}$ represent the occurrences of literals in the clauses of formula~\formula, and set~\ClausesTrue represents the aim to satisfy all of them. The new sets are defined as follows.
	\begin{itemize}
		\item For each proposition~$x_i$, set $\VarNeg{i}$ contains atoms $\RelNeg_i(x_i,\nx_i,\wproj)$ and~$\RelNeg_i(\nx_i,x_i,\wproj)$, intended to enforce the mapping of $(x_i,\nx_i)$ to either $(\wfalse,\wtrue)$ or $(\wtrue,\wfalse)$, that is, to \enquote{complementary truth values}.

		\item For each positive literal~$x_i$ and each clause~$\formula_j$ that contains literal~$x_i$, set~\LitPos{i} contains an atom \relClause{j}{x_i,\nx_i}.

		\item For each negated literal~$\nx_i$ and each clause~$\formula_j$ that contains literal~$\neg x_i$, set~\LitNeg{i} contains an atom \relClause{j}{\nx_i,x_i}.
	\end{itemize}
	Thus defined, the view is also hierarchical as shown in the following.
	 Every atom in every set contains variable~\wproj. Therefore, it follows that $\atoms(x) \subseteq \atoms(\wproj)$ for all $x \in \varsof{\V}$. Furthermore, all atoms in~$\{\relForm\}$ and~\ClausesFalse additionally contain both variables~\wfalse and~\wtrue and they are the only atoms to contain these variables. Thus, we have that $\atoms(\wfalse)  = \atoms(\wtrue)$  and $\atoms(y) \cap \atoms(\wfalse) = \emptyset$  and $\atoms(y) \cap \atoms(\wtrue) = \emptyset$  for all $y \in \varsof{V}\setminus\{u,\wfalse,\wtrue\}$.
	 Similarly, for each~$i \in \{1,\dots,n\}$, all atoms in $\VarNeg{i} \cup \LitPos{i} \cup \LitNeg{i}$ contain both variables~$x_i$ and~$\nx_i$, which do not occur in any other atom. It follows that $\atoms(x_i) = \atoms(\nx_i)$ and $\atoms(\nx_i) \cap \atoms(y) = \emptyset$ and  $\atoms(x_i) \cap \atoms(y) = \emptyset$  for all $y \in \varsof{V}\setminus\{x_i,\nx_i,u\}$ and $i \in \{1,\ldots,n\}$.
  Therefore, we obtain that each pair of variables in $\V$ satisfies at least one of the conditions in \autoref{definition:qhierarchical} and thus $\V$ is hierarchical---and acyclic in particular.

	For the correctness argument, however, the structure of the join tree of view~\V, depicted in \autoref{fig:hardness:view-structure} is probably more helpful.
	We close the description of the construction by remarking that the query and view can be computed in polynomial time for any given propositional formula in~3CNF.

\paragraph*{Correctness.}
	We prove now that formula~\formula is satisfiable if and only if
	there are mappings~\app and~$\psi$ such that $(\bodyof{\Q},V,\app,\psi)$ is a cover description for~\Q.
		\smallskip\noindent
		\textit{%
			First, we show that if formula~\formula is satisfiable,
			then there are mappings~\app and~$\psi$ such that $(\bodyof{Q},V,\app,\psi)$ is a cover description for~\Q.
		}

		To this end, let us assume that~\formula is satisfiable and that this is witnessed by a satisfying truth assignment~\tass. From this assignment for propositions $x_1,\dots,x_n$, we derive the application~\app of~\V. Let~\app be the mapping that is the identity on variables~\wfalse, \wtrue and~\wproj and behaves as follows on the literal variables, for every $i \in \{1,\dots,n\}$.
		\begin{displaymath}
			\app(x_i,\nx_i) =
			\begin{cases}
				(\wfalse,\wtrue)
				& \text{if $\tass(x_i)=0$}
				\\%
				(\wtrue,\wfalse)
				& \text{if $\tass(x_i)=1$}
			\end{cases}
		\end{displaymath}
		We choose mapping~$\psi$ to be the identity on the variables of~$\app(V)$. Then $(\bodyof{Q},V,\app,\psi)$ is a cover description for~\Q because
		application~\app establishes the following relationships between the atoms of view~\V (on the left-hand side) and the atoms of query~\Q (on the right-hand side).
			\begin{enumerate}[label={(\roman*)}]%
			\item\label{prop:clauses-false}%
				$\app(\ClausesFalse) = \ClausesFalse$
			\item\label{prop:neg}
				$\app(\VarNeg{1}) = \Neg{1}$, \dots, $\app(\VarNeg{n}) = \Neg{n}$
			\item\label{prop:correct}
				$\app(\LitPos{1} \cup \LitNeg{1} \cup \dots \cup \LitPos{n} \cup \LitNeg{n}) \subseteq \ClausesFalse \cup \ClausesTrue$
			\item\label{prop:complete}
				$\app(\LitPos{1} \cup \LitNeg{1} \cup \dots \cup \LitPos{n} \cup \LitNeg{n}) \supseteq \ClausesTrue$
		\end{enumerate}
			Property~\ref{prop:clauses-false} is rather obvious since the application~\app behaves like the identity on the variables~\wfalse, \wtrue and~\wproj in~\ClausesFalse.
		Next, let us consider the set $\VarNeg{i} = \{\relNeg{i},\relNegA{i}\}$ for an arbitrary $i \in \{1,\dots,n\}$. By definition, application~\app maps $(x_i,\nx_i)$ either to $(\wfalse,\wtrue)$ or $(\wtrue,\wfalse)$. In each case, we get
		\begin{displaymath}
			\app(\VarNeg{i}) =
			\{
				\RelNeg_i(\wfalse,\wtrue,\wproj),
				\RelNeg_i(\wtrue,\wfalse,\wproj)
			\}
			= \Neg{i}.
		\end{displaymath}
			Hence, Property~\ref{prop:neg} holds too.
			Again by definition of~\app, the atoms in $\LitPos{1} \cup \dots \cup \LitPos{n}$, which are of the form $\relClause{j}{x_i,\nx_i}$, are mapped to \relClauseF{j} or \relClauseT{j}, contained in $\ClausesFalse \cup \ClausesTrue$. An analogous argument holds for the atoms in $\LitNeg{1} \cup \dots \cup \LitNeg{n}$. Thus, Property~\ref{prop:correct} is also satisfied.
			Finally, we show that Property~\ref{prop:complete} holds. To this end, let \relClauseT{j} be an arbitrary atom in~\ClausesTrue. By our assumption, clause~$\formula_j$ in formula~\formula is satisfied by truth assignment~\tass. Hence, there is a literal~\lit{j,h} in clause~$\formula_j$ such that $\tass \models \lit{j,h}$. Let us assume $\lit{j,h} = x_i$
		for some $i \in \{1,\dots,n\}$; the argument for a negated literal is analogous. This assumption implies $\app(x_i) = \wtrue$ by the definition of~\app, which depends on~\tass. Since literal~$x_i$ is a literal in clause~$\formula_j$, we have $\relClause{j}{x_i,\nx_i} \in \LitPos{i}$ and thus $\app(\LitPos{i}) \ni \relClauseT{j}$.

			Since Properties~\ref{prop:clauses-false}~--~\ref{prop:complete} consider all body atoms of query~\Q and view~\V, they imply that the bodies are identical: $\bodyof{\Q} = \app(\bodyof{\V})$. For $\psi=\id$, this trivially implies $\bodyof{\Q} \subseteq \app(\bodyof{\V})$ and $\psi(\app(\bodyof{\V})) \subseteq \bodyof{\Q}$, namely Conditions~\ref{def:cover-desc:cond:phi-bhomo} and~\ref{def:cover-desc:cond:psi-bhomo} in \autoref{definition:cover-desc}. Furthermore, since query~\Q is Boolean, set $\calA = \bodyof{\Q}$ has \emph{no} bridge variables. Thus, Conditions~\ref{def:cover-desc:cond:identity} and~\ref{def:cover-desc:cond:bvars} are trivially satisfied.

		Therefore, $(\calA,V,\app,\psi)$ is a cover description for~\Q.

		\smallskip\noindent
		\textit{%
			Now, we show that formula~\formula is satisfiable if there are mappings~\app and~$\psi$ such that $(\bodyof{\Q},V,\app,\psi)$ is a cover description for~\Q.
		}

		Consider the set $\calB = \app(\bodyof{\V})$ of atoms. From \autoref{definition:cover-desc}, we know that $\bodyof{\Q} \subseteq \calB$ and $\psi(\calB) \subseteq \bodyof{\Q}$ hold.
		Since both sets~\bodyof{\Q} and~\calB contain only one \RelForm-atom $B=\relForm$, we can conlude $\app(B) = B$ by Condition~\ref{def:cover-desc:cond:phi-bhomo} and $\psi(\app(B)) = \psi(B) = B$ by Condition~\ref{def:cover-desc:cond:psi-bhomo}.
		Therefore, each mapping~\app and~$\psi$ is the identity on variables~\wfalse, \wtrue and~\wproj.
		Hence, application~\app induces unambiguously a truth assignment~\tass that maps proposition $x_i \mapsto 1$ if $\app(x_i) = \wtrue$; and maps $x_i \mapsto 0$ otherwise.

		It suffices to show that~\tass satisfies formula~\formula. Let~$\formula_j$ be an arbitrary clause of the formula.
			From \(\bodyof{\Q} \subseteq \app(\bodyof{\V})\), we know that atom \(\relClauseT{j}\) is in \(\app(\bodyof{\V})\) because \(\relClauseT{j}\) is in \(\ClausesTrue\) and, therefore, in \(\bodyof{\Q}\).
			Since \(\bodyof{\V}\) does not contain atoms from \(\ClausesTrue\), it has to contain a different \(\RelClause_{j}\)-atom which is mapped to \(\relClauseT{j}\) by \(\app\).
			This atom cannot be \(\relClauseF{j}\) from \(\ClausesFalse\) because \(\app\) is the identity on \(\wfalse, \wtrue\), and \(\wproj\).
			Hence, there has to be an atom~$\relClause{j}{x_i,\nx_i} \in \LitPos{i}$ or $\relClause{j}{\nx_i,x_i} \in \LitNeg{i}$ for some $i \in \{1,\dots,n\}$ in $\bodyof{\V}$ that is mapped to \(\relClauseT{j}\) by \(\app\).
			In the case that an atom \(\relClause{j}{x_i,\nx_i} \in \LitPos{i}\) is mapped to \(\relClauseT{j}\), it follows by construction that the literal \(x_i\) occurs in clause \(\formula_j\).
			Furthermore, from $\app(\relClause{j}{x_i,\nx_i}) = \relClauseT{j}$, we can infer $\app(x_i) = \wtrue$.
		Hence, truth assignment~\tass satisfies literal~$x_i$ in clause~$\formula_j$ because $\tass(x_i) = 1$ if $\app(x_i) = \wtrue$.
			The other case is analogous.

		Thus, truth assignment~\tass satisfies every clause of formula~\formula and formula~\formula is satisfiable if a rewriting of query~\Q in terms of view~\V exists.
	This concludes our proof.
 \end{proof}
The proof of \autoref{thm:arewriting-np-hard} can now be stated easily.
\begin{proof}[Proof of \autoref{thm:arewriting-np-hard}]
	The upper bound follows from
        \autoref{theo:np-complete}. %

        For the lower bound, we show that the reduction of
        \autoref{theorem:acyclic-np-hard} can be adapted to a
        reduction from \ThreeSat to  $\rewrprobk\HCQ\HCQ\CQ$.
   To this end, we show  that the query~\Q constructed in the proof of
   \autoref{theorem:acyclic-np-hard}   has a \(\{V\}\)-rewriting
   \emph{if} the formula~\formula is satisfiable.

		If the formula~\formula is satisfiable then there are mappings \app and \(\psi\) such that \(C = (\bodyof{Q},V,\app,\psi)\) is a cover description.
		Due to \autoref{theorem:char-ext} it follows that \(\Q\) is \(\{\V\}\)-rewritable, since \(C\) trivially constitutes a cover partition.

		If \(\Q\) is \(\{\V\}\)-rewritable then there is a cover partition \(\calC\) for \(\Q\) over \(\{\V\}\) due to \autoref{theorem:char-ext}.
		We will show that \(\calC\) consists of a single cover description \(C\).
		Since \(C\) then necessarily has the shape \((\bodyof{Q},V,\app',\psi')\), this implies that the formula~\formula is satisfiable.

		Suppose for the sake of contradiction that \(\calC\) consists of at least two cover descriptions.
		Let \(C_1\) be a cover description from \(\calC\).
		Since \V is the only view and \(C_1\) is not the only cover description in \(\calC\), \(C_1\) has the shape \(C_1 = (\calA_1,V,\app_1,\psi_1)\) with \(\calA_1 \subsetneq \bodyof{Q}\).
		We observe that the variable \(\wproj\) is a bridge variable of \(\calA_1\) since it occurs in every atom of \(Q\), and, thus, in \(\calA_1\) and outside \(\calA_1\).
		Since $u$ is always the last variable in every atom in $Q$ and $V$, it follows from \autoref{definition:cover-desc}\ref{def:cover-desc:cond:phi-bhomo} that  \(\app_1(\wproj) = \wproj\).
		But then \(\bvarsofrel{\calA_1}\nsubseteq\app_1(\varsof{\headof{V}})\), since \(\wproj\) does not occur in the head of \(V\) and \(\app_1\) does not unify quantified variables.
		This contradicts \autoref{definition:cover-desc}\ref{def:cover-desc:cond:bvars}, and, thus, \(C_1\) being a cover description.
		Therefore, \(\calC\) consists of a single cover description.

		To conclude, we have established that \(\Q\) is
                \(\{\V\}\)-rewritable if and only of formula~\formula
                is satisfiable.
	Hence, the \classNP-hardness shown for
        $\coverdescprob\HCQ\HCQ$ also holds for
        $\rewrprobk\HCQ\HCQ\CQ$.
        It trivially transfers to
        $\rewrprobk\ACQ\ACQ\CQ$ and by \autoref{theorem:exacyclicrewr} also to $\rewrprobk\ACQ\ACQ\ACQ$.
\end{proof}

	The fact that, for \autoref{thm:arewriting-np-hard}, a
        single view application is sufficient if there is any
        rewriting, allows to draw yet another conclusion for a related
        problem, defined next.

\begin{defi}%
	\label{def:select-full-project-equiv}%
	Given a query class $\QueryClass \subseteq \CQ$, the \emph{select-full-project equivalence} problem for~\QueryClass, denoted $\selprojequiv\QueryClass$, asks, upon input of a Boolean query $Q\in \QueryClass$ and a query $Q'\in \QueryClass$ whether
		there is a Boolean query \(Q''\) which is equivalent to \(Q\) and whose body can be obtained from the body of \(Q'\) by unifying head variables of \(Q'\) in its body.\footnote{The problem is called the select-full-project equivalence problem because the query \(Q''\) can, if it exists, be expressed in the relational algebra by applying select-operators and a (full) project-operator to an expression for \(Q'\).}
\end{defi}

	\begin{exa}\label{example:selprojequiv}
		Consider the Boolean query \(Q\) defined by
		\[ H() \gets R(x), S(y,x,y), T(y,z), T(z,x) \]
		and the query \(Q'\) defined by
		\[ H'(y,z,u,v) \gets R(x), S(y,x,v), S(y,x,w), T(y,z), T(u,x).\]
		The body of the Boolean query \(Q''\) given by
		\[ H() \gets R(x), S(y,x,y), S(y,x,w), T(y,z), T(z,x)\]
		can be obtained by unifying the head variables \(y\) and \(v\) as well as \(z\) and \(u\) in \(\bodyof{Q'}\).
		The query \(Q''\) is equivalent to \(Q\).
		Thus, \((Q,Q')\in \selprojequiv\CQ\).
	\end{exa}

	Interpreting the query \(Q'\) in \autoref{def:select-full-project-equiv} as a view, the unification of head variables can be realised by an application \app which does not unnecessarily rename variables.
	Whether the desired Boolean query \(Q''\) exists then boils down to whether there is a \(\{Q'\}\)-rewriting for \(Q\) whose body consists of a single view atom, i.e.\ \(\app(\headof{Q'})\).

	\begin{exa}[Continuation of \autoref{example:selprojequiv}]
		Recall that the body of the Boolean query \(Q''\) in \autoref{example:selprojequiv} is obtained from \(\bodyof{Q'}\) by unifying the head variables \(y\) and \(v\) as well as \(z\) and \(u\).
		Let \app be the application which maps \(v\) to \(y\), \(u\) to \(z\), and is the identity on every other variable.
		The Boolean query with body \(\{\app(\headof{Q'})\} = \{H'(y,z,z,y)\}\) is a \(\{Q'\}\)-rewriting of \(Q\).
		In fact, its expansion is \(Q''\) which is equivalent to \(Q\).
	\end{exa}

Due to the relationship between the select-full-project equivalence problem and rewritability explained above and \autoref{theorem:acyclic-np-hard} the following holds.%

\begin{cor}\label{corollary:select-full-project-hcq}
	$\selprojequiv\ACQ$ and $\selprojequiv\HCQ$ are \classNP-hard.
\end{cor}
\begin{proof}
	The proof is essentially the same as for \autoref{thm:arewriting-np-hard}.

	Let $Q$ and $V$ be the Boolean query and the view constructed in the proof for \autoref{thm:arewriting-np-hard}, respectively.
	If $(\bodyof{Q},V,\app,\psi)$, for some mappings $\app$ and $\psi$, is a cover description for $Q$, then the Boolean query with body $\app(\headof{V})$ is a rewriting of $Q$ since the cover description constitutes a consistent cover partition for $Q$ (a cover partition consisting of a single cover description is consistent by definition).

		Thus, \(Q\) is equivalent to the Boolean query \(Q''\) whose body is obtained from \(\bodyof{V}\) by unifying exactly the (head) variables in \(\bodyof{V}\) the application \app unifies.
		In other words, \(Q''\) equals \(\app(V)\) up to renaming variables.

		Conversely, if \(Q\) is equivalent to a Boolean query \(Q''\) whose body can be obtained from \(\bodyof{V}\) by unifying head variables of \(V\) in \(\bodyof{V}\), then there is an application \app with \(\app(\bodyof{V}) = \bodyof{Q''}\) and, thus, \(Q''\) is the expansion of a \(V\)-rewriting, namely the Boolean query with body \(\app(\headof{V})\), for \(Q\).

	Hence, by choosing the second input query to be $Q' = V$, \autoref{corollary:select-full-project-hcq} follows because $Q$ and $Q'$ are hierarchical. \end{proof}

			We note that the restriction of the
                        select-full-project equivalence problem to
                        Boolean queries  $Q'$ is just the equivalence problem for Boolean queries which is in $\classP$ for acyclic queries~\cite{DBLP:journals/tcs/ChekuriR00}.

	But, surprisingly, the select-full-project equivalence problem is \classNP-hard not only for hierarchical but even for \emph{q}-hierarchical queries---while the rewriting problem for q-hierarchical views is in \classP{} (\autoref{corollary:qhcq-views-tractable}).
\begin{cor}\label{corollary:select-full-project}
	$\selprojequiv\QHCQ$ is \classNP-hard, even over database schemas with fixed arity.
\end{cor}

\begin{proof}
	To lift \autoref{corollary:select-full-project-hcq} to q-hierarchical queries, we have to adapt the query $Q'$ in the proof of \autoref{corollary:select-full-project-hcq} since it is not q-hierarchical.
	Adding the variable $u$ to the head of $Q'$ yields the desired q-hierarchical query, since every full hierarchical query is q-hierarchical.
	We note that the query \(Q\) is a hierarchical Boolean query and, therefore, trivially q-hierarchical.

	Note that, for the proof of \autoref{theorem:acyclic-np-hard}, which builds upon the proof for \autoref{thm:arewriting-np-hard}, it is required that $u$ is \emph{not} in the head of $V$. This ensures that the whole query has to be covered by a single application of the view $V$.
	In other words, if there is a rewriting of $Q$, there is a rewriting that consists only of a single atom.
	Since this restriction is not necessary in the setting of the select-full-project equivalence problem, we can indeed simply add\footnote{In fact, we could also remove $u$ from $Q'$ altogether. In this case, we also have to remove it from $Q$.} $u$ to the head of $Q'$. \end{proof}
\section{A Tractable Case}
\label{section:tractable:free-connex}

In this section, we first show that the acyclic rewriting problem becomes tractable for free-connex acyclic views and acyclic queries over database schemas of bounded arity. We then define a slightly larger class of views, for which this statement holds as well.
For the first result, we mainly show that rewritability with respect to a set $\calV$ of free-connex acyclic views can be reduced to rewritability  with respect to a set $\calW$ of views of bounded arity.

\begin{prop}\label{prop:tractable:free-connex}
  There is a polynomial-time algorithm that computes from each set $\calV$ of free-connex acyclic views a set $\calW$ of acyclic\footnote{In fact, the views in $\calW$ are even free-connex acyclic again. However, we do not claim it here, since it is not needed in the following, and it does not need to hold for the subsequent generalisation.} views such that
  \begin{enumerate}[(a)]
  \item the arity of the views of $\calW$ is bounded by the arity of the underlying  schema, and
  \item every conjunctive query $Q$ is $\calV$-rewritable if and only if it is $\calW$-rewritable.
  \end{enumerate}
		Furthermore, given a $\calW$-rewriting of $Q$, a $\calV$-rewriting of $Q$ can be computed in polynomial time.
\end{prop}
The proof splits each view $V$ into a set of views obtained by the subtrees of the root $\headof{V}$ of a join tree of $\bodyof{V}\cup\{\headof{V}\}$. The arities of the children of $\headof{V}$ yield the desired arity bound.
\begin{proof}
  Let $\calV$ be a set of free-connex acyclic views over a schema $\schema$ with arity at most $k$.
Let $V\in\calV$ and $J$ be a join tree for $V$ including its head atom \headof{V}.
	Since a join tree is undirected, we can assume that the root of $J$ is the node labelled with $\headof{V}$. %
	Let $A_1,\ldots,A_n$ be the labels of the children of the root node. Furthermore, let, for each $1\le i\le n$,  $\calB_i$  be the set of atoms in the subtree of $J$ with root $A_i$. %
        Since $J$ is a join tree, each variable that occurs in  some atom of a set $\calB_i$ and in $\headof{V}$ also occurs in $A_i$.        Furthermore, variables that occur in two sets $\calB_i$, $\calB_j$, $i\not=j$, necessarily occur also in $\headof{V}$. Thus, the following two conditions hold,
for every $i \in \{1,\dots,n\}$ and every $j<i$,
	\begin{enumerate}[label={(\arabic*)}]
		\item\label{cond:weak-head-bounded1-in-proof} $|\varsof{\calB_i} \cap \varsof{\headof{V}}| \le k$, and
		\item\label{cond:weak-head-bounded2-in-proof} $\varsof{\calB_i}\cap\varsof{\calB_j} \subseteq \headof{V}$.
 	\end{enumerate}

 	For each $1\le i\le n$ we define the view $V_i$ as the projection of $V$ to the variables that occur in $\headof{V}$ and in $\calB_i$. That is,  the body of $V_i$ is just the body of $V$ and the head of $V_i$ contains precisely those head variables of $V$ that occur in $\calB_i$.
	By construction, the views $V_i$ have arity $\ar(A_i) \le k$.
	Furthermore, the views $V_i$ are acyclic since they have the same body as $V$ which is acyclic.\footnote{We observe that, since all head variables of $V_i$ occur in $A_i$, each $V_i$ is even free-connex acyclic.}

	The desired set $\calW$ of views  can thus be obtained by replacing each view $V\in\calV$ by views $V_1,\ldots,V_n$ as constructed above. It is easy to see that $\calW$ can be computed in polynomial time.

	It remains to show that an arbitrary CQ $Q$ is $\calV$-rewritable if and only if it is $\calW$-rewritable.
	For the direction from right to left, let $\calC_\calW$ be a cover partition witnessing that $Q$ is $\calW$-rewritable.
	Consider a cover description $(\calA,V_i,\app,\psi)$ where $V_i$ is a view constructed as above, originating from a view $V\in\calV$.
	Since the only difference between $V$ and $V_i$ is that $V$ has more head variables, replacing $V_i$ with $V$ yields a cover description $(\calA,V,\app,\psi)$.
	Analogous replacements in all cover descriptions in $\calC_\calW$ yield a cover partition witnessing $\calV$-rewritability of $Q$.
	Moreover, given a $\calW$-rewriting of $Q$, a $\calV$-rewriting of $Q$ can be obtained by replacing every $V_i$-atom in the $\calW$-rewriting by a $V$-atom where $V$ is, as above, the view $V_i$ originated from.
	Variables not occurring in $\calB_i$ but in the head of $V$ are replaced by fresh variables (i.e.\ variables not occurring anywhere else) in the $V$-atom.

	For the direction from left to right, let $\calC_\calV$ be a consistent cover partition witnessing $\calV$-rewritability of $Q$.
	We replace the cover descriptions in $\calC_\calV$ to obtain a cover partition witnessing $\calW$-rewritability of $Q$.
	To this end, let $C = (\calA,V,\app,\psi)$ be a cover description in $\calC_\calV$.
	Furthermore, let $\calB_1,\ldots,\calB_n$  and $V_1,\ldots,V_n$ be   as in the construction of $\calW$ above.
	For each $1\le i\le n$, let $\calA_i$ be the set of all atoms of $\calA$ which are in $\app(\calB_i)$ but in no $\app(\calB_j)$, for $j<i$. Since $\varsof{\calA} \subseteq \app(\bodyof{V})$, this yields a partition of $\calA$.
	We claim that, for each $1\le i\le n$, $C_i = (\calA_i,V_i,\app,\psi)$ is a cover description.
	Since $V_i$ and $V$ have the same body, Conditions~\ref{def:cover-desc:cond:phi-bhomo} and~\ref{def:cover-desc:cond:psi-bhomo} of \autoref{definition:cover-desc} hold.
	Condition~\ref{def:cover-desc:cond:identity} of \autoref{definition:cover-desc} holds since $\varsof{\calA_i} \subseteq \varsof{\calA}$.

	In the remainder, we prove that Condition~\ref{def:cover-desc:cond:bvars} of \autoref{definition:cover-desc} holds.
	Let $x\in\bvarsof{\calA_i}$. 	Then $x$ is either in $\bvarsof{A}$ or it is a \enquote{new} bridge variable that also occurs in some $\calA_j\subseteq \calA$, $j\neq i$. In the former case, $x\in \app(\headof{V})$ since Condition~\ref{def:cover-desc:cond:bvars} holds for $C$. Thus, there are variables $y\in \headof{V}$ and $y'$ in $\calB_i$ such that $\app(y)=x=\app(y')$. Since, thanks to quantified variable disjointness, $\app$ maps  quantified and head variables disjointly, it follows that $y'$ must be from   $\headof{V}$ as well.
	But then $y'$ occurs in $\headof{V_i}$ since it is in $\headof{V}$ and in $\calB_i$.
	Therefore, $x = \app(y') \in \app(\headof{V_i})$.

	In the other case, let $j\not=i$ be such that $x$ occurs in $\calA_j$ and, hence, in $\app(\calB_j)$.
	Let $y,z$ be  variables from $\calB_i$ and $\calB_j$, respectively, such that $\app(y) = x = \app(z)$.
	If $y = z$, then $y$ is a head variable of $V$, since  $\calB_i$ and $\calB_j$ have only head variables of $V$ in common.
	If $y \neq z$, then both $y$ and $z$ occur in the head of $V$ thanks to quantified variable disjointness.
	In both cases, we can conclude that $x=\app(y)$ occurs in $\app(\headof{V_i})$.
	Thus, Condition~\ref{def:cover-desc:cond:bvars} of \autoref{definition:cover-desc} holds for $C_i$.

	Since the $\calA_i$ form a partition of $\calA$, replacing $C$ in $\calC_\calV$ with $C_1,\ldots,C_n$ yields a cover partition (with respect to $\calV\cup\calW$) and iterating this process for all cover descriptions in the original partition $\calC_\calV$ yields a cover partition witnessing $\calW$-rewritability of $Q$.
 \end{proof}

The main result of this section is now a simple corollary to \autoref{prop:tractable:free-connex}.
\begin{thm}\label{theorem:tractable:free-connex}
For every fixed $k$,	$\rewrprobk{\CCQ}{\ACQ}{\ACQ}$ is in polynomial time and
	an acyclic rewriting can be computed in polynomial time, if it exists.
      \end{thm}
      \begin{proof}
        	Let $Q\in\ACQ$ be an acyclic query and $\calV\subseteq \CCQ$ be a set of free-connex acyclic views over a schema $\schema$ with arity at most $k$. Thanks to \autoref{prop:tractable:free-connex}, from $\calV$ an equivalent set $\calW$ of acyclic views of arity at most $k$ can be computed in polynomial time.
 The statements of the theorem thus follow immediately from \autoref{corollary:arewrprobk-tractable}.
\end{proof}
We leave the complexity of $\rewrprob{\CCQ}{\ACQ}{\ACQ}$ as an open problem.

      A closer inspection of the proof of \autoref{prop:tractable:free-connex} reveals that it does not exactly require that the views are free-connex acyclic. In fact, it suffices that each view has a partition $\calB_1 \uplus \dots \uplus \calB_m$ of its body that obeys Conditions~\ref{cond:weak-head-bounded1-in-proof} and~\ref{cond:weak-head-bounded2-in-proof} defined in the proof of \autoref{prop:tractable:free-connex}.
      Therefore, we turn these two requirements into a new notion. We formulate and study this notion for conjunctive queries \Q, but we emphasise that we will use it for queries that define views only.

\begin{defi}\label{def:weak-head-arity}
	The \emph{weak head arity} %
        of a query $Q$ is the smallest $k$ for which there is a partition $\calB_1 \uplus \dots \uplus \calB_n$ of $\bodyof{Q}$ such that, for every $i \in \{1,\dots,n\}$ and every $j<i$,
	\begin{enumerate}[label={(\arabic*)}]
		\item\label{cond:weak-head-bounded1}  $|\varsof{\calB_i} \cap \varsof{\headof{Q}}| \le k$, and
		\item\label{cond:weak-head-bounded2} $\varsof{\calB_i}\cap\varsof{\calB_j} \subseteq \headof{Q}$.
 	\end{enumerate}
\end{defi}

	From the proof of \autoref{prop:tractable:free-connex} it follows that free-connex acyclic queries over a fixed schema have bounded weak head arity.
The following example illustrates that there are indeed views over a fixed schema that have bounded weak head arity but are not free-connex acyclic.
\begin{exa}\label{example:weak-head-arity}
	Let us consider the family ${\big(V_n\big)}_{n\in\N}$ of views with
	\begin{itemize}
		\item head $V_n(x, y_1,\ldots, y_n, z_1, \ldots, z_n)$, and
		\item body $\{ R(x, u_i, y_i), S(x, u_i, z_i), T(y_i) \mid 1 \le i \le n \}$.
	\end{itemize}
	For $n\ge 1$ the view $V_n$ is acyclic but \emph{not} free-connex acyclic.
	It has, however, weak head arity $3$.
	This is witnessed by the sets $\calB_i = \{R(x, u_i, y_i), S(x, u_i, z_i), T(y_i)\}$ for $1\le i\le n$ which form a partition satisfying the conditions of \autoref{def:weak-head-arity}.
\end{exa}

To generalise \autoref{theorem:tractable:free-connex} for views of bounded weak head arity, we thus only need to show that partitions obeying Conditions~\ref{cond:weak-head-bounded1} and~\ref{cond:weak-head-bounded2} of \autoref{def:weak-head-arity} can be efficiently computed.
In the remainder of this section we design an algorithm that determines the weak head arity   of a given conjunctive query \Q and computes a corresponding  partition. %

The algorithm relies on the concept of a \covergraph for \Q.

\begin{defi}\label{definition:covergraph-weakhead}
	The \covergraph $\graphof{\Q}$ of a conjunctive query $\Q$ is the undirected graph with node set $\bodyof{\Q}$ and edges $(A,A')$ for atoms $A$ and $A'$ that share a variable that does not belong to $\headof{\Q}$.
\end{defi}
\begin{exa}
	The \covergraph of the view $V_3$ from \autoref{example:weak-head-arity} is depicted in \autoref{figure:example:cover-graph}.
\end{exa}
\begin{figure}
	\centering
	\begin{tikzpicture}[node distance=.5cm]
		\node (R1) {$R(x,u_1,y_1)$};
		\node[below=of R1] (S1) {$S(x,u_1,z_1)$};
		\node[right=of R1] (T1) {$T(y_1)$};

		\node[right=of T1] (R2) {$R(x,u_2,y_2)$};
		\node[below=of R2] (S2) {$S(x,u_2,z_2)$};
		\node[right=of R2] (T2) {$T(y_2)$};

		\node[right=of T2] (R3) {$R(x,u_3,y_3)$};
		\node[below=of R3] (S3) {$S(x,u_3,z_3)$};
		\node[right=of R3] (T3) {$T(y_3)$};

		\path[-]
			(R1) edge (S1)
			(R2) edge (S2)
			(R3) edge (S3)
			;
	\end{tikzpicture}%
	\caption{The \covergraph $\graphof{V_3}$ of the query $V_n$ for $n = 3$ defined in \autoref{example:weak-head-arity}.}%
	\label{figure:example:cover-graph}%
\end{figure}
The following lemma states the relationship between weak head arity and \covergraph.

\begin{lem}\label{lemma:weakhead-to-covergraph}
	Let $\Q$ be a conjunctive query and $\calB_1,\ldots,\calB_n$ the connected components of its \covergraph.
The weak head arity of $\Q$ is the maximal number $\ell$ of head variables of $\Q$ in a set $\calB_i$.
	Moreover, the connected components $\calB_1,\ldots,\calB_n$ witness that $\Q$ has weak head arity~$\ell$.
\end{lem}

\begin{proof}
  Let $\Q$ be a conjunctive query and let $\calB'_1 \uplus \cdots \uplus \calB'_m$ be a partition of $\bodyof{\Q}$ witnessing that $\Q$ has weak head arity $k$.

Since by definition of the \covergraph, two different sets $\calB_i$ and $\calB_j$ only share head variables, $\calB_1 \uplus \cdots \uplus \calB_n$ is a partition of $\bodyof{\Q}$ witnessing that $\Q$ has weak head arity at most $\ell$, hence it holds $k\le \ell$.

To show that $\ell\le k$, it suffices to show that every connected component $\calB$ in $\graphof{\Q}$ is contained in some atom set $\calB'_i$.
	Towards a contradiction we assume that there is a connected component $\calB$ with atoms from two different subsets of the partition. Since $\calB$ is connected there must be some $A_1,A_2\in\calB$, connected by an edge, such that $A_1\in\calB'_i$ and $A_2\in\calB'_j$, for some $i\not=j$.
	By definition of $\graphof{\Q}$,  $A_1$ and $A_2$ share a variable that is not part of the head of $\Q$. But that contradicts Condition~\ref{cond:weak-head-bounded2} in \autoref{def:weak-head-arity}.
	Thus, we can conclude $\ell\le k$.
\end{proof}

\autoref{lemma:weakhead-to-covergraph} offers an algorithm to compute the weak head arity of a conjunctive query  \Q and a witness partition for it.   It simply computes the \covergraph $\graphof{\Q}$ of $\Q$ and its connected components.
Then the weak head arity is the maximum number of head variables that occur in any connected component.
Furthermore, the connected components form the desired partition satisfying the conditions of \autoref{def:weak-head-arity}.
We thus have the following corollary.

\begin{cor}\label{corollary:algo-comput-weak-arity}
	There is an algorithm that, upon input of a conjunctive query $Q$,
	computes in polynomial time the weak head arity of $Q$  and a partition of $\bodyof{Q}$ that witnesses it.
\end{cor}

By combining \autoref{corollary:algo-comput-weak-arity} with the proofs of \autoref{prop:tractable:free-connex} and \autoref{theorem:tractable:free-connex}, we obtain the following generalisation of \autoref{theorem:tractable:free-connex}.
\begin{thm}\label{thm:quasi-bounded-rewriting}\label{theorem:tractable:weak-head-arity}
	For each fixed $k\in\N$,
	the acyclic rewriting problem for acyclic queries and acyclic views with weak head arity $k$ is in polynomial time.
\end{thm}

\section{The Existence of Hierarchical and q-hierarchical Rewritings}\label{section:tractable}\label{section:hierarchical}

In \autoref{section:towards-acyclic-rewritings}, we have shown that every \emph{acyclic} query has an \emph{acyclic} rewriting if it has a rewriting at all.
This gives us a guarantee that there is a rewriting that has the same complexity benefits for query evaluation as the original query.

It is natural to ask whether other, stronger properties transfer in the same fashion.
In this section, we consider this question for hierarchical and q-hierarchical queries.

	The following example illustrates that, as for acyclic rewritings, even if a hierarchical rewriting for a hierarchical query exists, the canonical rewriting is not necessarily hierarchical.

\begin{exa}\label{example:compute-minimal:decomposable}
	Consider the hierarchical query $H(x,y) \gets
        R(x),S(y),T(x),T(y)$ and the views $V_1(x_1,y_1) \gets R(x_1),S(y_1)$
        and $V_2(z_2) \gets T(z_2)$.
        The canonical rewriting  \[H(x,y) \gets V_1(x,y),V_2(x),V_2(y)\] is not hierarchical, since $\atoms(x)$ and $\atoms(y)$
        are neither disjoint nor subsets of one
        another.
	However, a hierarchical rewriting exists, for instance \[H(x,y) \gets V_1(x,y'),V_1(x',y),V_2(x),V_2(y)\] is a hierarchical rewriting.
\end{exa}

It turns out that \autoref{theorem:exacyclicrewr} also holds for hierarchical and q-hierarchical queries.
\begin{thm}\label{theorem:hierarchical-rewriting}
	Let \(Q\) be a conjunctive query and \(\calV\) be a set of views.
	\begin{enumerate}[(a)]
		\item\label{part:hierarchical-rewriting:hier} If \(Q\)
                  is hierarchical and \(\calV\)-rewritable, then it has a hierarchical \(\calV\)-rewriting.
		\item\label{part:hierarchical-rewriting:qhier} If
                  \(Q\) is q-hierarchical and \(\calV\)-rewritable,
                  then it has a q-hierarchical \(\calV\)-rewriting.
	\end{enumerate}
\end{thm}

	  Similarly as for \autoref{theorem:exacyclicrewr}, the proof of \autoref{theorem:hierarchical-rewriting} partitions cover descriptions. However, the strategy for doing so is different here: instead of defining the partition as the connected components of an atom set with respect to a join tree of the query at hand, here the partition guaranteed by the following lemma is used.

	\begin{lem}\label{lemma:partition-hierarchical}
          Let $\Q$ be a hierarchical query and $(\calA,V,\app,\psi)$ a cover description for $\Q$. There is a partition $\calA_1 \uplus \cdots \uplus \calA_n = \calA$ such that the following conditions hold.
          \begin{enumerate}[(A)]
			  \item Each variable $y \notin \varsof{\app(\headof{V})}$ appears in at most one set $\calA_i$.\label{lemma:partition-hierarchical:condA}
            \item Each $\calA_i$ is\label{lemma:partition-hierarchical-B}
            	\begin{enumerate}[({B}a)]
                	\item a singleton set or
                  	\item there is a variable $x \notin \varsof{\app(\headof{V})}$ that appears in every atom in $\calA_i$.
                 \end{enumerate}
           \end{enumerate}
	\end{lem}

\begin{proof}
The algorithm uses an undirected graph $G$ similar to the cover graph of \autoref{section:tractable:free-connex}.
The graph $G$ has vertex set $\calA$ and an edge labelled by variable $x$ between two atoms $A_1,A_2$, if $x \notin \varsof{\headof{\app(V)}}$ and $x$ occurs in $A_1$ and $A_2$.  We note that there can be more than one edge between  two atoms.

Clearly, for each variable $x \notin \varsof{\headof{\app(V)}}$, the atoms that contain $x$ constitute a clique in $G$ with edges labelled by $x$.  In particular, each variable $x \notin \varsof{\headof{\app(V)}}$ can occur in at most one connected component.
Since $\Q$ is hierarchical, for two cliques $C_x,C_y$ induced by different variables $x$ and $y$ it holds that either they are disjoint or one contains the other. It readily follows that for each connected component $H$ of $G$  there is a variable $x$, such that $C_x$ contains all atoms of $H$. 

Therefore the partition $\calA_1 \uplus \cdots \uplus \calA_n = \calA$ given by  the connected components of $G$ fulfils the conditions of the Lemma.
 \end{proof}

Now we are prepared  to prove \autoref{theorem:hierarchical-rewriting}.

\begin{proof}[Proof of \autoref{theorem:hierarchical-rewriting}]
Towards (a),	let~\Q be a \calV-rewritable hierarchical query.
W.l.o.g.\ we can assume that~\Q is minimal (cf.\ \autoref{remark:minimality}).
Thanks to \autoref{theorem:char-ext}, there is a cover partition $\calC' = C'_1,\dots,C'_k$ for~\Q over $\calV$.
	We replace every cover description $(\calA,V,\app,\psi)$ in $\calC'$ by $(\calA_1,V,\app,\psi),\ldots,(\calA_n,V,\app,\psi)$ were $\calA_1 \uplus \cdots \uplus \calA_n = \calA$ is the partition of $\calA$ from \autoref{lemma:partition-hierarchical} and let $\calC$ be the collection we obtain after the replacement. For each $(\calA_i,V,\app,\psi)$ Condition~\ref{def:cover-desc:cond:bvars} of \autoref{definition:cover-desc} is satisfied thanks to Condition~\ref{lemma:partition-hierarchical:condA} of \autoref{lemma:partition-hierarchical}.
	It is easy to verify that the other conditions in \autoref{definition:cover-desc} are satisfied for each $(\calA_i,V,\app,\psi)$. All in all, the collection $\calC$ is a cover partition for $\Q$ over $\calV$.
	Thanks to \autoref{theorem:char-ext} we can assume that $\calC$ is a \consistent one with the same partition of $\bodyof{\Q}$.  

	Let $\R$ be the rewriting obtained from $\calC$, i.e., the query $\Q_{\calC}$.
	In the remainder of the proof we show that $\R$ is hierarchical, i.e., every pair of  variables in $x$ and $y$ in $\R$ satisfies the conditions in \autoref{definition:qhierarchical}. Let $x$ and $y$ be two variables in $\R$.

    	If \(x\) occurs in only one atom of \(\bodyof{R}\) then there are two cases: \begin{enumerate*}[label=(\roman*)]
    		\item if in the only atom in which \(x\) occurs, \(y\) occurs as well, then \(\atoms_R(x)\subseteq \atoms_{R}(y)\), and
    		\item otherwise, \(\atoms_R(x)\cap\atoms_R(y)=\emptyset\) holds.
    	\end{enumerate*}
    	Analogously, if \(y\) occurs in only one atom of \(\bodyof{R}\) then either \(\atoms_{R}(y)\subseteq\atoms_{R}(x)\) or \(\atoms_R(x)\cap\atoms_R(y)=\emptyset\) holds.

		Let us finally assume that both \(x\) and \(y\) occur in at least two atoms of \(\bodyof{R}\) each.
		Thanks to \(\calC\) being a \consistent cover partition and Condition~\ref{def:cover-desc:cond:bvars} of \autoref{definition:cover-desc}, \(x\) and \(y\) are bridge variables, and occur in atom sets of at least two different cover descriptions of \(\calC\) each.

    Since $\Q$ is hierarchical we have the following three cases.
	\begin{enumdesc}
		\item\label{proof:hierarchical-case1}
			$\atoms_{\Q}(x)\subseteq\atoms_{\Q}(y)$.
			Let  \((\calA,V,\app,\psi)\) in \(\calC\) be a cover description which fulfils  \(x\in\app(\headof{V})\). In particular, \(x\) occurs in an atom \(A\in \calA\).	Therefore, \(y\) also occurs in \(A\), since \(\atoms_{\Q}(x)\subseteq\atoms_{\Q}(y)\). Since $y$ is a bridge variable by assumption, it follows that  \(y\in\app(\headof{V})\) thanks to Condition~\ref{def:cover-desc:cond:bvars} of \autoref{definition:cover-desc}. We conclude that \(\atoms_{R}(x)\subseteq\atoms_{R}(y)\) holds.

		\item\label{proof:hierarchical-case2}
			$\atoms_{\Q}(y)\subseteq\atoms_{\Q}(x)$. This case is analogous to the first case.
		\item\label{proof:hierarchical-case3}
			$\atoms_{\Q}(x)\cap\atoms_{\Q}(y)=\emptyset$.
                  If there is no cover description in $\calC$, in which $x$ and $y$ occur together, then $\atoms_{\R}(x) \cap \atoms_{\R}(y)=\emptyset$ holds.
                  Let us thus assume that there is a cover description $(\calA,V,\app,\psi)$ in $\mathcal{C}$ in which both $x$ and $y$ occur. Thanks to $\atoms_{\Q}(x)\cap\atoms_{\Q}(y)=\emptyset$, set~$\calA$ contains at least two atoms.
                  Thanks to \autoref{lemma:partition-hierarchical}, there is a variable $u \notin \varsof{\headof{\app(V)}}$ that appears in all atoms of $\calA$. Since $Q$ is hierarchical, we must have $\atoms_{\Q}(x) \subsetneq \atoms_{\Q}(u)$. However, this yields a contradiction,
                  because \(x\) occurs in at least two cover descriptions, and therefore outside \(\calA\), whereas \(u\) does not.
	\end{enumdesc}
	This concludes the proof of Statement~\ref{part:hierarchical-rewriting:hier}.

		Towards~\ref{part:hierarchical-rewriting:qhier}, let us assume that  $\Q$ is q-hierarchical. Our goal is to show that for all variables $x,y\in\varsof{R}$,
	if $\atoms_R(x)\subsetneq\atoms_R(y)$ holds and $x$ is in the head of $R$, then $y$ is also in the head of $R$.
		Note that \(x\) and \(y\) are both bridge variables because \(x\) occurs in the head of \(Q\) which is the same as the head of \(R\), and \(y\) occurs in at least two atoms of \(\bodyof{R}\) due to \(\atoms_R(x)\subsetneq\atoms_R(y)\).
		In particular, \(x\) and \(y\) both occur in \(Q\).
	Since the heads of $Q$ and $R$ are the same, whenever  $\atoms_Q(x)\subsetneq\atoms_Q(y)$ holds, we can conclude that if $x$ is in the head of $R$, then $y$ is also in the head of $R$.

	In the remainder we assume, for the sake of a contradiction, that $\atoms_Q(x)\subsetneq\atoms_Q(y)$ does not hold.
	Since \(\Q\) is hierarchical this means that either \(\atoms_Q(x)\supseteq\atoms_Q(y)\) or \(\atoms_Q(x) \cap \atoms_Q(y) = \emptyset\) holds.\footnote{Note, that the case $\atoms_Q(x) = \atoms_Q(y)$ is a special case of $\atoms_Q(x)\supseteq\atoms_Q(y)$.}

	If \(x\) also occurs in at least two atoms of \(\bodyof{R}\), then the preconditions for \ref{proof:hierarchical-case2} and \ref{proof:hierarchical-case3} in the proof for Statement~(a) above are met.
	Thus, \(\atoms_Q(x)\supseteq\atoms_Q(y)\) and \(\atoms_Q(x) \cap \atoms_Q(y) = \emptyset\) imply \(\atoms_R(x)\supseteq\atoms_R(y)\) and \(\atoms_R(x) \cap \atoms_R(y) = \emptyset\), respectively.
	But this is a contradiction to \(\atoms_R(x)\subsetneq\atoms_R(y)\).

	In the remainder, consider the case that   \(x\) occurs in exactly one atom \(B\) of \(\bodyof{R}\). Variable \(y\) occurs in \(B\) and outside \(B\).
    Since \(x\) and \(y\) are bridge variables, it follows that \(x\) occurs in exactly one atom set \(\calA\) of \(\calC\) and \(y\) occurs in and outside \(\calA\).
    Hence, \(\atoms_Q(x) \cap \atoms_Q(y) = \emptyset\) holds because \(\atoms_Q(x)\supseteq\atoms_Q(y)\) cannot.
    This implies that \(\calA\) consists of at least two atoms, since \(x\) and \(y\) co-occur in \(\calA\).
    Therefore, there is a non-bridge variable \(u\) that occurs in all atoms of \(\calA\) thanks to \autoref{lemma:partition-hierarchical}\ref{lemma:partition-hierarchical-B}.
    But then \(\atoms_Q(x)\subsetneq\atoms_Q(u)\) holds, and, since \(u\) is a non-head variable, \(x\) is a non-head variable as well, because \(Q\) is q-hierarchical.
    This is a contradiction to \(x\) being a head variable.

	All in all, we can conclude that \(\atoms_Q(x)\subsetneq\atoms_Q(y)\) holds, and therefore, that \(y\) is a head variable.
	Thus, \(R\) is q-hierarchical.
 \end{proof}

Similarly to \autoref{theorem:exacyclicrewr}, \autoref{theorem:hierarchical-rewriting} delivers good news as well as bad news.
The good news is that, since $\QHCQ\subseteq \CCQ$ and $\QHCQ\subseteq
\HCQ\subseteq \ACQ$ hold, the rewriting problem for q-hierarchical
views and hierarchical queries over a fixed schema is tractable thanks to
\autoref{theorem:tractable:free-connex} and \autoref{theorem:hierarchical-rewriting}.
  The bad news is that \autoref{thm:arewriting-np-hard}
  and \autoref{theorem:hierarchical-rewriting} imply
  \classNP-completeness for hierarchical queries and views.
\begin{cor}\label{corollary:qhcq-views-tractable}
	$\rewrprobk{\QHCQ}{\HCQ}{\HCQ}$ and $\rewrprobk{\QHCQ}{\QHCQ}{\QHCQ}$ are in polynomial time for every $k\in\N$.
\end{cor}
\begin{cor}\label{corollary:hcq-np-complete}
	$\rewrprobk{\HCQ}{\HCQ}{\HCQ}$ is \classNP-complete for every $k\ge 3$.
\end{cor}

Of course, \autoref{corollary:hcq-np-complete} implies
\classNP-hardness of \(\rewrprobk\ViewClass\QueryClass\ACQ\), for all pairs \(\ViewClass, \QueryClass\) of classes with \(\HCQ\subseteq\ViewClass\subseteq\CQ\) and \(\HCQ\subseteq\QueryClass\subseteq\CQ\).
\section{Related Work}
\label{section:relationship-characterization}
We already mentioned that our notion of cover partitions is similar to various notions from the literature. We mention three of them here.
In \cite{DBLP:series/synthesis/2019Afrati} algorithms for finding exact rewritings with a minimal number of atoms and (maximally) contained rewritings  are presented.
For this purpose triples $(S,S',h)$ are considered which are comparable to cover descriptions \cite[Definition~3.12]{DBLP:series/synthesis/2019Afrati}.
Namely, $S'$ corresponds to the set $\calA$ in a cover description, $S = \bodyof{\Q}$, and $h$ is a homomorphism from $S'$ into a view $V$.
In our characterisation we can assume $h=\id$ (and, thus, omit it in the specification of a cover description) thanks to the view application $\app$.
We note that $h$ can be assumed to be one-to-one for (equivalent) $\calV$-rewritings \cite[Theorem~3.15]{DBLP:series/synthesis/2019Afrati}.
Furthermore, for (equivalent) rewritings, only candidates whose body is a proper subset of the canonical rewriting's body are considered (cf.\ \cite[Section~3.2.3]{DBLP:series/synthesis/2019Afrati}).
Note that the existence of body homomorphisms $\psi_i$ mapping the $\app_i(V_i)$ into $Q$ is guaranteed for such candidates and $\psi_i$ determines the application $\app_i$ (in particular, the unification of variables in the head of a view).
This is, however, too restrictive for our purposes (cf., for instance, \autoref{example:cyclic-rewriting}) which is why $\app$ and $\psi$ are part of a cover description.
We emphasise that Condition~\ref{def:cover-desc:cond:bvars} is equivalent to the \emph{shared-variable property}~\cite[Definition~3.12]{DBLP:series/synthesis/2019Afrati}.
An analogue of the implication~\ref{theorem:char-ext:rewriting} $\Rightarrow$ \ref{theorem:char-ext:partition} of \autoref{theorem:char-ext} is proven \cite[Theorem~3.15]{DBLP:series/synthesis/2019Afrati} and, based on that, an algorithm \textsc{CoreCover} for finding (equivalent) rewritings is derived.
This algorithm, however, considers only triples with maximal sets $S'$ (called \emph{tuple cores}) and they are allowed to overlap, i.e.\ they do not have to form a partition.
In contrast, we consider non-maximal sets, for instance in the proof of \autoref{theorem:exacyclicrewr} which depends upon the possibility to split these sets.
For (maximally) contained rewritings, an analogue of~\ref{theorem:char-ext:partition} $\Rightarrow$ \ref{theorem:char-ext:rewriting} is implied by \cite[Theorem~4.19]{DBLP:series/synthesis/2019Afrati}.
Interestingly, the proof of this result (and the associated algorithm) exploits triples with minimal sets $S'$ (that still satisfy the shared-variable property) and a partition property.
We discuss the relation of this minimality constraint with our results below in more detail.
An analogue of~\ref{theorem:char-ext:partition} $\Rightarrow$ \ref{theorem:char-ext:rewriting} for (equivalent) $\calV$-rewritings, and, therefore, a characterisation for $\calV$-rewritability, is not stated (nor implied).

In \cite{DBLP:conf/sigmod/GouKC06} a characterisation for $\calV$-rewritability is employed to find rewritings efficiently.
It is in terms of \emph{tuple coverages} and a \emph{partition condition} corresponding to cover descriptions and partitions, respectively \cite[Theorem~5, Theorem~6]{DBLP:conf/sigmod/GouKC06}.
A tuple coverage, denoted $s(t_V, Q)$, is a (non-empty) set $G$, where $t_V$ corresponds to $\app(V)$ and $G$ to the set $\calA$ of a cover description.
Similar to \cite{DBLP:series/synthesis/2019Afrati} the most crucial differences in comparison with our characterisation are that only rewritings whose body is contained in the canonical rewriting, i.e.\ $\varsof{\app(V)} \subseteq \varsof{Q}$ holds for all tuple coverages, are considered.
Also the mappings $\psi$ and $\app$ and their associated conditions are not denoted explicitly, instead it is required that $G$ is isomorphic to a subset of $\app(V)$ and Condition~\ref{def:cover-desc:cond:bvars} holds. This is equivalent to our conditions if restricted to the rewritings considered in~\cite{DBLP:conf/sigmod/GouKC06}.

Lastly, in \cite{DBLP:journals/vldb/PottingerH01} \emph{MiniCon descriptions (MCDs)} are used to compute (maximally) contained rewritings (which are unions of conjunctive queries, in general).
A MCD is a tuple of the form $(h, V, \varphi, G)$ which relates to a cover description $(\calA, V, \app, \psi)$ as follows: $h$ is called a head homomorphism and is basically the restriction of an application $\app$ to the head variables of the view $V$, $G$ corresponds to the set $\calA$, and $\varphi$ is mapping embedding $G$ into $h(V)$.
The component $\varphi$ has no counterpart in a cover description because, thanks to $\app$ being able to rename any variable in $V$, we can assume $\varphi=\id$.
We note that the applications $\app_i$ of a consistent cover partition $\calC$ also allow us to conveniently denote the expansion of the associated rewriting $\Q_\calC$.
On the other hand, there is no counterpart for the body-homomorphism $\psi$ in a MCD (since the $\psi_i$ in a consistent cover partition ensure that the query is contained in the associated expansion of the rewriting, there is also no need for them if contained rewritings are considered).
Condition~\ref{def:cover-desc:cond:bvars} is stated as \cite[Property~1]{DBLP:journals/vldb/PottingerH01} and the idea of a cover partition in \cite[Property~2]{DBLP:journals/vldb/PottingerH01}.

\paragraph*{Minimal cover descriptions}%
In \cite{DBLP:series/synthesis/2019Afrati}, \enquote{minimal} triples are used to construct maximally contained rewritings. In terms of cover descriptions, an analogous definition of minimality is as follows.

\begin{defi}[Minimal Cover Descriptions]
	\label{def:minimal-cover-description-alt}%
		A cover description~$(\calA,V,\app,\psi)$ for a query $Q$ is called \emph{minimal} if there is no partition $\calA_1 \uplus \dots \uplus \calA_k = \calA$ for $k \geq 2$ with nonempty subsets $\calA_1,\dots,\calA_k$ such that cover descriptions $C_1,\dots,C_k$ with $C_i=(\calA_i,V,\app_i,\psi_i)$ for $Q$ exist.
\end{defi}

It seems obvious to exploit such minimal cover description for the constructions in \autoref{theorem:exacyclicrewr} and \autoref{theorem:hierarchical-rewriting}, instead of partitioning the set $\calA$ of a cover description \enquote{manually}.
However, in contrast to our constructions, this would (possibly) not result in efficient algorithms to compute an acyclic or hierarchical rewriting from an arbitrary one, since it turns out that deciding whether a cover description is minimal is \classcoNP-hard.

\begin{prop}\label{proposition:np-hard-minimal}
	Let $\Q$ be a hierarchical conjunctive query.
	It is \classcoNP-hard to decide whether a cover description $(\mathcal{A},\Q,\alpha,\psi)$ is minimal.
\end{prop}

\begin{proof}

\begingroup
\newcommand{\AtomsOfQuery}{\ensuremath{\mathcal{A}_{\Q}^{\plus}}}
\newcommand{\BigView}{\ensuremath{W}}
\newcommand{\QueryDesc}{\ensuremath{Q'}}
\newcommand{\AtomExp}[2]{#1^{#2}}
\newcommand{\ExpView}[1]{\ensuremath{\AtomExp{V_{#1}}{\plus}}}
\newcommand{\AtomsOfView}[1]{\ensuremath{\AtomExp{\bodyof{V_{#1}}}{v_{#1}}}}
\newcommand{\newid}{\ensuremath{\widetilde{\id}}}
\newcommand{\CoverDesc}{\ensuremath{C_{\calV}^{\Q}}}

	We reduce the problem $\rewrprob\HCQ\HCQ\CQ$, which is \classNP-hard due to \autoref{thm:arewriting-np-hard}, to the \emph{complement} of the minimisation problem.
	For a hierarchical query $\Q$ and a set of hierarchical views $\calV$, we describe how we can derive a cover description $\CoverDesc$ and a query $\QueryDesc$ such that $\CoverDesc$ is \emph{not} minimal for $\QueryDesc$ if and only if there is a $\calV$-rewriting for $\Q$.

	For convenience, we introduce some notation.
	For an atom $A = R(x_1,\ldots,x_r)$ and a variable $u$ we denote by $\AtomExp{A}{u}$ the atom $R(u,x_1,\ldots,x_r)$ resulting from extending $A$ by the variable $u$.
	We lift this notation to sets~\calA of atoms in the natural way, i.e.\
	\[
	\AtomExp{\calA}{u}\ =\ \left\{R(u,x_1,\ldots,x_r) \mid R(x_1,\ldots,x_r)\in \calA\right\}.
	\]
	If a set of atoms or a query consists of atoms extended by (possibly different) variables as described above, we often signify this with a \enquote{\plus}-symbol in the superscript, e.g.\ we write $\AtomExp{\calA}{\plus}$.

	Let $V_1,\ldots,V_n$ be the views in $\calV$.
	We assume that the views in~\calV  and the query \Q refer to distinct variables, which is no restriction since the variables can be renamed accordingly in polynomial time.

	\paragraph*{Construction.}
	We start with the construction of the cover description $\CoverDesc$.
	To define a view and a query for the cover description,
	we consider views $\ExpView{1},\ldots,\ExpView{n}$ that are copies of the input views $V_1,\ldots,V_n$ where each atom will be expanded by a special variable, that only appears in the associated view, in the first component of each body atom and of the head, i.e. for every $i \in \{1,\ldots,n\}$ we define a view $\ExpView{i}$ with $\headof{\ExpView{i}} = \AtomExp{\headof{V_i}}{v_i}$
	and $\bodyof{\ExpView{i}} = \AtomExp{\bodyof{V_i}}{v_i}$ where  $v_1,\ldots,v_n$ are distinct variables that do not occur in the query or in any of the views in $\calV$.
	In the same fashion, we define a set of atoms for the atoms in the query, but we add a new atom whose relation symbol does not occur in $\Q$ or in any of the views in $\calV$. For a new variable $u$ and a new relation symbol $S$, we define $\AtomsOfQuery = \AtomExp{\bodyof{\Q}}{u} \cup \{S(u)\}$.

	Let us now turn to the definition of the view $\BigView$ that is used in the cover description. The view $\BigView$ contains the atoms in $\AtomsOfQuery$, the atoms in $\bodyof{\ExpView{i}}$ for all $i \in \{1,\ldots,n\}$, and a special atom $S(v_{n+1})$ with a new variable $v_{n+1}$ in the body. The head of $\BigView$ contains the variables in the head of $\Q$ and variables in the heads of the views $V_i$ for all $i \in \{1,\ldots,n\}$ as well as the variables $v_1,\ldots,v_{n+1}$ , but not $u$.
	Let us emphasize that, for each view $\ExpView{i}$, the set $\bodyof{\ExpView{i}}$ is contained in $\bodyof{\BigView}$.

	The query $\QueryDesc$ is defined by the rule ${\headof{\Q} \gets \bodyof{\BigView}}$, that is $\QueryDesc$ has the head variables of $\Q$ and the body of $\QueryDesc$ has the same atoms as the body of $\BigView$.

	The cover description $\CoverDesc$ is defined as $(\AtomsOfQuery,\BigView,\id,\id)$. Note that the bridge variables of $\AtomsOfQuery$ with respect to $\QueryDesc$ are contained in the head variables in $\BigView$. Hence, $\CoverDesc$ is a cover description for $\QueryDesc$.

	Note that the views $\ExpView{i}$ are hierarchical since for the new variables $v_i$ we have that $\atoms_{\ExpView{i}}(v_i) \supseteq \atoms_{\ExpView{i}}(y)$ for all $y \in \varsof{V_i}$ and the input views are hierarchical.
	The same is true for the set \AtomsOfQuery{} (viewed as a Boolean query here).
	Moreover, $\BigView$ and $\QueryDesc$ are also hierarchical since the views and the query, and, hence, the $\ExpView{i}$ and $\AtomsOfQuery$ do not share any variable.

	\paragraph*{Correctness.}
	In the following, we show that $\Q$ has a $\calV$-rewriting if and only if $\CoverDesc$ for $\QueryDesc$ is not minimal.

	For the only if-direction assume that $Q$ has a $\calV$-rewriting. Then there is a consistent cover partition $\mathcal{C}$ for $\Q$. Note that by construction each cover description $(\calA,V_j,\app,\psi)$ in $\mathcal{C}$ can be turned into a cover description $(\AtomExp{\calA}{u},\ExpView{j},\app\cup\{v_j\mapsto u\}, \psi \cup \{u \mapsto u\})$ for $\QueryDesc$ and, furthermore into a cover description $(\AtomExp{\calA}{u},\BigView,\app',\psi')$ for $\QueryDesc$ where $\app'$ and $\psi'$ coincide with $\app\cup\{v_j\mapsto u\}$ and $\psi \cup \{u \mapsto u\}$ on their domain, respectively, and are the identity on all other variables.
	Let $\mathcal{C}'$ be the collection of cover descriptions we obtain by transforming every cover description in $\mathcal{C}$ to a cover description as described above and the additional cover description $(\{S(u)\},\BigView,\{v_{n+1}\mapsto u\},\id)$ for $\QueryDesc$.

	Since $\calC$ is a cover partition for $\Q$ and $\Q$ does not contain an $S$-atom, the atom sets of the cover descriptions in $\calC'$ form a partition of $\AtomsOfQuery = \AtomExp{\bodyof{\Q}}{u} \cup \{S(u)\}$.
	But then $(\AtomsOfQuery,\BigView,\id,\id)$ is not minimal for $\QueryDesc$ because $\calC'$ consists of at least two cover descriptions.

	For the if-direction assume that $\CoverDesc$ is not minimal for $\QueryDesc$. Let
	\[\calC' = (\calA_1^{\plus},\BigView,\app_1,\psi_1),\ldots,(\calA_k^{\plus},\BigView,\app_k,\psi_k)\]
	be a collection of cover descriptions witnessing that $\CoverDesc = (\AtomsOfQuery,\BigView,\id,\id)$ is not minimal for $\QueryDesc$.
	From $\calC'$ we will derive a cover partition $\calC$ for $\Q$ witnessing that $\Q$ is indeed $\calV$-rewritable.
	For this purpose, we first analyse to which atoms in $\app_i(W)$ the atoms of a set $\calA_i^{\plus}$ are mapped to and then associate views $\ExpView{j}$ with (subsets of) the sets $\calA_i^{\plus}$.

	We can assume that the $\app_i$ fulfil the quantified variable disjointness.
	Note, that $k \ge 2$ since $\CoverDesc$ is not minimal, and $u$ is a bridge variable of each set $\calA_i^{\plus}$ since it occurs in every atom in $\AtomsOfQuery$. Therefore, no atom in $\calA_i^{\plus}$ can be mapped into the copy of $\app_i(\AtomsOfQuery)$ in $\alpha_i(\BigView)$ because the variable $u$ is not a head variable of $\BigView$.
	Hence, $\calA_i^{\plus}$ is a subset of $\app_i(\AtomsOfView{1}) \cup \ldots \cup \app_i(\AtomsOfView{n}) \cup \{\app_i(S(v_{n+1}))\}$.

	Since the sets $\app_i(\AtomsOfView{j})$ do not share any variable that is not in the $\headof{\app_i(\BigView)}$, each cover description in $\calC'$ can be partitioned into cover descriptions
	\[(\calB_{i,1}^{\plus},\BigView,\app_i,\psi_i),\ldots,(\calB_{i,n}^{\plus},\BigView,\app_i,\psi_i)\]
	for $\QueryDesc$ where $\calB_{i,j}^{\plus} = \calA_i^{\plus} \cap \app_i(\AtomsOfView{j})$ and, in case $S(u)\in\calA_i^{\plus}$, a cover description $(\{S(u)\},\BigView,\app_i,\psi_i)$.
	For the sake of readability we assume that $\calB_{i,j}^{\plus} \neq \emptyset$ holds for all $i,j$.
	If not, the respective cover descriptions can just be removed from the sequence.

	Note that the view $\BigView$ in a cover description $(\calB_{i,j}^{\plus},\BigView,\app_i,\psi_i)$ can be replaced by $\ExpView{j}$, because $\calB_{i,j}^{\plus}\subseteq \app_{i}(\bodyof{\ExpView{j}})$ by definition.
	Hence, each cover description  $(\calB_{i,j}^{\plus},\BigView,\app_i,\psi_i)$ can be transformed into a cover description $(\calB_{i,j}^{\plus},\ExpView{j},\app_{i,j},\psi_{i,j})$ where $\alpha_{i,j}$ and $\psi_{i,j}$ are the restriction of $\app_i$ and $\psi_i$ resp.\ on $\varsof{\ExpView{j}}$.
	A cover description $(\calB_{i,j}^{\plus},\ExpView{j},\app_{i,j},\psi_{i,j})$ can, in turn, be transformed into a cover description $(\calB_{i,j},V_{j},\widehat{\app}_{i,j},\widehat{\psi}_{i,j})$ for $\Q$ where $\calB_{i,j}$ is the atom set with $\AtomExp{\calB_{i,j}}{v_{j}} = \calB_{i,j}^{\plus}$, and $\widehat{\app}_{i,j}$ and $\widehat{\psi}_{i,j}$ are the restriction of $\app_{i,j}$ and $\psi_{i,j}$ on $\varsof{V_{j}}$.

	Let $\calC$ be the collection of cover descriptions we obtain by
	applying the transformations described above to all cover descriptions in $\calC'$ and
	removing cover descriptions with atom set $\{S(u)\}$.
	Note, that by construction, the atom sets in $\calC'$ form a partition of $\bodyof{\Q}$.
	Thus $\calC'$ is a cover partition and hence, $\Q$ is $\calV$-rewritable.
\endgroup%
 \end{proof}

\paragraph*{Applications of structurally simple queries.}%

	It is well known that many problems are tractable for acyclic conjunctive queries but (presumably) not for conjunctive queries in general.
	Notably, the evaluation, minimisation, and the containment problem are tractable for acyclic queries \cite{DBLP:conf/vldb/Yannakakis81, DBLP:journals/tcs/ChekuriR00, DBLP:journals/jacm/GottlobLS01} but \classNP-complete for the class of conjunctive queries \cite{DBLP:conf/stoc/ChandraM77}.

	The class of free-connex conjunctive queries plays a central role in the enumeration complexity of conjunctive queries.
	In \cite{DBLP:conf/csl/BaganDG07}, Bagan, Durand and Grandjean showed that the result of a free-connex acyclic conjunctive query can be enumerated with constant delay after a linear time preprocessing phase.
	Moreover, they also showed that the result of an acyclic conjunctive query without self-joins that is not free-connex cannot be enumerated with constant delay after a linear time preprocessing, unless $n \times n$ matrices can be multiplied in time $O(n^2)$.

Hierarchical queries play a central role in different contexts.
On the one hand, Dalvi and Suciu~\cite{DBLP:conf/pods/DalviS07a}  showed that the class of hierarchical Boolean conjunctive queries without self-joins characterises precisely
the Boolean CQs without self-joins that can be
answered in polynomial time on probabilistic
databases. This has been extended by Fink and Olteanu  \cite{DBLP:conf/pods/FinkO14} to the notion of non-Boolean queries and queries with negation.
	On the other hand, Koutris and Suciu \cite{DBLP:conf/pods/KoutrisS11}
	studied hierarchical  join  queries in the context of query evaluation on massively parallel architectures.
	We refer to \cite{DBLP:journals/tods/FinkO16} for further applications of hierarchical queries.

The notion of q-hierarchical queries has played a central role in the evaluation of conjunctive queries under single tuple
updates \cite{DBLP:conf/pods/BerkholzKS17}.
	In \cite{DBLP:conf/pods/BerkholzKS17} it is shown that the result of a q-hierarchical conjunctive query
	can be evaluated (by answering a Boolean CQ in constant time,
	enumerating the result of a non-Boolean CQ with constant delay,
	or outputting the number of results in constant time)  with constant update time
	after a linear time preprocessing. Moreover, they showed that the result of a conjunctive query without self-joins that is not q-hierarchical cannot be evaluated with constant update time after a linear time preprocessing, unless some algorithmic conjectures on Online Matrix-Vector multiplication (see \cite{DBLP:conf/stoc/HenzingerKNS15} for more information about the conjecture) do not hold.
The notion of q-hierarchical queries is also related to factorised databases \cite{DBLP:phd/dnb/Keppeler20}.
The notion of factorised databases has already been considered in
various contexts \cite{DBLP:journals/tods/OlteanuZ15,DBLP:journals/pvldb/BakibayevKOZ13,DBLP:conf/sigmod/SchleichOC16}.
A further recent source of information on structurally simple queries is \cite{DBLP:conf/pods/0002NOZ20}.

\section{Conclusion}\label{section:conclusion}
We studied rewritability by acyclic queries or queries from \CCQ, \HCQ, or \QHCQ. Based on a new characterisation of (exact) rewritability, we showed that acyclic queries have  acyclic rewritings, if they have any CQ rewriting.
The same holds for the other three query classes.

We showed that for acyclic queries and views the decision problem, whether an acyclic rewriting exists, is intractable, even  for schemas with bounded arity, but becomes tractable if views have a bounded arity (even with unbounded schema arity) or are free-connex acyclic.

We leave the case of free-connex acyclic views and unbounded schemas open.
 Another interesting open question is the complexity of rewriting problems \rewrprob\ViewClass\QueryClass\RewritingClass with $\RewritingClass \subsetneq \QueryClass$, e.g.\ \rewrprob\ACQ\ACQ\QHCQ.
 So far we have only \classNP-hardness results for problems \rewrprob\ViewClass\QueryClass\ACQ and \rewrprob\ViewClass\QueryClass\HCQ where \ViewClass and \QueryClass encompass all acyclic or all hierarchical queries, respectively.

Finally, it would be interesting to study whether our results can be extended  to other classes of queries that can be evaluated efficiently like conjunctive queries of bounded treewidth.

\bibliographystyle{alphaurl}
\bibliography{main}

\begin{thebibliography}{CGLP20b}

\bibitem[AC19]{DBLP:series/synthesis/2019Afrati}
Foto~N. Afrati and Rada Chirkova.
\newblock {\em Answering Queries Using Views, Second Edition}.
\newblock Synthesis Lectures on Data Management. Morgan {\&} Claypool
  Publishers, 2019.
\newblock \href {https://doi.org/10.2200/S00884ED2V01Y201811DTM054}
  {\path{doi:10.2200/S00884ED2V01Y201811DTM054}}.

\bibitem[AHV95]{DBLP:books/aw/AbiteboulHV95}
Serge Abiteboul, Richard Hull, and Victor Vianu.
\newblock {\em Foundations of Databases}.
\newblock Addison-Wesley, 1995.
\newblock URL: \url{http://webdam.inria.fr/Alice/}.

\bibitem[BB13]{HAL:braultbaron:tel-01081392}
Johann Brault-Baron.
\newblock {\em {De la pertinence de l'{\'e}num{\'e}ration : complexit{\'e} en
  logiques propositionnelle et du premier ordre}}.
\newblock Theses, {Universit{\'e} de Caen}, 2013.
\newblock URL: \url{https://hal.archives-ouvertes.fr/tel-01081392}.

\bibitem[BDG07]{DBLP:conf/csl/BaganDG07}
Guillaume Bagan, Arnaud Durand, and Etienne Grandjean.
\newblock On acyclic conjunctive queries and constant delay enumeration.
\newblock In Jacques Duparc and Thomas~A. Henzinger, editors, {\em Computer
  Science Logic, 21st International Workshop, {CSL} 2007, 16th Annual
  Conference of the EACSL, Lausanne, Switzerland, September 11-15, 2007,
  Proceedings}, volume 4646 of {\em Lecture Notes in Computer Science}, pages
  208--222. Springer Berlin Heidelberg, 2007.
\newblock \href {https://doi.org/10.1007/978-3-540-74915-8_18}
  {\path{doi:10.1007/978-3-540-74915-8_18}}.

\bibitem[BKOZ13]{DBLP:journals/pvldb/BakibayevKOZ13}
Nurzhan Bakibayev, Tom{\'{a}}s Kocisk{\'{y}}, Dan Olteanu, and Jakub Zavodny.
\newblock Aggregation and ordering in factorised databases.
\newblock {\em Proceedings of the {VLDB} Endowment}, 6(14):1990--2001, 2013.
\newblock \href {https://doi.org/10.14778/2556549.2556579}
  {\path{doi:10.14778/2556549.2556579}}.

\bibitem[BKS17]{DBLP:conf/pods/BerkholzKS17}
Christoph Berkholz, Jens Keppeler, and Nicole Schweikardt.
\newblock Answering conjunctive queries under updates.
\newblock In Emanuel Sallinger, Jan~Van den Bussche, and Floris Geerts,
  editors, {\em Proceedings of the 36th {ACM} {SIGMOD-SIGACT-SIGAI} Symposium
  on Principles of Database Systems, {PODS} 2017, Chicago, IL, USA, May 14-19,
  2017}, pages 303--318. {ACM}, 2017.
\newblock \href {https://doi.org/10.1145/3034786.3034789}
  {\path{doi:10.1145/3034786.3034789}}.

\bibitem[BPR17]{DBLP:journals/sigmod/BarceloPR17}
Pablo Barcel{\'{o}}, Andreas Pieris, and Miguel Romero.
\newblock Semantic optimization in tractable classes of conjunctive queries.
\newblock {\em {ACM} {SIGMOD} Record}, 46(2):5--17, 2017.
\newblock \href {https://doi.org/10.1145/3137586.3137588}
  {\path{doi:10.1145/3137586.3137588}}.

\bibitem[CGLP20a]{DBLP:conf/ijcai/ChenGLP20}
Hubie Chen, Georg Gottlob, Matthias Lanzinger, and Reinhard Pichler.
\newblock Semantic width and the fixed-parameter tractability of constraint
  satisfaction problems.
\newblock In Christian Bessiere, editor, {\em Proceedings of the Twenty-Ninth
  International Joint Conference on Artificial Intelligence, {IJCAI} 2020},
  pages 1726--1733. International Joint Conferences on Artificial Intelligence
  Organization, 2020.
\newblock \href {https://doi.org/10.24963/ijcai.2020/239}
  {\path{doi:10.24963/ijcai.2020/239}}.

\bibitem[CGLP20b]{DBLP:journals/corr/abs-2007-14169}
Hubie Chen, Georg Gottlob, Matthias Lanzinger, and Reinhard Pichler.
\newblock Semantic width and the fixed-parameter tractability of constraint
  satisfaction problems.
\newblock {\em CoRR}, abs/2007.14169, 2020.
\newblock URL: \url{https://arxiv.org/abs/2007.14169}, \href
  {http://arxiv.org/abs/2007.14169} {\path{arXiv:2007.14169}}.

\bibitem[CGLV05]{DBLP:conf/icdt/CalvaneseGLV05}
Diego Calvanese, Giuseppe~De Giacomo, Maurizio Lenzerini, and Moshe~Y. Vardi.
\newblock View-based query processing: On the relationship between rewriting,
  answering and losslessness.
\newblock In Thomas Eiter and Leonid Libkin, editors, {\em Database Theory -
  {ICDT} 2005, 10th International Conference, Edinburgh, UK, January 5-7, 2005,
  Proceedings}, volume 3363 of {\em Lecture Notes in Computer Science}, pages
  321--336. Springer Berlin Heidelberg, 2005.
\newblock \href {https://doi.org/10.1007/978-3-540-30570-5_22}
  {\path{doi:10.1007/978-3-540-30570-5_22}}.

\bibitem[CM77]{DBLP:conf/stoc/ChandraM77}
Ashok~K. Chandra and Philip~M. Merlin.
\newblock Optimal implementation of conjunctive queries in relational data
  bases.
\newblock In John~E. Hopcroft, Emily~P. Friedman, and Michael~A. Harrison,
  editors, {\em Proceedings of the 9th Annual {ACM} Symposium on Theory of
  Computing, May 4-6, 1977, Boulder, Colorado, {USA}}, pages 77--90. {ACM}
  Press, 1977.
\newblock \href {https://doi.org/10.1145/800105.803397}
  {\path{doi:10.1145/800105.803397}}.

\bibitem[CR00]{DBLP:journals/tcs/ChekuriR00}
Chandra Chekuri and Anand Rajaraman.
\newblock Conjunctive query containment revisited.
\newblock {\em Theor. Comput. Sci.}, 239(2):211--229, 2000.
\newblock \href {https://doi.org/10.1016/S0304-3975(99)00220-0}
  {\path{doi:10.1016/S0304-3975(99)00220-0}}.

\bibitem[CY12]{DBLP:journals/ftdb/ChirkovaY12}
Rada Chirkova and Jun Yang.
\newblock Materialized views.
\newblock {\em Foundations and Trends{\textregistered} in Databases},
  4(4):295--405, 2012.
\newblock \href {https://doi.org/10.1561/1900000020}
  {\path{doi:10.1561/1900000020}}.

\bibitem[DS07]{DBLP:conf/pods/DalviS07a}
Nilesh~N. Dalvi and Dan Suciu.
\newblock The dichotomy of conjunctive queries on probabilistic structures.
\newblock In Leonid Libkin, editor, {\em Proceedings of the Twenty-Sixth {ACM}
  {SIGACT-SIGMOD-SIGART} Symposium on Principles of Database Systems, June
  11-13, 2007, Beijing, China}, pages 293--302. {ACM}, 2007.
\newblock \href {https://doi.org/10.1145/1265530.1265571}
  {\path{doi:10.1145/1265530.1265571}}.

\bibitem[FO14]{DBLP:conf/pods/FinkO14}
Robert Fink and Dan Olteanu.
\newblock A dichotomy for non-repeating queries with negation in probabilistic
  databases.
\newblock In Richard Hull and Martin Grohe, editors, {\em Proceedings of the
  33rd {ACM} {SIGMOD-SIGACT-SIGART} Symposium on Principles of Database
  Systems, PODS'14, Snowbird, UT, USA, June 22-27, 2014}, pages 144--155.
  {ACM}, 2014.
\newblock \href {https://doi.org/10.1145/2594538.2594549}
  {\path{doi:10.1145/2594538.2594549}}.

\bibitem[FO16]{DBLP:journals/tods/FinkO16}
Robert Fink and Dan Olteanu.
\newblock Dichotomies for queries with negation in probabilistic databases.
\newblock {\em {ACM} Transactions on Database Systems}, 41(1):4:1--4:47, 2016.
\newblock \href {https://doi.org/10.1145/2877203} {\path{doi:10.1145/2877203}}.

\bibitem[GKC06]{DBLP:conf/sigmod/GouKC06}
Gang Gou, Maxim Kormilitsin, and Rada Chirkova.
\newblock Query evaluation using overlapping views: completeness and
  efficiency.
\newblock In Surajit Chaudhuri, Vagelis Hristidis, and Neoklis Polyzotis,
  editors, {\em Proceedings of the {ACM} {SIGMOD} International Conference on
  Management of Data, Chicago, Illinois, USA, June 27-29, 2006}, pages 37--48.
  {ACM}, 2006.
\newblock \href {https://doi.org/10.1145/1142473.1142479}
  {\path{doi:10.1145/1142473.1142479}}.

\bibitem[GKSS22]{geck_et_al:LIPIcs.ICDT.2022.8}
Gaetano Geck, Jens Keppeler, Thomas Schwentick, and Christopher Spinrath.
\newblock {Rewriting with Acyclic Queries: Mind Your Head}.
\newblock In Dan Olteanu and Nils Vortmeier, editors, {\em 25th International
  Conference on Database Theory (ICDT 2022)}, volume 220 of {\em Leibniz
  International Proceedings in Informatics (LIPIcs)}, pages 8:1--8:20,
  Dagstuhl, Germany, 2022. Schloss Dagstuhl -- Leibniz-Zentrum f{\"u}r
  Informatik.
\newblock \href {https://doi.org/10.4230/LIPIcs.ICDT.2022.8}
  {\path{doi:10.4230/LIPIcs.ICDT.2022.8}}.

\bibitem[GLS01]{DBLP:journals/jacm/GottlobLS01}
Georg Gottlob, Nicola Leone, and Francesco Scarcello.
\newblock The complexity of acyclic conjunctive queries.
\newblock {\em Journal of the {ACM}}, 48(3):431--498, 2001.
\newblock \href {https://doi.org/10.1145/382780.382783}
  {\path{doi:10.1145/382780.382783}}.

\bibitem[Hal01]{DBLP:journals/vldb/Halevy01}
Alon~Y. Halevy.
\newblock Answering queries using views: {A} survey.
\newblock {\em The {VLDB} Journal}, 10(4):270--294, 2001.
\newblock \href {https://doi.org/10.1007/s007780100054}
  {\path{doi:10.1007/s007780100054}}.

\bibitem[HKNS15]{DBLP:conf/stoc/HenzingerKNS15}
Monika Henzinger, Sebastian Krinninger, Danupon Nanongkai, and Thatchaphol
  Saranurak.
\newblock Unifying and strengthening hardness for dynamic problems via the
  online matrix-vector multiplication conjecture.
\newblock In Rocco~A. Servedio and Ronitt Rubinfeld, editors, {\em Proceedings
  of the Forty-Seventh Annual {ACM} on Symposium on Theory of Computing, {STOC}
  2015, Portland, OR, USA, June 14-17, 2015}, pages 21--30. {ACM}, 2015.
\newblock \href {https://doi.org/10.1145/2746539.2746609}
  {\path{doi:10.1145/2746539.2746609}}.

\bibitem[HY19]{DBLP:conf/pods/Hu019}
Xiao Hu and Ke~Yi.
\newblock Instance and output optimal parallel algorithms for acyclic joins.
\newblock In Dan Suciu, Sebastian Skritek, and Christoph Koch, editors, {\em
  Proceedings of the 38th {ACM} {SIGMOD-SIGACT-SIGAI} Symposium on Principles
  of Database Systems, {PODS} 2019, Amsterdam, The Netherlands, June 30 - July
  5, 2019}, pages 450--463. {ACM}, 2019.
\newblock \href {https://doi.org/10.1145/3294052.3319698}
  {\path{doi:10.1145/3294052.3319698}}.

\bibitem[IUV17]{DBLP:conf/sigmod/IdrisUV17}
Muhammad Idris, Mart{\'{\i}}n Ugarte, and Stijn Vansummeren.
\newblock The dynamic yannakakis algorithm: Compact and efficient query
  processing under updates.
\newblock In Semih Salihoglu, Wenchao Zhou, Rada Chirkova, Jun Yang, and Dan
  Suciu, editors, {\em Proceedings of the 2017 {ACM} International Conference
  on Management of Data, {SIGMOD} Conference 2017, Chicago, IL, USA, May 14-19,
  2017}, pages 1259--1274. {ACM}, 2017.
\newblock \href {https://doi.org/10.1145/3035918.3064027}
  {\path{doi:10.1145/3035918.3064027}}.

\bibitem[Kep20]{DBLP:phd/dnb/Keppeler20}
Jens Keppeler.
\newblock {\em Answering Conjunctive Queries and {FO+MOD} Queries under
  Updates}.
\newblock PhD thesis, Humboldt University of Berlin, Germany, 2020.
\newblock \href {https://doi.org/10.18452/21483} {\path{doi:10.18452/21483}}.

\bibitem[KNOZ20]{DBLP:conf/pods/0002NOZ20}
Ahmet Kara, Milos Nikolic, Dan Olteanu, and Haozhe Zhang.
\newblock Trade-offs in static and dynamic evaluation of hierarchical queries.
\newblock In Dan Suciu, Yufei Tao, and Zhewei Wei, editors, {\em Proceedings of
  the 39th {ACM} {SIGMOD-SIGACT-SIGAI} Symposium on Principles of Database
  Systems, {PODS} 2020, Portland, OR, USA, June 14-19, 2020}, pages 375--392.
  {ACM}, 2020.
\newblock \href {https://doi.org/10.1145/3375395.3387646}
  {\path{doi:10.1145/3375395.3387646}}.

\bibitem[KS06]{DBLP:conf/ngits/KimelfeldS06}
Benny Kimelfeld and Yehoshua Sagiv.
\newblock Incrementally computing ordered answers of acyclic conjunctive
  queries.
\newblock In Opher Etzion, Tsvi Kuflik, and Amihai Motro, editors, {\em Next
  Generation Information Technologies and Systems, 6th International Workshop,
  {NGITS} 2006, Kibbutz Shefayim, Israel, July 4-6, 2006, Proceedings}, volume
  4032 of {\em Lecture Notes in Computer Science}, pages 141--152. Springer
  Berlin Heidelberg, 2006.
\newblock \href {https://doi.org/10.1007/11780991_13}
  {\path{doi:10.1007/11780991_13}}.

\bibitem[KS11]{DBLP:conf/pods/KoutrisS11}
Paraschos Koutris and Dan Suciu.
\newblock Parallel evaluation of conjunctive queries.
\newblock In Maurizio Lenzerini and Thomas Schwentick, editors, {\em
  Proceedings of the 30th {ACM} {SIGMOD-SIGACT-SIGART} Symposium on Principles
  of Database Systems, {PODS} 2011, June 12-16, 2011, Athens, Greece}, pages
  223--234. {ACM}, 2011.
\newblock \href {https://doi.org/10.1145/1989284.1989310}
  {\path{doi:10.1145/1989284.1989310}}.

\bibitem[LMSS95]{DBLP:conf/pods/LevyMSS95}
Alon~Y. Levy, Alberto~O. Mendelzon, Yehoshua Sagiv, and Divesh Srivastava.
\newblock Answering queries using views.
\newblock In Mihalis Yannakakis and Serge Abiteboul, editors, {\em Proceedings
  of the Fourteenth {ACM} {SIGACT-SIGMOD-SIGART} Symposium on Principles of
  Database Systems, May 22-25, 1995, San Jose, California, {USA}}, pages
  95--104. {ACM} Press, 1995.
\newblock \href {https://doi.org/10.1145/212433.220198}
  {\path{doi:10.1145/212433.220198}}.

\bibitem[NSV10]{DBLP:journals/tods/NashSV10}
Alan Nash, Luc Segoufin, and Victor Vianu.
\newblock Views and queries: Determinacy and rewriting.
\newblock {\em {ACM} Transactions on Database Systems}, 35(3):21:1--21:41,
  2010.
\newblock \href {https://doi.org/10.1145/1806907.1806913}
  {\path{doi:10.1145/1806907.1806913}}.

\bibitem[OZ15]{DBLP:journals/tods/OlteanuZ15}
Dan Olteanu and Jakub Z{\'{a}}vodn{\'{y}}.
\newblock Size bounds for factorised representations of query results.
\newblock {\em {ACM} Transactions on Database Systems}, 40(1):2:1--2:44, 2015.
\newblock \href {https://doi.org/10.1145/2656335} {\path{doi:10.1145/2656335}}.

\bibitem[PH01]{DBLP:journals/vldb/PottingerH01}
Rachel Pottinger and Alon~Y. Halevy.
\newblock Minicon: {A} scalable algorithm for answering queries using views.
\newblock {\em The {VLDB} Journal}, 10(2-3):182--198, 2001.
\newblock \href {https://doi.org/10.1007/s007780100048}
  {\path{doi:10.1007/s007780100048}}.

\bibitem[SOC16]{DBLP:conf/sigmod/SchleichOC16}
Maximilian Schleich, Dan Olteanu, and Radu Ciucanu.
\newblock Learning linear regression models over factorized joins.
\newblock In Fatma {\"{O}}zcan, Georgia Koutrika, and Sam Madden, editors, {\em
  Proceedings of the 2016 International Conference on Management of Data,
  {SIGMOD} Conference 2016, San Francisco, CA, USA, June 26 - July 01, 2016},
  pages 3--18. {ACM}, 2016.
\newblock \href {https://doi.org/10.1145/2882903.2882939}
  {\path{doi:10.1145/2882903.2882939}}.

\bibitem[Yan81]{DBLP:conf/vldb/Yannakakis81}
Mihalis Yannakakis.
\newblock Algorithms for acyclic database schemes.
\newblock In {\em Very Large Data Bases, 7th International Conference,
  September 9-11, 1981, Cannes, France, Proceedings}, pages 82--94. {IEEE}
  Computer Society, 1981.

\end{thebibliography}

\end{document}